\documentclass[twocolumn,twocolappendix]{aastex63b}
\pdfoutput=1 
\usepackage{amsmath,amstext}
\usepackage{mathtools}
\usepackage[T1]{fontenc}
\usepackage{apjfonts}
\usepackage[figure,figure*,table,table*]{hypcap}
\usepackage[figuresright]{rotating}
\usepackage{xspace}
\usepackage{verbatim}
\usepackage{needspace}
\usepackage{changepage}
\usepackage{appendix}

\graphicspath{{figures/}{./}}

\newcommand{\lb}[1]{\textcolor{blue}{LB: #1}}

\newcommand{\E}[1]{\ensuremath{\langle #1 \rangle}}
\newcommand{\EF}{\ensuremath{\mathcal{E}}}
\newcommand{\Ei}{\ensuremath{\mathrm{Ei}}}
\newcommand{\Eop}[1]{\ensuremath{\mathrm{E}\left[#1\right]}}
\newcommand{\Arg}[1]{\ensuremath{\mathrm{Arg}\left[#1\right]}}

\newcommand{\cc}[2]{\ensuremath{#1^{\vphantom{\ast}} #2^{\ast}}}
\newcommand{\gamgam}[2]{\ensuremath{\gamma_#1^{\vphantom{\ast}} \gamma_#2^{\ast}}}
\newcommand{\transpose}{\top}

\newcommand{\bhi}{Black Hole Initiative, Harvard University, 20 Garden Street, Cambridge, MA 02138, USA}

\newcommand{\cfa}{Center for Astrophysics $|$ Harvard \& Smithsonian, 60 Garden Street, Cambridge, MA 02138, USA}
\newcommand{\arizona}{University of Arizona, 933 North Cherry Avenue, Tucson, AZ 85721, USA}
\newcommand{\princeton}{Princeton Center for Theoretical Science, Jadwin Hall, Princeton University, Princeton, NJ 08544, USA}
\newcommand{\einstein}{NASA Hubble Fellowship Program, Einstein Fellow}

\usepackage{etoolbox}
\makeatletter
\patchcmd{\hyper@makecurrent}{%
    \ifx\Hy@param\Hy@chapterstring
        \let\Hy@param\Hy@chapapp
    \fi
}{%
    \iftoggle{inappendix}{
        \@checkappendixparam{chapter}%
        \@checkappendixparam{section}%
        \@checkappendixparam{subsection}%
        \@checkappendixparam{subsubsection}%
        \@checkappendixparam{paragraph}%
        \@checkappendixparam{subparagraph}%
    }{}%
}{}{\errmessage{failed to patch}}

\newcommand*{\@checkappendixparam}[1]{%
    \def\@checkappendixparamtmp{#1}%
    \ifx\Hy@param\@checkappendixparamtmp
        \let\Hy@param\Hy@appendixstring
    \fi
}
\makeatletter

\newtoggle{inappendix}
\togglefalse{inappendix}

\apptocmd{\appendix}{\toggletrue{inappendix}}{}{\errmessage{failed to patch}}
\apptocmd{\subappendices}{\toggletrue{inappendix}}{}{\errmessage{failed to patch}}

\usepackage{bm}

\newcommand{\R}{\ensuremath{\textbf{R}}\xspace}

\newcommand{\gcn}{\ensuremath{N}\xspace}  
\newcommand{\vcn}{\ensuremath{B}\xspace}  
\newcommand{\gpn}{\ensuremath{N}\xspace}  
\newcommand{\vpn}{\ensuremath{B}\xspace}  
\newcommand{\cpn}{\ensuremath{T}\xspace}  
\newcommand{\gan}{\ensuremath{N}\xspace}  
\newcommand{\van}{\ensuremath{B}\xspace}  
\newcommand{\can}{\ensuremath{Q}\xspace}  
\newcommand{\lgan}{\ensuremath{N}\xspace} 
\newcommand{\lvan}{\ensuremath{B}\xspace} 
\newcommand{\lcan}{\ensuremath{Q}\xspace} 

\newcommand{\gcs}[1][i]{\ensuremath{\gamma_{#1}}\xspace}
\newcommand{\vcs}[1][ij]{\ensuremath{V_{#1}}\xspace}
\newcommand{\gps}[1][i]{\ensuremath{\theta_{#1}}\xspace}
\newcommand{\vps}[1][ij]{\ensuremath{\phi_{#1}}\xspace}
\newcommand{\cps}[1][ijk]{\ensuremath{\psi_{#1}}\xspace}
\newcommand{\gas}[1][i]{\ensuremath{G_{#1}}\xspace}
\newcommand{\vas}[1][ij]{\ensuremath{A_{#1}}\xspace}
\newcommand{\cas}[1][ijkl]{\ensuremath{C_{#1}}\xspace}
\newcommand{\lgas}[1][i]{\ensuremath{g_{#1}}\xspace}
\newcommand{\lvas}[1][ij]{\ensuremath{a_{#1}}\xspace}
\newcommand{\lcas}[1][ijkl]{\ensuremath{c_{#1}}\xspace}

\newcommand{\gcv}{\ensuremath{\dots}\xspace}
\newcommand{\vcv}{\ensuremath{\dots}\xspace}
\newcommand{\gpv}{\ensuremath{\bm{\theta}}\xspace}
\newcommand{\vpv}{\ensuremath{\bm{\phi}}\xspace}
\newcommand{\cpv}{\ensuremath{\bm{\psi}}\xspace}
\newcommand{\mpv}{\ensuremath{\bm{\psi}_+}\xspace}
\newcommand{\gav}{\ensuremath{\dots}\xspace}
\newcommand{\vav}{\ensuremath{\dots}\xspace}
\newcommand{\cav}{\ensuremath{\dots}\xspace}
\newcommand{\lgav}{\ensuremath{\textbf{\textit{g}}}\xspace}
\newcommand{\lvav}{\ensuremath{\textbf{\textit{a}}}\xspace}
\newcommand{\lmav}{\ensuremath{\textbf{\textit{c}}_+}\xspace}
\newcommand{\lcav}{\ensuremath{\textbf{\textit{c}}}\xspace}

\newcommand{\gcms}[1][i]{\ensuremath{\hat{\gamma}_{#1}}\xspace}
\newcommand{\vcms}[1][ij]{\ensuremath{\hat{V}_{#1}}\xspace}
\newcommand{\gpms}[1][i]{\ensuremath{\hat{\theta}_{#1}}\xspace}
\newcommand{\vpms}[1][ij]{\ensuremath{\hat{\phi}_{#1}}\xspace}
\newcommand{\cpms}[1][ijk]{\ensuremath{\hat{\psi}_{#1}}\xspace}
\newcommand{\gams}[1][i]{\ensuremath{\hat{G}_{#1}}\xspace}
\newcommand{\vams}[1][ij]{\ensuremath{\hat{A}_{#1}}\xspace}
\newcommand{\cams}[1][ijkl]{\ensuremath{\hat{C}_{#1}}\xspace}
\newcommand{\lgams}[1][i]{\ensuremath{\hat{g}_{#1}}\xspace}
\newcommand{\lvams}[1][ij]{\ensuremath{\hat{a}_{#1}}\xspace}
\newcommand{\lcams}[1][ijkl]{\ensuremath{\hat{c}_{#1}}\xspace}

\newcommand{\gcmv}{\ensuremath{\dots}\xspace}
\newcommand{\vcmv}{\ensuremath{\dots}\xspace}
\newcommand{\gpmv}{\ensuremath{\mathbf{\skew{+2}\hat{\bm{\theta}}}}\xspace}
\newcommand{\vpmv}{\ensuremath{\mathbf{\skew{+2}\hat{\bm{\phi}}}}\xspace}
\newcommand{\cpmv}{\ensuremath{\mathbf{\skew{+2}\hat{\bm{\psi}}}}\xspace}
\newcommand{\gamv}{\ensuremath{\dots}\xspace}
\newcommand{\vamv}{\ensuremath{\dots}\xspace}
\newcommand{\camv}{\ensuremath{\dots}\xspace}
\newcommand{\lgamv}{\ensuremath{\mathbf{\hat{\textbf{\textit{g}}}}}\xspace}
\newcommand{\lvamv}{\ensuremath{\mathbf{\hat{\textbf{\textit{a}}}}}\xspace}
\newcommand{\lcamv}{\ensuremath{\mathbf{\hat{\textbf{\textit{c}}}}}\xspace}

\newcommand{\gcrs}[1][i]{\ensuremath{\tilde{\gamma}_{#1}}\xspace}
\newcommand{\vcrs}[1][ij]{\ensuremath{\tilde{V}_{#1}}\xspace}
\newcommand{\gprs}[1][i]{\ensuremath{\tilde{\theta}_{#1}}\xspace}
\newcommand{\vprs}[1][ij]{\ensuremath{\tilde{\phi}_{#1}}\xspace}
\newcommand{\cprs}[1][ijk]{\ensuremath{\tilde{\psi}_{#1}}\xspace}
\newcommand{\gars}[1][i]{\ensuremath{\tilde{G}_{#1}}\xspace}
\newcommand{\vars}[1][ij]{\ensuremath{\tilde{A}_{#1}}\xspace}
\newcommand{\cars}[1][ijkl]{\ensuremath{\tilde{C}_{#1}}\xspace}
\newcommand{\lgars}[1][i]{\ensuremath{\tilde{g}_{#1}}\xspace}
\newcommand{\lvars}[1][ij]{\ensuremath{\tilde{a}_{#1}}\xspace}
\newcommand{\lcars}[1][ijkl]{\ensuremath{\tilde{c}_{#1}}\xspace}

\newcommand{\gcrv}{\ensuremath{\dots}\xspace}
\newcommand{\vcrv}{\ensuremath{\dots}\xspace}
\newcommand{\gprv}{\ensuremath{\bm{\skew{+2}\tilde{\theta}}}\xspace}
\newcommand{\vprv}{\ensuremath{\bm{\skew{+2}\tilde{\phi}}}\xspace}
\newcommand{\cprv}{\ensuremath{\bm{\skew{+2}\tilde{\psi}}}\xspace}
\newcommand{\mprv}{\ensuremath{\bm{\skew{+2}\tilde{\psi}}_+}}
\newcommand{\garv}{\ensuremath{\dots}\xspace}
\newcommand{\varv}{\ensuremath{\dots}\xspace}
\newcommand{\carv}{\ensuremath{\dots}\xspace}
\newcommand{\lgarv}{\ensuremath{\bm{\tilde{\textbf{\textit{g}}}}}\xspace}
\newcommand{\lvarv}{\ensuremath{\bm{\tilde{\textbf{\textit{a}}}}}\xspace}
\newcommand{\lcarv}{\ensuremath{\bm{\tilde{\textbf{\textit{c}}}}}\xspace}

\newcommand{\gcvar}{\ensuremath{\dots}\xspace}
\newcommand{\vcvar}[1][ij]{\ensuremath{\sigma_{V,#1}^2}\xspace} 
\newcommand{\gpvar}[1][i]{\ensuremath{\sigma_{\theta,#1}^2}}
\newcommand{\vpvar}[1][ij]{\ensuremath{\sigma_{#1}^2}\xspace}
\newcommand{\cpvar}[1][ijk]{\ensuremath{\sigma_{#1}^2}\xspace}
\newcommand{\gavar}{\ensuremath{\dots}\xspace}
\newcommand{\vavar}{\ensuremath{\dots}\xspace}
\newcommand{\cavar}{\ensuremath{\dots}\xspace}
\newcommand{\lgavar}[1][i]{\ensuremath{\sigma_{g,#1}^2}\xspace}
\newcommand{\lvavar}[1][ij]{\ensuremath{\sigma_{#1}^2}\xspace}
\newcommand{\lcavar}[1][ijkl]{\ensuremath{\sigma_{#1}^2}\xspace}

\newcommand{\gccov}{\ensuremath{\dots}\xspace}
\newcommand{\vccov}{\ensuremath{\dots}\xspace}
\newcommand{\gpcov}{\ensuremath{\bm{\Sigma_{\theta}}}\xspace}
\newcommand{\vpcov}{\ensuremath{\bm{\Sigma_{\phi}}}\xspace}
\newcommand{\cpcov}{\ensuremath{\bm{\Sigma_{\psi}}}\xspace}
\newcommand{\mpcov}{\ensuremath{\bm{\Sigma_{\psi+}}}\xspace}
\newcommand{\gacov}{\ensuremath{\dots}\xspace}
\newcommand{\vacov}{\ensuremath{\dots}\xspace}
\newcommand{\cacov}{\ensuremath{\dots}\xspace}
\newcommand{\lgacov}{\ensuremath{\bm{\Sigma}_{\textbf{\textit{g}}}}\xspace}
\newcommand{\lvacov}{\ensuremath{\bm{\Sigma}_{\textbf{\textit{a}}}}\xspace}
\newcommand{\lcacov}{\ensuremath{\bm{\Sigma}_{\textbf{\textit{c}}}}\xspace}
\newcommand{\lcacovdiag}{\ensuremath{\bm{\Sigma}_{\textbf{\textit{c}}\mathrm{,diag}}}\xspace}

\newcommand{\gcdm}{\ensuremath{\dots}\xspace}
\newcommand{\vcd}{\ensuremath{\dots}\xspace}
\newcommand{\gpd}{\ensuremath{\dots}\xspace}
\newcommand{\vpd}{\ensuremath{\raisebox{-0.25pt}{\fontsize{10pt}{0}\selectfont $\bm{{\Phi}}$}}\xspace}
\newcommand{\cpd}{\ensuremath{\raisebox{-0.25pt}{\fontsize{10pt}{0}\selectfont $\bm{{\Psi}}$}}\xspace}
\newcommand{\mpd}{\ensuremath{\bm{\Psi}_+}}
\newcommand{\gad}{\ensuremath{\dots}\xspace}
\newcommand{\vad}{\ensuremath{\dots}\xspace}
\newcommand{\cad}{\ensuremath{\dots}\xspace}
\newcommand{\lgad}{\ensuremath{\dots}\xspace}
\newcommand{\lvad}{\ensuremath{\textbf{A}}\xspace}
\newcommand{\lcad}{\ensuremath{\textbf{C}}\xspace}
\newcommand{\lmad}{\ensuremath{\textbf{C}_+}\xspace}

\newcommand{\vpchisq}{\ensuremath{\chi_{\phi}^2}\xspace}
\newcommand{\cpchisq}{\ensuremath{\chi_{\psi}^2}\xspace}
\newcommand{\lvachisq}{\ensuremath{\chi_a^2}\xspace}
\newcommand{\lcachisq}{\ensuremath{\chi_c^2}\xspace}

\shorttitle{Closure statistics in interferometric data}

\begin{document}

\title{Closure statistics in interferometric data}

\author[0000-0002-9030-642X]{Lindy Blackburn}
\affiliation{\cfa}
\affiliation{\bhi}
\author[0000-0002-5278-9221]{Dominic W. Pesce}
\affiliation{\cfa}
\affiliation{\bhi}
\author[0000-0002-4120-3029]{Michael D. Johnson}
\affiliation{\cfa}
\affiliation{\bhi}
\author[0000-0002-8635-4242]{Maciek Wielgus}
\affiliation{\cfa}
\affiliation{\bhi}
\author[0000-0003-2966-6220]{Andrew A. Chael}
\affiliation{\princeton}
\affiliation{\einstein}
\author[0000-0001-6820-9941]{Pierre Christian}
\affiliation{\arizona}
\author[0000-0002-9031-0904]{Sheperd S. Doeleman}
\affiliation{\cfa}
\affiliation{\bhi}

\begin{abstract}

Interferometric visibilities, reflecting the complex correlations between signals recorded at antennas in an interferometric array, carry information about the angular structure of a distant source. While unknown antenna gains in both amplitude and phase can prevent direct interpretation of these measurements,
certain combinations of visibilities called closure phases and closure amplitudes are independent of antenna gains and provide a convenient set of robust observables.
However, these closure quantities have subtle noise properties and are generally both linearly and statistically dependent. 
These complications have obstructed the proper use of closure quantities in interferometric analysis, and they have obscured the relationship between analysis with closure quantities and other analysis techniques such as self calibration.
We review the statistics of closure quantities, noting common pitfalls that arise when approaching low signal-to-noise due to the nonlinear propagation of statistical errors.
We then develop a strategy for isolating and fitting to the independent degrees of freedom captured by the closure quantities through explicit construction of linearly independent sets of quantities along with their noise covariance in the Gaussian limit, valid for moderate signal-to-noise, and we demonstrate that model fits have biased posteriors when this covariance is ignored. Finally, we introduce a unified procedure for fitting to both closure information and partially calibrated visibilities, and we demonstrate both analytically and numerically the direct equivalence of inference based on closure quantities to that based on self calibration of complex visibilities with unconstrained antenna gains.

\end{abstract}


\keywords{techniques: interferometric --- methods: statistical --- techniques: high angular resolution}

\tableofcontents

\clearpage

\setcounter{footnote}{5}

\Needspace*{4\baselineskip}
\section{Introduction}

Interferometric observations allow diffraction-limited resolution on angular scales that are inaccessible to single-element systems. However, interferometers have the limitation of only sparsely sampling information in the so-called visibility domain. While measured visibilities have simple and deterministic thermal noise, they also have complex systematic errors. These systematic errors manifest as variations in visibility amplitudes and phases on many timescales, representing limitations imposed by a broad range of sources including the constituent interferometer elements, reference frequencies, and atmosphere.

The dominant systematic errors are station-based effects, corresponding to multiplicative complex gain factors. In this case, ``closure'' quantities can be constructed, which are independent of the station-based systematic calibration errors. Specifically, closure phases consist of a directed sum of visibility phases around a closed triangle joining three stations \citep{Jennison1958,rogers1974}, while closure amplitudes are the quotient of two visibility products involving four stations \citep{Twiss_1960,Readhead1980}. These quantities have found particular utility in very long baseline radio interferometry (VLBI) where array sparsity motivates the use of model independent observables. For an array with $N$ stations, one can form ${\sim}N^3$ closure phases and ${\sim}N^4$ closure amplitudes from the original set of ${\sim}N^2$ visibilities. However, there are at most $(N-1)(N-2)/2$ degrees of freedom in the closure phases and $N(N-3)/2$ degrees of freedom in the closure amplitudes.
The necessary degeneracy between the full sets of closure quantities is captured by the structure of their covariance.

Closure quantities are useful for interferometric analysis, especially model fitting and imaging \citep[e.g.,][]{Readhead1978,Chael_2018}, because they eliminate the need to model station gain systematics, and their error budget can be determined from first principles. Yet, despite the fundamental importance of closure quantities for interferometry, there is widespread variation in the literature concerning their properties, best practices when utilizing closure quantities, their relationship with standard analyses such as self calibration, and the role of linearly independent sets of closure quantities (which are not necessarily statistically independent).
Moreover, most analyses to date ignore the covariance between closure quantities, which can be significant; although, covariance of closure phases has been studied in the optical interferometry community \citep[e.g.,][]{kulkarni1991,martinache2010,ireland2013}.

Here, we provide a rigorous foundation for analysis using closure quantities, and we give procedures for selecting nonredundant sets of closure phases and amplitudes. We demonstrate that, when covariance is correctly accounted for, these nonredundant sets carry the full information of the complete sets of closure quantities. Moreover, in the limit of completely unconstrained station gains, we show that analysis of closure quantities is identical to analysis of complex visibilities with gain marginalization. We also give procedures for selecting nonredundant sets that minimize covariance, and we demonstrate the effects of covariance among closure products using simulated data from simple models.
 
We begin, in Section~\ref{sec:closurequantitieserrors}, by discussing thermal and systematic errors in interferometric measurements, and we assess the conditions under which errors on closure quantities can be approximated as Gaussian. Next, in Section~\ref{sec::Independence_of_Closure}, we evaluate the covariance among closure quantities and give prescriptions for selecting nonredundant sets of closure quantities.  Then, in Section~\ref{sec:MonteCarlo}, we apply our results to simple model fits using closure quantities and demonstrate the role of nonredundant sets and covariance among closure products. We summarize our results in Section~\ref{sec:summary}. The notation used throughout the paper is described in \autoref{tab:notation}.

\section{Closure quantities and errors}
\label{sec:closurequantitieserrors}

\subsection{Inteferometric visibility and gain}
\label{sec:interferometricvisibilities}

An interferometric array aims to measure the complex coherence function of the electric field, or {\em
visibility}
$V_{ij} = \Eop{\cc{\EF_i}{\EF_j}}$
(represented here in the frequency domain, with $\EF$ in units such that expectation value $S_\nu \sim \Eop{|\EF_i|^2}$ is the electromagnetic flux spectral density), from a distant
source at two locations $i$ and $j$ in the plane of propagation. $V_{ij}$ samples a Fourier component of the brightness distribution on the sky \citep[via the van Cittert--Zernike theorem,][]{vancittert,zernike,tms}, with spatial frequency corresponding to the projected baseline length in units of observing wavelength.
In the idealized case, the field
is measured without any attenuation or propagation delays (e.g., through
atmosphere). In practice, the measured complex signal $v_i$ at an antenna $i$ can be modeled
as the
idealized incident aligned electric field
$\EF_i$ subject to a linear complex gain factor $\gamma_i$, plus additive zero-mean circularly symmetric complex Gaussian noise $n_i$
\begin{equation}
\label{eqn:vi}
v_i = \gamma_i \, \EF_i + n_i.
\end{equation}

In continuum-VLBI, the noise power typically exceeds signal power by a large
factor: $\Eop{|n|^2} \gg \Eop{|\gamma \EF|^2}$. The gain for a particular antenna feed is a
function of time and frequency $\gamma=\gamma(t, f)$ and while a variety of simplifying
assumptions and factorizations can be made, the gain is often not known {\em a
priori} to a high degree of precision. Thus, the fundamental observable is not
source visibility $V_{ij}$, but cross-covariance $r_{ij}$ between pairs of antennas:
\begin{equation}
\label{eqn:rij}
r_{ij} = \Eop{\cc{v_i}{v_j}} = \cc{\gamma_i}{\gamma_j}\,V_{ij}.
\end{equation}
If the signals $v_i$ and $v_j$ are normalized by their noise power such that
$\Eop{|n_i|^2} = 1$, $r_{ij}$ is the correlation coefficient, and
$\text{SEFD}_i = 1 / |\gamma_i|^2$ represents the {\em system-equivalent flux
density} (noise power in units of flux above the atmosphere).

Relating the measured correlation coefficients $r_{ij}$ to source visibilities
$V_{ij}$ is the process of {\em calibration} and may include estimating the
magnitude of $|\gamma_i|$ through the observation of bright flux calibrators,
measuring differential phase $\Arg{\cc{\gamma_i}{\gamma_j}}$ by observing phase
calibrators with known structure, or by process of {\em self calibration}
where the gains $\gamma_i$ are solved simultaneously with unknown source model parameters. For a VLBI array at millimeter wavelengths,
calibration is made difficult by the strong and
rapidly changing atmospheric effects and by the lack of bright compact calibration
sources of known structure. Amplitude and phase gain systematics often dominate
over the thermal (statistical) noise that arises from estimating $r_{ij}$ over
finite time and bandwidth.
\subsection{Closure phase and closure amplitude}

Closure quantities are special combinations of correlation measurements taken
over closed loops in an antenna network. They are able to cancel out
station-based gains $\gamma_i$, giving observables that depend only on intrinsic
source parameters.

A {\em closure phase} is the sum of measured phases around a closed triangle of
baselines,
\begin{equation}
\label{eqn:cphase}
\phi_{123} = \phi_{12} + \phi_{23} + \phi_{31},
\end{equation}
where $\phi_{12}$ is the phase on baseline 1--2,
\begin{equation}
\phi_{12} = \Arg{r_{12}}. \label{eqn:VisibilityPhase}
\end{equation}
Written as the phase of the complex {\em bispectrum} (triple product) $V_{123} = V_{12}V_{23}V_{31}$, we
see that a closure phase is independent of arbitrary phase gain $\Arg{\gamma_i}$,
\begin{align}
\Arg{r_{12}r_{23}r_{31}} &= \Arg{\gamgam{1}{2} V_{12} \, \gamgam{2}{3} V_{23} \, \gamgam{3}{1} V_{31}} \\
 &= \Arg{V_{12}V_{23}V_{31}},
\end{align}
since every gain term on the right-hand side is multiplied by its complex
conjugate.

A set of correlation coefficients connecting four sites in a closed quadrangle
can be used to calculate a {\em closure amplitude},
\begin{equation}
\label{eqn:camp}
\frac{|r_{12}||r_{34}|}{|r_{13}||r_{24}|} =
\frac{|\gamgam{1}{2}V_{12}||\gamgam{3}{4}V_{34}|}{|\gamgam{1}{3}V_{13}||\gamgam{2}{4}V_{24}|} =
\frac{|V_{12}||V_{34}|}{|V_{13}||V_{24}|}.
\end{equation}
In this case, we see that closure amplitude is independent of arbitrary
amplitude gain $|\gamma_i|$ since each station gain amplitude term appears in both the
numerator and denominator.\footnote{This quadrangle can be easily cast as a complex closure quantity, but doing so provides no phase information beyond the set of complex bispectra.
When taken over the four polarization feeds of a single baseline however, e.g., $(V_\mathrm{LR}\,V_\mathrm{RL})/(V_\mathrm{LL}\,V_\mathrm{RR})$, such a construction can provide some information about delay closure and/or polarization fraction.
}

By canceling station gains, the closure quantities are able to isolate
measurement degrees of freedom that are independent from gains,
and they provide observables that are accurate to the thermal-noise
limit or to residual baseline errors \citep{massi1991}, which are typically much smaller than the station errors. Thus, they are particularly valuable when systematic gain uncertainty is
much larger than statistical uncertainty. We will see in subsequent sections that the closure phases and closure amplitudes capture all of the
gain-invariant degrees of freedom from the baseline visibilities, at the cost of removing any prior information about the gains.

\subsection{Statistical thermal noise}
\label{sec:thermalnoise}

When estimated from actual data, the closure quantities and associated correlation coefficients from Equations~\mbox{\ref{eqn:rij}--\ref{eqn:camp}} must be averaged
over a finite time and bandwidth in order to accumulate signal-to-noise ratio (S/N). A measurement of $r_{ij}$ taken over integration time $\Delta t$ and bandwidth $\Delta\nu$ averages $\Delta t\,\Delta\nu$ independent complex samples (finite average denoted with \E{}), and includes contributions from both the source and the independent zero-mean (and normalized) thermal noise at each antenna,
\begin{equation}
    r_{ij} = \breve{r}_{ij} + \E{\cc{n_i}{n_j}}.
    \label{eqn:rij_measured}
\end{equation}
We have introduced a breve accent $\breve{r}_{ij} = \breve{\gamma}_i \breve{\gamma}_j^\ast \breve{V}_{ij}$ to distinguish underlying (ground-truth) values from those that are subject to statistical or systematic errors.\footnote{In principle, the source contribution $\mathrm{E}\big[\cc{\EF_i}{\EF_j}\big]$ is also subject to statistical fluctuations, i.e. the {\em self noise} of \citet{Kulkarni_1989}, but these are strongly subdominant to uncertainties from thermal noise in the weak-signal limit $\Eop{|n|^2} \gg \Eop{|\gamma \EF|^2}$ (Section~\ref{sec:interferometricvisibilities}).} Under the previously adopted normalization $\Eop{|n_i|^2} = 1$ (see \autoref{eqn:rij}), the variance for one sample of correlated complex noise $\Eop{|\cc{n_i}{n_j}|^2} = 1$, and the variance in one component (real or imaginary) of the averaged complex noise correlation $\E{\cc{n_i}{n_j}}$ is then
\begin{equation}
    \sigma_{r,\,ij}^2 = \frac{\Eop{|\E{\cc{n_i}{n_j}}|^2}}{2} = \frac{1}{2\,\Delta t\,\Delta\nu},
    \label{eqn:sigma_ij}
\end{equation}
where the amount of time-frequency averaging to reduce $\sigma_r^2$ is ultimately constrained by assumptions regarding station gain variability and source model variability.

The underlying signal-to-noise $\breve{\rho}$ of a correlation coefficient amplitude $|\breve{r}|$ under time-frequency averaging, and assuming negligible evolution of model visibility due to source structure or residual systematics, is
\begin{equation}
\label{eqn:rhorij}
\breve{\rho} = |\breve{r}| / \sigma_r.
\end{equation}
This is taken in the moderate-to-high $\breve{\rho}$ limit where it is meaningful to measure noise along just one of the complex components. A correlated flux density of 1 Jy with typical geometric mean SEFD of $10^4$ Jy would give an expected correlation coefficient of $10^{-4}$ and an S/N of 4.5 over 1 GHz of bandwidth in 1 s of integration time. At $\breve{\rho} < 1$, the ability to measure correlation phase and amplitude degrades rapidly \citep{rogers1995}, so that the minimum acceptable integration time is fundamentally limited by a combination of source strength, bandwidth, and collecting area.

At the same time, any uncompensated complex gain $\breve{\gamma}$ (from \autoref{eqn:vi} and \ref{eqn:rij}) must be stable over the averaging timescale to both ensure a meaningful measurement and to avoid phase decoherence while vector averaging complex baseline visibility. At millimeter and submillimeter observing wavelengths, phase decoherence due to atmospheric turbulence occurs on timescales of seconds. The requirement that $\breve{\rho} > 1$ over an averaging time $\Delta t$ and bandwidth $\Delta\nu$ where gain variation remains negligible sets the observational constraints where the use of closure quantities is particularly effective. Gain variation over the frequency bandpass is generally a stable instrumental effect that can be well measured and calibrated out. Similarly, relative complex gain between two orthogonal feeds in an antenna, e.g., $\gamma_R/\gamma_L$ or $\gamma_X/\gamma_Y$, is generally stable and can be calibrated, so that synthesized combinations of correlation products such as Stokes $I = V_{RR} + V_{LL}$ are also characterized by station-based residual gains that close.

For observations at high radio frequencies, rapid phase gain variability in time due to the atmosphere is a primary driver of efforts to expand the collecting area and instantaneous bandwidth of mm-VLBI arrays such as the Event Horizon Telescope \citep[EHT;][]{PaperII,PaperIII}.
So long as there exists at least one high-S/N baseline to a given site connecting it to the array phase center, the atmospheric phase variations can generally be solved for and removed. This allows for longer coherent integration on weak baselines to that site \citep{Blackburn2019}. In the case of the EHT, this condition is generally satisfied by the presence of the highly sensitive ALMA in the array.

In the following subsections, we discuss the consequences of low S/N on characterization of errors in phase and amplitude, and also the propagation of these errors across derived closure quantities.
However, we do not explore optimal averaging strategies for the generation of closure phase and closure amplitudes. This would require assuming a prior model for gain variability. Rather, we assume there exist some $\Delta t$ and $\Delta \nu$ such that $\breve{\rho} > 1$ is maintained on all baselines, and over which $\breve{\gamma}$ can be made reasonably stable.

\Needspace*{4\baselineskip}
\subsection{Non-Gaussian errors at low S/N}
\label{sec:nongaussian}

The observed correlation coefficients (Equations~\ref{eqn:rij_measured} and \ref{eqn:sigma_ij}) are subject to measurement noise that is complex and independent in real and imaginary components.
While this implies that complex visibility is
the natural measurement space for correlation observables, gain systematics are
largely separable into amplitude factors (e.g., aperture efficiency) and
phase factors (e.g., variable path delay), and this is reflected in the way
closure amplitude and closure phase are formed.

The transformation from errors in real and imaginary
coefficients to errors in amplitude and phase is only effectively linear for $\breve{\rho} \gg 1$. A consequence is that the statistical
error budget of closure quantities becomes progressively non-Gaussian as $\breve{\rho}$ becomes small. This is particularly severe for the case of reciprocal amplitude, which is a necessary component of closure amplitude (\autoref{eqn:camp}). Heavy tails in the distribution for reciprocal amplitude are one motivation to move to log-closure amplitudes which place the numerator and denominator of a closure amplitude on equal footing.

\begin{figure}
\centering
\includegraphics[width=3.2in]{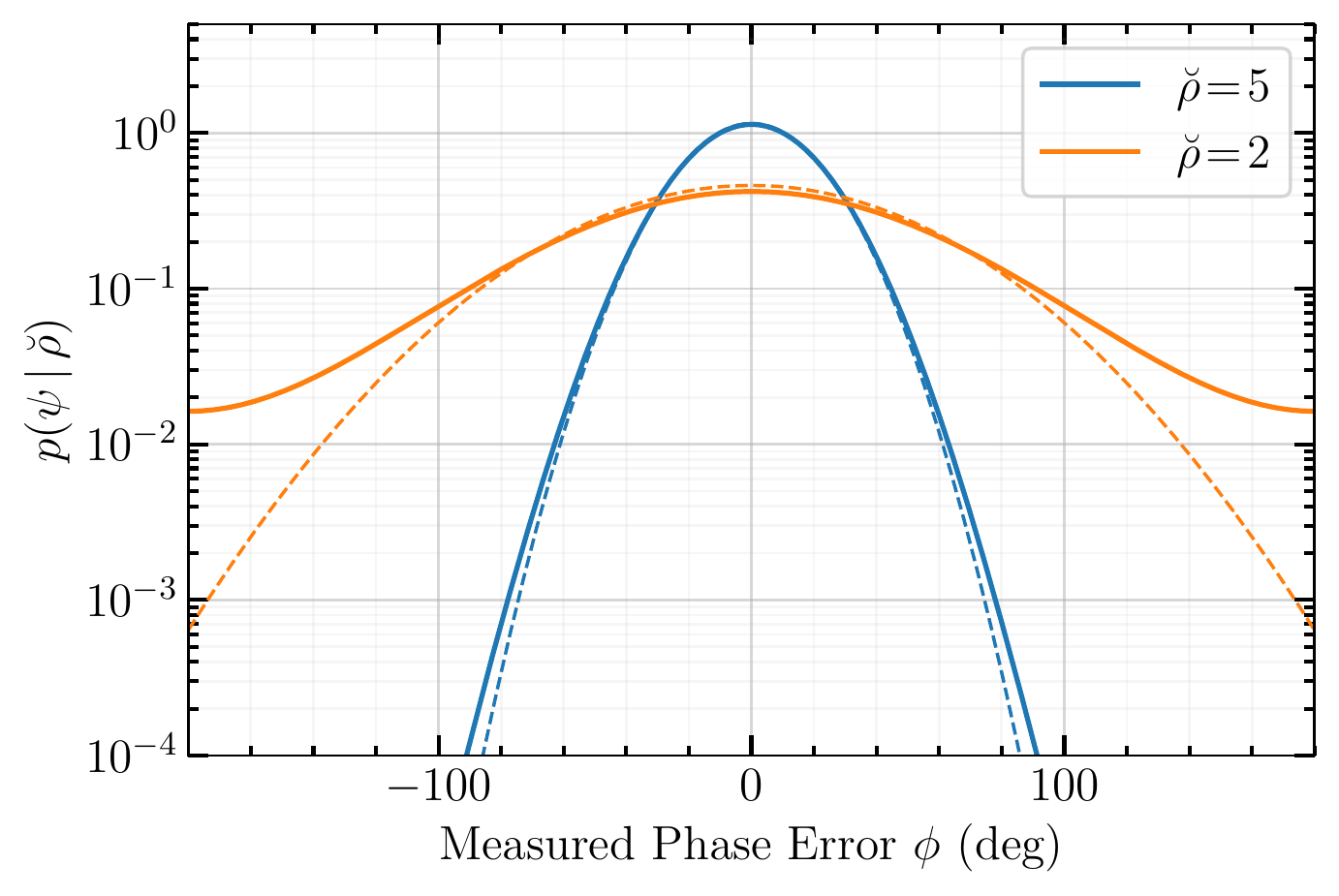}
\caption{
Distribution of closure phases versus Gaussian approximation
from the high-S/N theoretical limit (dashed lines). For each closure phase, all three baseline visibilities are drawn from a complex
normal distribution with mean value $\breve{\rho}$ and unity variance in
each complex component. 
}
\label{fig:closure_phase_distr}
\end{figure}

\begin{figure}
\centering
\includegraphics[width=3.2in]{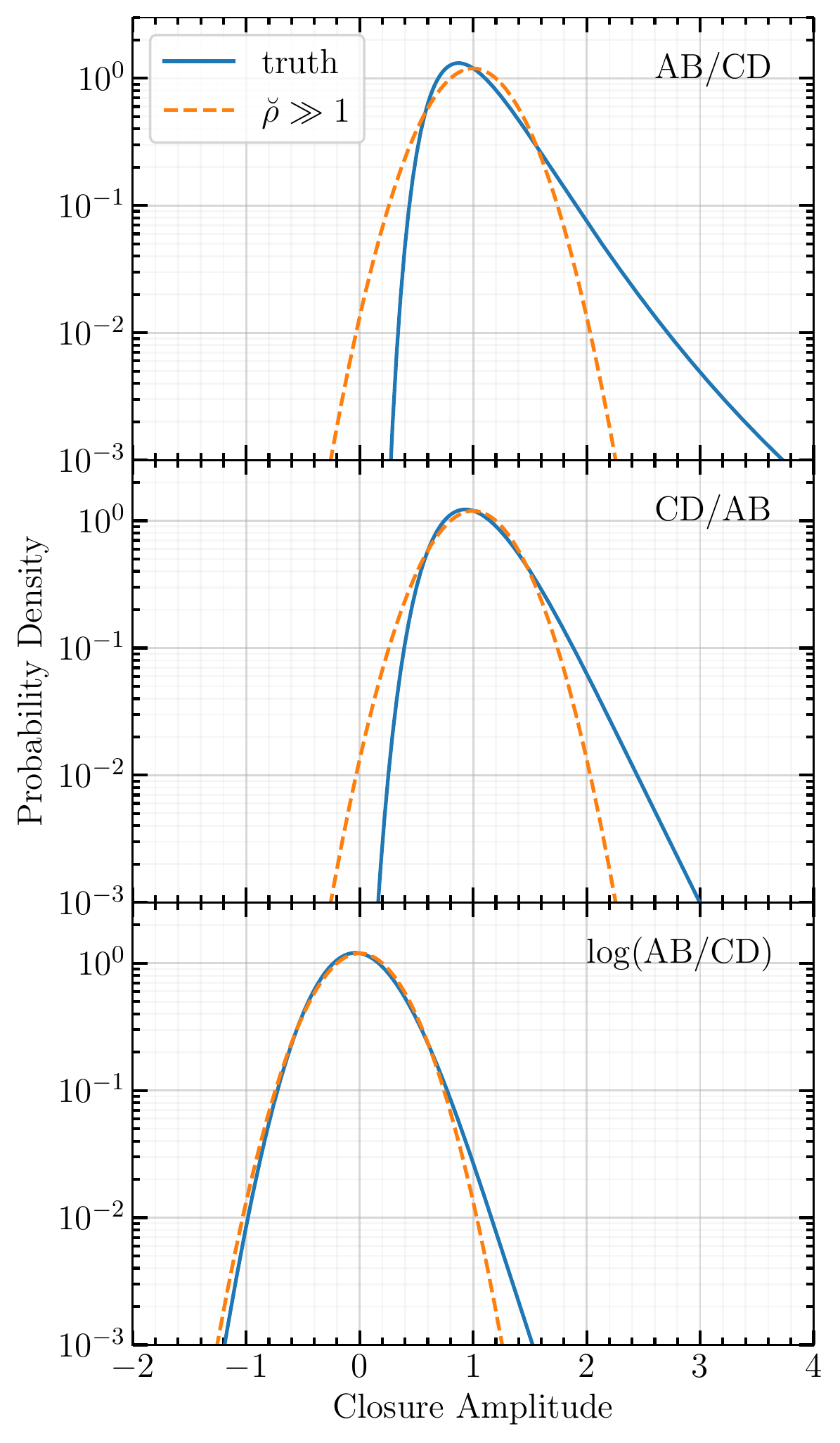}
\caption{
Distributions of closure amplitudes versus Gaussian approximations
from the high-S/N theoretical limit. Baseline amplitudes A, B,
C, D are drawn from a Rice distribution with noncentral amplitude 1 and
$\breve{\rho}_A, \breve{\rho}_B, \breve{\rho}_C, \breve{\rho}_D$ = (8, 8, 5, 5). There are large tails in the
standard closure amplitude ratio due to amplitudes in the denominator that
approach zero (top panel). The tail is mitigated somewhat by placing the lower-S/N measurements in the numerator (middle panel). However, using log-closure
amplitude provides a better-behaved distribution overall (bottom panel).}
\label{fig:closure_ampl_distr}
\end{figure}

We will primarily assume that measured amplitudes, log amplitudes, and phases for visibilities and for closure quantities can each be approximately, yet adequately, characterized as a Gaussian random process with assumed model mean and variance. This is the case for $\breve{\rho} \gtrsim \text{few}$, which is typically achieved in continuum radio interferometry through sufficient time-frequency averaging in the weak-signal limit. Examples of closure phase and closure amplitude distributions, along with corresponding high-S/N normal distribution approximations, are shown in Figures \ref{fig:closure_phase_distr}-\ref{fig:closure_ampl_distr}.

\label{sec:ignorance}

These ensemble distributions for measured phase and amplitude are exactly calculable
for a given model $\breve{\rho}$,
even in the low $\breve{\rho}$ limit where the distributions become non-Gaussian. However, in practice, the underlying intrinsic signal-to-noise $\breve{\rho}$ is generally not known, which means the distribution from which a single measured $\rho$ is drawn is also not known precisely.
Unless $\breve{\rho}$ is either assumed under a complete forward model (incorporating model visibility and all forward gains) or based on additional averaging beyond the single measurement of $r$, any estimate will be subject to thermal noise. In addition to a general mischaracterization of errors, this can also lead to a self-selection bias if realizations that are randomly low amplitude are assigned larger errors or if they are preferentially flagged from the data.

An expanded description of phase and amplitude distributions is given in \autoref{app:distributions}.
The distributions for phase, amplitude, and log amplitude can be reasonably approximated as Gaussian for $\breve{\rho}$ above 2--5. For log amplitude, a full characterization of the distribution under incoherent averaging of amplitudes is given in terms of moments. This is useful for estimating the a priori amplitude noise bias that becomes significant at low S/N. However, if a significant amount of informative data has low S/N, it may be advantageous to forward model complex gains and explicitly marginalize over their uncertainties, at least for the affected stations. This keeps data in the complex domain and their errors Gaussian.

\Needspace*{6\baselineskip}
\section{Independence of closure quantities}
\label{sec::Independence_of_Closure}

The ${\sim}N^3$ possible closure phases and ${\sim}N^4$ closure amplitudes are formed using the original ${\sim}N^2$ baseline visibilities and
become highly redundant at large $N$, where a much smaller subset of nonredundant quantities captures all source degrees of freedom \citep{Readhead1980,pearson1984}. The codependence of redundant closure quantities and their initial construction from common baseline quantities leads to a general lack of statistical independence in their residual thermal noise \citep{Kulkarni_1989}.\footnote{Although we focus here on the covariance of thermal noise, which contributes to covariance in the residual measured quantities under a true source model, we note that the same relationships also hold for non-closing baseline errors \citep{massi1991}. Such errors are particularly straightforward to incorporate into the analysis if modeled as additional independent Gaussian systematic error in baseline quantities. The same covariance relationships also hold for variations in structure closure phases, and the analysis is relevant for isolating independent structural variability degrees of freedom that are measured across the array.} For closure phases and log-closure amplitudes in the Gaussian limit,\footnote{The Gaussian limit is appropriate for S/N > $\sim$few (\autoref{sec:nongaussian}). Even at low S/N, closure quantities formed from a single set of baseline visibilities will be dependent by construction, but their statistical dependence can not be characterized by a multivariate Gaussian due to nonlinear effects. However, in the special case of coherent ensemble averages over a very large number of closure quantities that have more than one low-S/N baseline, the closure quantities become approximately statistically independent. This is often the case for bispectral averaging in optical interferometry where the atmospheric coherence time is extremely short \citep[e.g., ][]{kulkarni1991}.} the statistical dependence is fully characterized by a nonzero covariance.

In the following subsections, we detail the covariance structure for closure phases and log-closure amplitudes, and we demonstrate the relationship of the covariance to the unique and statistically independent degrees of freedom present in the quantities. We then present strategies for the construction of nonredundant but complete sets of quantities, and we discuss proper accounting of the number of gain-invariant degrees of freedom.

\Needspace*{4\baselineskip}
\subsection{Closure covariance due to thermal noise}
\label{sec:ClosureCovariance}

\begin{figure}
\centering
\includegraphics[width=1.75in]{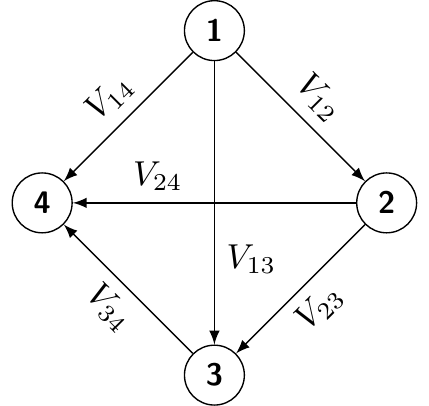}
\caption{Network of four sites. There are six baselines, three nonredundant closure phases,
and two nonredundant closure amplitudes.
}
\label{fig:4sites}
\end{figure}

Closure phases and log-closure amplitudes are formed from sums and differences
of shared baseline quantities, so that the closure quantities do not have
independent noise. Under the approximation that baseline observables are
Gaussian random variables, the joint distribution of $\cpn$ nonredundant closure phases $\cps$, for example, is
characterized by a multivariate Gaussian distribution,
\begin{equation}
    G(\cpv;\cpmv,\cpcov) = \frac{1}{\sqrt{(2\pi)^{\cpn}\det(\cpcov)}}
    \exp\left[-\frac{1}{2}\cprv^{\transpose}\cpcov^{-1}\cprv\right] ,
\label{eqn:cphase_likelihood}
\end{equation}
where residual closure phases $\cprv = \cpv - \cpmv$ are taken about model values $\cpmv = \{\cpms\}$ and have covariance matrix $\cpcov$. This corresponds to the likelihood of observing the residuals $\cprv$ under the model hypothesis.

For a collection of all baseline phases measured among four sites, $\vpv = \{\vps[12],\vps[13],\vps[14],\vps[23],\vps[24],\vps[34]\}$ (\autoref{fig:4sites}), the first three closure phases are,
\begin{equation}
\begin{gathered}
\cps[123] = \vps[12]+\vps[23]-\vps[13] \\
\cps[124] = \vps[12]+\vps[24]-\vps[14] \\
\cps[134] = \vps[13]+\vps[34]-\vps[14]
\end{gathered}
\end{equation}
The final closure phase is redundant with the other three,
\begin{align}
\cps[234] = \vps[23]+\vps[34]-\vps[24] = \cps[123]+\cps[134]-\cps[124] .
\end{align}

We can represent the generation of closure phases as a linear operator (closure phase design
matrix $\cpd$) applied to the baseline phases: ${\cpv = \cpd \vpv}$,
\begin{equation}
\begin{pmatrix}
\cps[123] \\ \cps[124] \\ \cps[134]
\end{pmatrix} =
\begin{pmatrix}
1 & -1 & 0 & 1 & \phantom{-}0 & 0 \\
1 & 0 & -1 & 0 & 1 & 0 \\
0 & 1 & -1 & 0 & 0 & 1
\end{pmatrix}
\begin{pmatrix}
\vps[12] \\ \vps[13] \\ \vps[14] \\ \vps[23] \\ \vps[24] \\ \vps[34]
\end{pmatrix} .
\end{equation}
This closure phase design matrix is equivalent to the ``phase closure operator'' of \cite{Lannes_1990b} and the ``phase compilation operator'' of \cite{Lannes_1991}.

The covariance matrix for the nonredundant set is ${\cpcov = \cpd
\vpcov \cpd^{\transpose}}$, where $\vpcov$ is the covariance of the measured
baseline phases. In general, $\vpcov$ has a diagonal contribution from $\vpn$ independent
baseline thermal-noise contributions ${\mathbf{S} = \mathrm{diag}(\vpvar[00], \dots, \vpvar[BB])}$,
plus diagonal and off-diagonal contributions from common systematic gain errors $\gpvar$. However, the common gain
errors are ultimately eliminated through the formation of closure quantities. Therefore, $\cpd \vpcov \cpd^{\transpose} = \cpd\,\mathbf{S}\,\cpd^{\transpose}$, and
\begin{equation}
\cpcov =
\left( \begin{array}{ccc}
\vpvar[12] + \vpvar[23] + \vpvar[13] &
\vpvar[12] &
-\vpvar[13] \\[0.4em]
\vpvar[12] &
\vpvar[12]+ \vpvar[24] + \vpvar[14] &
\vpvar[14] \\[0.4em]
-\vpvar[13] &
\vpvar[14] &
\vpvar[13]+ \vpvar[34] + \vpvar[14] \\
\end{array}\right).
\label{eqn:cphase_cov_example3}
\end{equation}
The cross terms of $\cpcov$ are nonzero and are based on the sign of the shared baseline components
of each closure phase.

For the same network of four sites, the first two log-closure amplitudes are
also based on sums and differences of log-baseline amplitudes $\lvav = \{\lvas[12],\lvas[13],\lvas[14],\lvas[23],\lvas[24],\lvas[34]\}$,
\begin{equation}
\begin{gathered}
\lcas[1234] = \lvas[12] + \lvas[34] - \lvas[13] - \lvas[24] \\
\lcas[1243] = \lvas[12] + \lvas[34] - \lvas[14] - \lvas[23]
\end{gathered}
\end{equation}
with a third closure amplitude that is redundant,
\begin{equation}
\lcas[1342] = \lvas[13] + \lvas[24] - \lvas[14] - \lvas[23].
\end{equation}
By using log amplitude, the redundancy in closure amplitudes can be cast in terms of linear dependence, as is already the case for closure phases.
Covariance terms are formed according to shared baselines, as was done for closure phases,
\begin{equation}
\lcacov =
\left( \begin{array}{cc}
\lvavar[12] + \lvavar[34] + \lvavar[13] + \lvavar[24] &
\lvavar[12] + \lvavar[34] \\ 
\lvavar[12] + \lvavar[34] &
\lvavar[12] + \lvavar[34] + \lvavar[14] + \lvavar[23]
\end{array}\right)
\end{equation}

The likelihood of observing a set of measured residual log-closure amplitudes $\lcarv = \lcav - \lcamv$, given measurements $\lcav$ and model hypothesis $\lcamv$, parallels \autoref{eqn:cphase_likelihood} for closure phases,
\begin{equation}
\mathcal{L} = \frac{1}{\sqrt{(2 \pi)^\lcan \text{det}\left( \lcacov \right)}} \exp\left[ - \frac{1}{2} \lcarv^{\transpose} \lcacov^{-1} \lcarv \right] .
\label{eqn:camp_likelihood}
\end{equation}
The covariance matrix
$\lcacov$ must be formed from a nonredundant set of $Q \leq Q_\text{minimal}$ closure
quantities---otherwise the matrix will be rank deficient and not invertible. $Q_\text{minimal}$ is the minimum size set that captures all available degrees of freedom, as well as the largest nonredundant set that can be formed (this is demonstrated in \autoref{sec:construction}).
The value
${\lcarv}^{\transpose} \lcacov^{-1} \lcarv$
will then follow a $\chi^2$ distribution with $Q$ degrees of freedom.

If we write the inverse covariance matrix as $\lcacov^{-1} = \mathbf{U}^{\transpose}
\lcacovdiag^{-1} \mathbf{U}$, we see that matrix $\mathbf{U}$ transforms a
nonredundant set of $Q_\text{minimal}$ closure quantities into a space of combinations of
closure quantities with independent noise, and characterized by diagonal
covariance matrix $\lcacovdiag$.
When applied to closure phases, this generates the so-called ``kernel phases,'' first noted by \citet{martinache2010}.
The closure basis formed in this
manner can be arbitrarily rotated by different choices of $\mathbf{U}$, but all rotations capture the
same $Q_\text{minimal}$ degrees of freedom. Additional redundant closure quantities to
this set will be perfectly degenerate with linear combinations of the closure
basis, and they will not add additional information to the likelihood of a set of
observations. Thus, the calculation of $\chi^2$ is unique and does not depend on
the particular set of nonredundant closure quantities used (specific examples of this invariance are provided in Appendix~\ref{app:WorkedExamples}).

In terms of the closure (log amplitude) design matrix $\lcad$, the factorization can also be written
\begin{equation}
\lcacov^{-1} = \lcacov^+ = (\lcad^+)^{\transpose} \mathbf{S}^{-1} \lcad^+
\label{eqn:designmatrix}
\end{equation}
where $\lcad^+$ is the pseudo-inverse of $\lcad$ and
$\mathbf{S}^{-1}$ is a diagonal matrix containing the reciprocal {\em baseline} thermal variances $\lvavar$.
$\lcad^+$ itself does not depend on the actual baseline noise and can be readily
computed via singular value decomposition~(SVD). Redundant degrees of freedom
will be reflected by singular values of zero and can be avoided by first
removing the redundant closure quantities by matrix reduction or explicit
construction (Section~\ref{sec:construction}). In that case, the pseudo-inverse
will be a true inverse. The advantage to inverting the design matrix rather
than the covariance matrix (as in \autoref{eqn:cphase_likelihood} or \ref{eqn:camp_likelihood}) is that the
operation on the design matrix can be done once and then applied to different
baseline noise prescriptions with little computational cost.

\subsection{Minimal complete sets}
\label{sec:construction}

\begin{figure*}
\centering
\includegraphics[width=0.675\textwidth,trim=4pt 0 4pt 0]{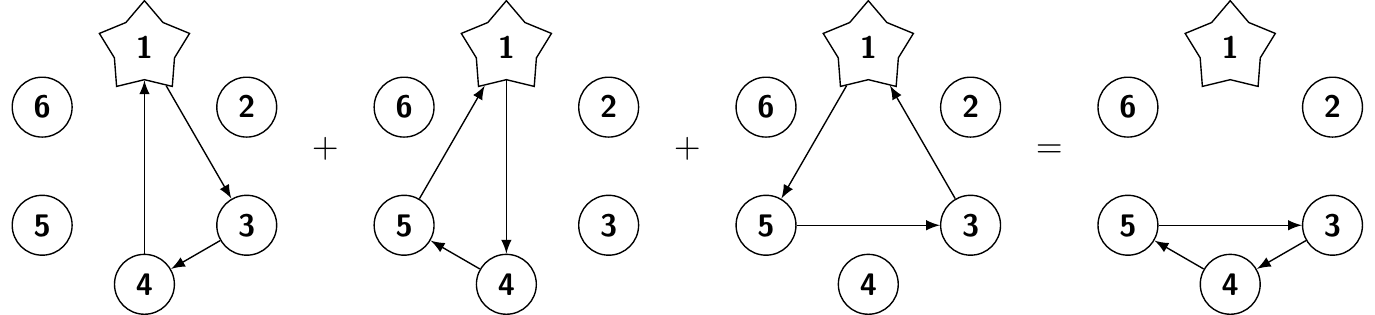}
\caption{Construction of a minimal set of closure phases. Site 1 is used
as a reference, from which there are ${(\gpn-1)(\gpn-2)/2}$ choices for the
other two sites that build the set of all triangles containing site 1.
Combinations of closure phases from this set can be used to form arbitrary
triangles that do not contain the reference station, proving that the set
is complete. This prescription is described in \cite{tms}.}
\label{fig:cphrecipe}
\end{figure*}

The total number, $\cpn$, of triangles that can be constructed from a fully connected
set of baselines across $\gpn$ sites is
\begin{equation}
    \cpn_\text{all} =\binom{N}{3} = \frac{\gpn(\gpn-1)(\gpn-2)}{3!},
\end{equation}
while the number of closure phase degrees of freedom is only the total
number of baseline phases ($\gpn(\gpn-1)/2$) minus the number of degrees of freedom
contained in site phase differences ($\gpn-1$). These degrees of freedom should be captured by a nonredundant subset of closure phases of size
\begin{equation}
\cpn_\text{minimal} = \frac{(\gpn-1)(\gpn-2)}{2}.
\end{equation}
For a large network, the set of all closure triangles will quickly outpace the
number of independent measurements, resulting in a highly redundant set. One
method for choosing a minimal set of closure triangles is given by
\cite{tms} and shown in \autoref{fig:cphrecipe}. It involves
selecting a single reference station and selecting the set of all triangles
that contain it. Triangles that do not contain the reference station are formed
as combinations of triangles from the minimal set.

\begin{figure*}
\centering
\includegraphics[width=0.6\textwidth]{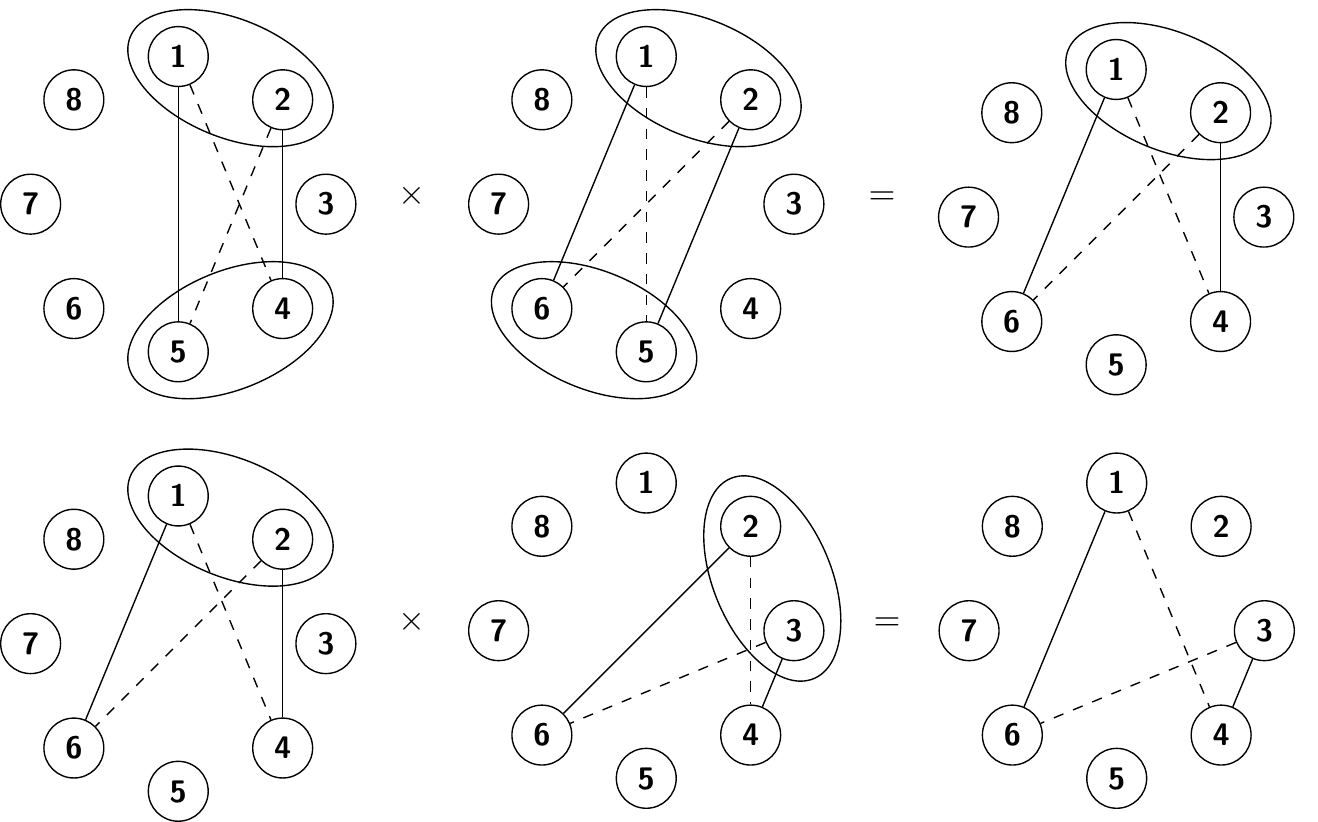}
\caption{Construction of a minimal set of closure amplitudes. We begin with
the set of closure amplitudes defined by choosing all sets of two
nonoverlapping pairs of adjacent sites and forming one closure amplitude from
each collection of four sites according to the baselines shown on the top left.
The solid and dashed lines determine which baselines go in the numerator and
denominator of the closure amplitude. Since there are $\gpn$ choices for the
placement of the first adjacent pair and $\gpn-3$ choices for the placement of
the second pair, there are $\gpn(\gpn-3)/2$ nonredundant closure amplitudes formed. By
multiplying closure amplitudes from our set, we can construct arbitrary closure
amplitudes containing nonadjacent sites. The set is therefore complete.}
\label{fig:camprecipe}
\end{figure*}

We introduce a corresponding diagrammatic procedure for selecting a minimal set of
closure amplitudes (\autoref{fig:camprecipe}). The independent closure
amplitudes are formed by arranging $\gpn$ sites on a ring and selecting all pairs
of two adjacent nonoverlapping sites. One closure amplitude is formed from
each four-site arrangement (quadrangle) from the baselines that span the pair. Because the
order of the pair does not matter, this results in the formation of
\begin{equation}
Q_\text{minimal} = \frac{\gpn(\gpn-3)}{2}
\label{eqn:qminimal}
\end{equation}
total closure amplitudes, equal to the ${\gpn(\gpn-1)/2}$ baseline degrees of
freedom minus the $\gpn$ unknown site gain factors. Combinations of closure
amplitudes from the basis can be used to construct all remaining possible
closure amplitudes, showing that the remaining closure amplitudes are redundant. Note that because adding a new station to the ring will necessarily break up one previous pair, the minimal set formed this way across $\gpn$ stations is not a proper subset of that formed across $\gpn+1$ stations.
An alternative strategy for building the set of closure amplitudes in a staged
matrix-driven approach is presented in \autoref{sec:CampDesign}.

\begin{figure}
\centering
\includegraphics[width=3.25in]{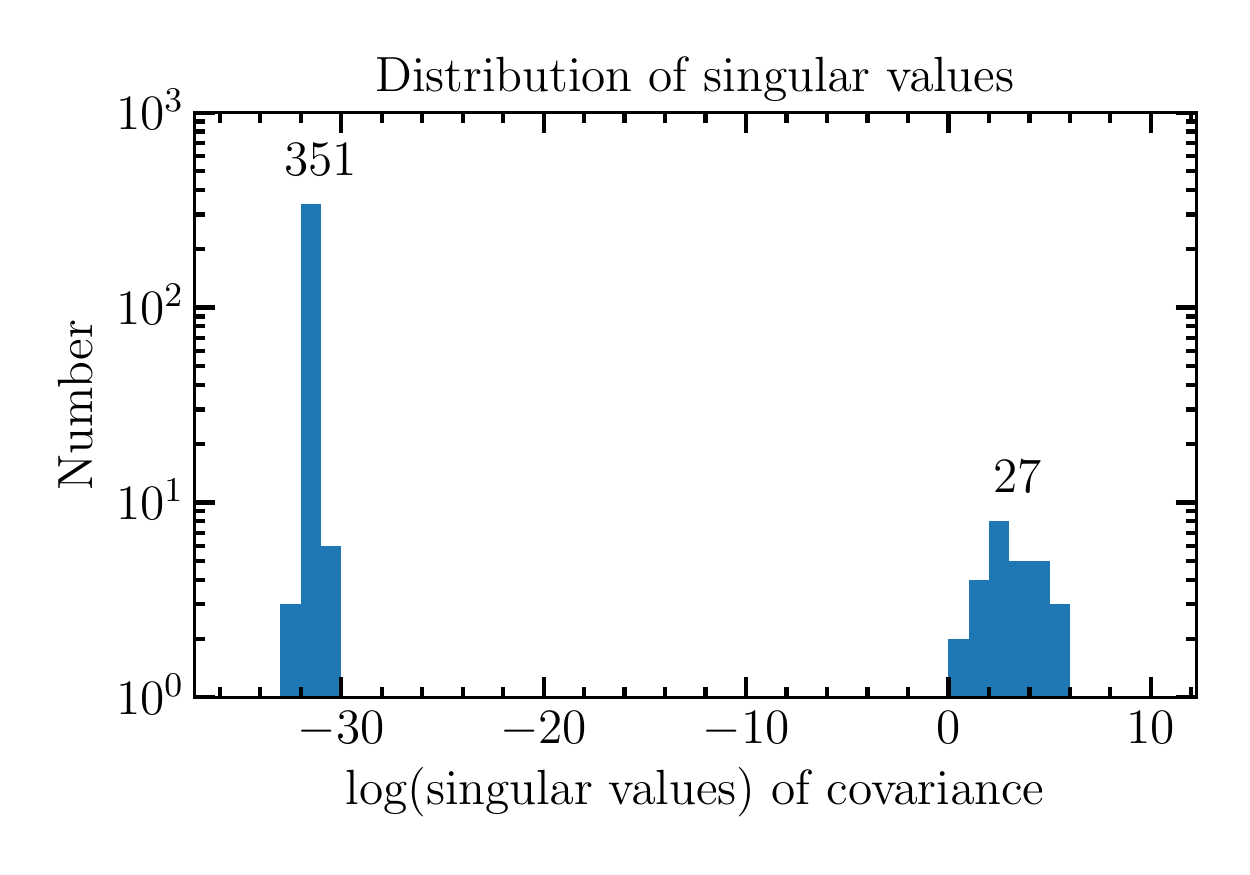}
\caption{Singular value decomposition of the covariance matrix formed from a full
set of 378 closure amplitudes over nine sites. The baseline noise prescription is
random. There are 27 nonzero singular values corresponding to the
$9\times(9-3)/2=27$ independent degrees of freedom represented in the closure
amplitudes. SVD is particularly useful in situations of arbitrary missing
baselines, which complicates the direct generation of a minimal set of
closure quantities.
}
\label{fig:svdcov}
\end{figure}

In practice, the full set of $\gpn(\gpn-1)/2$ baseline visibilities may not be
available due to processing issues or by choice, which
complicates the generation of a minimal set of closure quantities.
For closure phases, so long as a missing baseline does not include the
reference station (\autoref{fig:cphrecipe}), the closure triangle containing
the missing baseline can be excluded from the minimal set.
For closure amplitudes, missing exterior baselines between adjacent sites along the ring (\autoref{fig:camprecipe}) appear in only one
of the closure amplitude basis quadrangles. Removing the quadrangle with the missing exterior baseline
will correctly exclude all derivative closure amplitudes from the set.
For more complicated baseline unavailability, a site-based procedure for
forming nonredundant closure quantities may not work. One alternative method
for extracting the unique degrees of freedom from a
partially redundant set of closure quantities is through singular value
decomposition (SVD) of the covariance matrix or design matrix (\autoref{eqn:designmatrix}, \autoref{fig:svdcov}). Alternatively, a minimal set of the original closure quantities can be identified and extracted by matrix
reduction of the design matrix.

\label{sec:minimizingcovariance}

An upper limit on the number $n_{\psi}$ of different minimal closure phase subsets that exist for a fully connected array with $N$ stations is given by the binomial coefficient

\begin{equation}
n_{\psi} \leq \binom{\text{size of maximal set}}{\text{size of minimal set}} = \binom{\binom{N}{3}}{\binom{N-1}{2}} .
\end{equation}

\noindent This expression yields only an upper limit because for a given maximal set and $N > 4$, some selections of subsets with size equal to that of the minimal set will contain redundant closure phases, and so they will not themselves be valid minimal sets. An analogous upper limit holds for the number of nonredundant sets of closure amplitudes,

\begin{equation}
n_c \leq \binom{3 \binom{N}{4}}{\frac{N(N-3)}{2}} .
\end{equation}

Both $n_{\psi}$ and $n_c$ grow super-exponentially with $N$ (see \autoref{tab:UniqueMinimalSets}), and as the number of stations increases beyond a few, it quickly becomes prohibitive to search through all possible nonredundant subsets for the one that minimizes covariance.  The minimal-covariance subset for both closure phases and log-closure amplitudes will generically depend on the specific baseline S/N distribution of the array, and we do not know of a general-purpose algorithm for selecting the optimal set.  Instead, we consider rules of thumb for two limiting cases that approximate realistic array configurations: an array with a uniform S/N on all baselines, and an array with S/N dominated by strong baselines to a single station or with other means to clearly identify weak baselines.




We use the determinant $\lambda$ of the correlation matrix $\boldsymbol{\varrho}$
to quantify the degree of independence for any specific choice of minimal subset,
\begin{gather}
    \lambda \equiv \det\left( \boldsymbol{\varrho} \right) \\
    \varrho_{ij} = \frac{\Sigma_{ij}}{\sqrt{\Sigma_{ii} \Sigma_{jj}}} ,
\end{gather}
where the elements of $\boldsymbol{\varrho}$ are related to the elements of the covariance matrix $\boldsymbol{\Sigma}$.
\noindent The value of $\lambda$ varies between zero and one, with $\lambda = 1$ corresponding to no correlation and $\lambda = 0$ corresponding to complete correlation.

For an array with a uniform S/N on all baselines (e.g., a homogeneous array observing a point source), the covariance is minimized (i.e., $\lambda$ is maximized) when all stations are represented as nearly equally as possible\footnote{Note that it is almost never the case that strictly equal representation of all baselines is possible; for closure phases, only the $N=3$ and $N=6$ arrays can achieve perfect balance (with each baseline represented exactly once or twice, respectively), while log-closure amplitudes are limited to only the $N=5$ array (with each baseline represented exactly twice) and $N=9$ array (with each baseline represented exactly three times).} in the minimal set of either closure phases or log-closure amplitudes.  For example, an array with $N=6$ stations has a minimal closure phase set size of 10, but of the $n_{\psi} = 46{,}620$ different choices of minimal set, only 12 equally represent all baselines.\footnote{An example such set is $\{$\cps[123], \cps[124], \cps[135], \cps[146], \cps[156], \cps[236], \cps[245], \cps[256], \cps[345], \cps[346]$\}$.}  Similarly, an array with $N=5$ stations has a minimal log-closure amplitude set size of five, but only six out of $n_c = 1518$ minimal sets equally represent all baselines.\footnote{An example such set is $\{$\lcas[1234], \lcas[1245], \lcas[1352], \lcas[1453], \lcas[2345]$\}$.}

For an array with a high S/N on baselines to only one station (e.g., a heterogeneous array containing one highly sensitive station), the closure phase covariance is minimized when the minimal set is constructed using only triangles containing the reference station;
that is, using the minimal set construction algorithm described earlier in this section produces an optimal set when the reference station dominates the array sensitivity. This is because weak baselines between two non-reference stations are then used only once in the construction. For log-closure amplitudes, placing the lowest S/N baselines on the ring as adjacent sites (as in \autoref{fig:camprecipe}) accomplishes the same goal; the weakest baselines are used only once in the minimal set and, thus, do not contribute to the overall covariance.

\begin{deluxetable}{CCCCC}
\tablecolumns{5}
\tablewidth{\columnwidth}
\tablecaption{Unique minimal sets\label{tab:UniqueMinimalSets}}
\tablehead{\hphantom{00} & \colhead{$N$} & \colhead{\hphantom{00000000}$n_{\psi}$\hphantom{00000000}} & \colhead{$n_c$} & \hphantom{00} }
\startdata
& 3 & 1     & \ldots & \\
& 4 & 4     & 3 & \\
& 5 & 125   & 1518 & \\
& 6 & 46620 & 351117922 & \\
\enddata
\tablecomments{The number of unique minimal sets of closure phases and log-closure amplitudes for small arrays.}
\end{deluxetable}

\Needspace*{4\baselineskip}
\subsection{Redundant baselines}
\label{sec:redundantbaselines}

Some interferometric arrays have multiple baselines that are effectively redundant (dense arrays are often designed with this redundancy, to aid calibration). For instance, a common case in VLBI is to have multiple sites that can effectively be considered colocated. For example, the CSO, JCMT, and SMA are all on Maunakea and have participated in EHT experiments. Likewise, the APEX telescope is located within a few kilometers of the ALMA phased array center. Baselines to these redundant sites sample the same visibility and source structure, and they can be combined to reduce thermal noise and to improve calibration. For example, the addition of ALMA to the EHT including APEX does not provide new baselines. However, it significantly reduces the thermal noise of baselines to Chile.

We have so far focused on the unique statistical degrees of freedom contained in the closure quantities, which do not depend on array geometry. Baseline redundancy does have a dramatic effect, however, on the unique source structure degrees of freedom measured by the array.
For example, the addition of colocated sites to a VLBI network does not sample new nontrivial source information via closure phases even as the statistical degrees of freedom grow according to 
\autoref{eqn:qminimal}, but it does increase the amount of source information measured via closure amplitudes. In the limit where every site has a redundant partner, all source visibility amplitude information is sampled via closure amplitudes apart from a single unknown degree of freedom for the total flux density.

To assess the independent degrees of freedom for an array with baseline redundancy, we introduce a redundancy matrix $\R$ of dimensions $\vcn_\text{NR} \times \vcn$ that links multiple measurements from redundant baselines into a single degree of freedom, such that $\vcn_\text{NR} \leq \vcn$ is the number of nonredundant geometric baselines that sample unique source structure. For each row corresponding to a unique geometric baseline, $\R$ contains a ``1'' in each column for each matching station pair. If there are no redundant baselines, $\R$ is the identity matrix. For the four-site network in \autoref{fig:4sites}, if stations 1 and 2 are taken to be colocated, then of the six measured baselines $\{\vcs[12], \vcs[13], \vcs[14], \vcs[23], \vcs[24],\vcs[34]\}$, $\vcs[13] \sim \vcs[23]$ sample the same geometric baseline, as do $\vcs[14] \sim \vcs[24]$, so that
\begin{equation}
\R =
\begin{pmatrix}
1 & 0 & 0 & 0 & 0 & 0 \\
0 & 1 & 0 & 1 & 0 & 0 \\
0 & 0 & 1 & 0 & 1 & 0 \\
0 & 0 & 0 & 0 & 0 & 1 \\
\end{pmatrix}
\end{equation}
with the ``zero baseline'' $\vcs[12]$ serving as one of the four unique geometric baselines.

The number of unique source degrees of freedom captured by closure quantities is found by taking the rank of the compound design matrix, which converts nonredundant amplitudes to closure quantities. For the previous four-station example with one colocated pair, this gives ${\text{rank}(\cpd \R^\transpose)}=2$ gain-independent phase structure degrees of freedom, and $\text{rank}(\lcad \R^\transpose)=1$ gain-independent amplitude structure degrees of freedom.\footnote{The construction does not impose a trivial phase for the zero baseline, but it can be assumed by explicitly removing the corresponding row from $\R$. Doing so leaves one remaining phase structure degree of freedom corresponding to the single open triangle. There is also a trivial closure amplitude for the case of a colocated pair of sites where each baseline in the numerator has a matching baseline in the denominator with the same amplitude. In this case, the trivial behavior is already fully captured by the closure amplitude design matrix $\lcad$ and does not need to be taken {\em a priori}.}
A four-site array arranged in a square satisfies different constraints with $\vcs[12]\sim\vcs[34]$ and $\vcs[14]\sim\vcs[23]$. While there are still four unique geometric baselines, there are now ${\text{rank}(\cpd \R^\transpose)}=3$ structure closure phases and $\text{rank}(\lcad \R^\transpose)=2$ structure closure amplitudes, both equal to the corresponding number of linearly independent closure quantities.

The analysis indicates that by judicious use of redundancy, an interferometric array can reduce the overall complexity of the measurements (with $\text{rank}(\R)$ as an indication of complexity) while not sacrificing measured gain-independent structure degrees of freedom. For example, $\text{rank}(\R) - \text{rank}(\lcad \R^\transpose)$ can be taken as an indication of the number of ``amplitude gains'' that remain unconstrained. As some level of a priori gain information is generally available, a sparse array that samples the maximum number of unique geometric baselines is likely preferable over one that utilizes geometric redundancy for most situations. Colocated sites in particular can cause a significant loss in measured information. However, they do provide a link to zero baseline quantities (such as total flux), which are often known a priori and, thus, inform model independent calibration \citep{Blackburn2019}.

\section{Model fitting with unknown gains}
\label{sec:MonteCarlo}

In this section, we apply the closure construction procedures detailed in the appendices to perform a series of simple model fits to different simulated data products generated from the same underlying truth image.  The goal of these tests is to demonstrate that the same model parameter posteriors can be recovered using different representations of the data products, so long as covariances between measurements are properly accounted for.

\Needspace*{4\baselineskip}
\subsection{Visibility covariance due to gain error}
\label{sec:visibilitycovariance}

To connect model fitting to closure quantities with model fitting to baseline visibilities, we first introduce a parallel construction (to \autoref{sec:ClosureCovariance}) for the covariance in visibility measurements under the presence of uncertainty in station gain. In both cases we characterize the covariance in a residual quantity ($\cprv$ or $\lcarv$ for closure quantities, $\vprv$ or $\lvarv$ for visibilities -- see \autoref{tab:notation}), reflecting the difference between the measured quantity and the model prediction. However, while the covariance for residual closure quantities is due to thermal error on shared baselines, the baseline thermal noise is independent for visibility quantities in the weak source limit, and the covariance is due to systematic error in model gain over shared stations.
A visibility measurement \vcs[ij] contains contributions from both the source and from the station gains,

\begin{equation}
\vcs[ij] = \gcs[i] \gcs[j]^* r_{ij} .
\end{equation}

\noindent The multiplicative complex gains manifest as additive terms modifying the visibility phases and log visibility amplitudes,

\begin{subequations}
\begin{eqnarray}
\vps[ij] & = & \breve{\phi}_{ij} + \gps[i] - \gps[j] , \\
\lvas[ij] & = & \breve{a}_{ij} + \lgas[j] + \lgas[j] ,
\end{eqnarray}
\end{subequations}

\noindent where the sign differences in the second gain terms arise because complex conjugation negates phases but leaves amplitudes unchanged.

More generally, we can express the gain contributions to a collection of visibility phases or log visibility amplitudes in terms of design matrices \vpd or \lvad operating on the vector of gain phases or log gain amplitudes,

\begin{subequations}
\begin{eqnarray}
\vpv & = & \boldsymbol{\breve{\phi}} + \vpd \gpv , \\
\lvav & = & \boldsymbol{\breve{\textbf{\textit{a}}}} + \lvad \lgav .
\end{eqnarray}
\end{subequations}

\noindent For example, the visibility phases measured on the baselines in \autoref{fig:4sites} can be expressed using

\begin{equation}
\begin{pmatrix}
\vps[12] \\
\vps[13] \\
\vps[14] \\
\vps[23] \\
\vps[24] \\
\vps[34]
\end{pmatrix} = \begin{pmatrix}
\breve{\phi}_{12} \\
\breve{\phi}_{13} \\
\breve{\phi}_{14} \\
\breve{\phi}_{23} \\
\breve{\phi}_{24} \\
\breve{\phi}_{34}
\end{pmatrix} + \begin{pmatrix}
1 & -1 & 0 & 0 \\
1 & 0 & -1 & 0 \\
1 & 0 & 0 & -1 \\
0 & 1 & -1 & 0 \\
0 & 1 & 0 & -1 \\
0 & 0 & 1 & -1
\end{pmatrix} \begin{pmatrix}
\gps[1] \\
\gps[2] \\
\gps[3] \\
\gps[4]
\end{pmatrix} , \label{eqn:VisibilitPhaseVectorExample}
\end{equation}

\noindent while the log visibility amplitudes can be similarly expressed using

\begin{equation}
\begin{pmatrix}
\lvas[12] \\
\lvas[13] \\
\lvas[14] \\
\lvas[23] \\
\lvas[24] \\
\lvas[34]
\end{pmatrix} = \begin{pmatrix}
\breve{a}_{12} \\
\breve{a}_{13} \\
\breve{a}_{14} \\
\breve{a}_{23} \\
\breve{a}_{24} \\
\breve{a}_{34}
\end{pmatrix} + \begin{pmatrix}
1 & 1 & 0 & 0 \\
1 & 0 & 1 & 0 \\
1 & 0 & 0 & 1 \\
0 & 1 & 1 & 0 \\
0 & 1 & 0 & 1 \\
0 & 0 & 1 & 1
\end{pmatrix} \begin{pmatrix}
\lgas[1] \\
\lgas[2] \\
\lgas[3] \\
\lgas[4]
\end{pmatrix} . \label{eqn:LogVisibilitAmplitudeVectorExample}
\end{equation}

\noindent The visibility phase design matrix is equivalent to the ``phase aberration operator'' of \cite{Lannes_1990b}, while the log visibility amplitude design matrix matches the ``amplitude aberration operator'' of \cite{Lannes_1990a,Lannes_1991}.

This additivity makes it convenient to model the gain phases and log gain amplitudes as Gaussian distributed,
so that their variances simply add to those of the corresponding visibility quantities.  The baseline-based thermal variances are uncorrelated across baselines, and in the absence of gains, they would fully describe the visibility covariances via the diagonal matrices $\textbf{S}_{\boldsymbol{\phi}}$ for visibility phases and $\textbf{S}_{\textbf{a}}$ for log visibility amplitudes (see \autoref{tab:VisibilityPhaseTable} and \autoref{tab:VisibilityAmpTable} in \autoref{app:VisDesign}).  The station-based gain variances do drive covariances in the visibility residuals for baselines that share a station, with the design matrices serving to map stations to baselines.  The covariance matrices are then constructed as the sum of the baseline-based and station-based contributions,

\begin{subequations}
\begin{eqnarray}
\vpcov & = & \textbf{S}_{\boldsymbol{\phi}} + \vpd \gpcov \vpd^{\transpose} , \\
\lvacov & = & \textbf{S}_{\textbf{a}} + \lvad \lgacov \lvad^{\transpose} ,
\end{eqnarray}
\end{subequations}

\noindent with the off-diagonal elements consisting of only station-based terms while the diagonal elements combine both station-based and baseline-based terms.  The covariance matrix corresponding to the visibility phases in \autoref{eqn:VisibilitPhaseVectorExample} is given by

\begin{equation}
\hspace{-0.25cm} \vpcov = 
\begin{pmatrix}
\vpvar[12] + \gpvar[1] + \gpvar[2] & \gpvar[1] & \dots & 0 \\
\gpvar[1] & \vpvar[13] + \gpvar[1] + \gpvar[3] & \dots & -\gpvar[3] \\
\gpvar[1] & \gpvar[1]  & \dots & 0 \\
-\gpvar[2]         & \gpvar[3]  & \dots & -\gpvar[3] \\
-\gpvar[2]        & 0          & \dots & 0 \\
0         & -\gpvar[3] & \dots & \vpvar[34] + \gpvar[3] + \gpvar[4] \\
\end{pmatrix} ,
\end{equation}

\noindent while the covariance matrix corresponding to the log visibility amplitudes in \autoref{eqn:LogVisibilitAmplitudeVectorExample} is structurally identical except for the off-diagonal term signs,

\begin{equation}
\hspace{-0.25cm} \lvacov = 
\begin{pmatrix}
\lvavar[12] + \lgavar[1] + \lgavar[2] & \lgavar[1] & \dots & 0 \\
\lgavar[1] & \lvavar[13] + \lgavar[1] + \lgavar[3] & \dots & \lgavar[3] \\
\lgavar[1] & \lgavar[1]  & \dots & 0 \\
\lgavar[2]         & \lgavar[3]  & \dots & \lgavar[3] \\
\lgavar[2]        & 0          & \dots & 0 \\
0         & \lgavar[3] & \dots & \lvavar[34] + \lgavar[3] + \lgavar[4] \\
\end{pmatrix} .
\end{equation}

The likelihood of observing a collection of $\vpn = \gpn(\gpn-1)/2$ residual visibility phases under a given source and gain model is then
\begin{equation}
\mathcal{L} = \frac{1}{\sqrt{(2 \pi)^\vpn \text{det}(\vpcov)}} \exp\left[ - \frac{1}{2} \vprv^{\transpose} \vpcov^{-1} \vprv \right] , \label{eqn:VPhaseLikelihood}
\end{equation}
with a similar construction for log visibility amplitudes $\lvav$. This likelihood reduces to the simple case of statistically independent measured visibilities in the limit of zero systematic gain error (i.e., perfectly calibrated data).

\Needspace*{4\baselineskip}
\subsection{Model specifications}

\begin{figure*}
    \centering
    \includegraphics[width=1.00\textwidth]{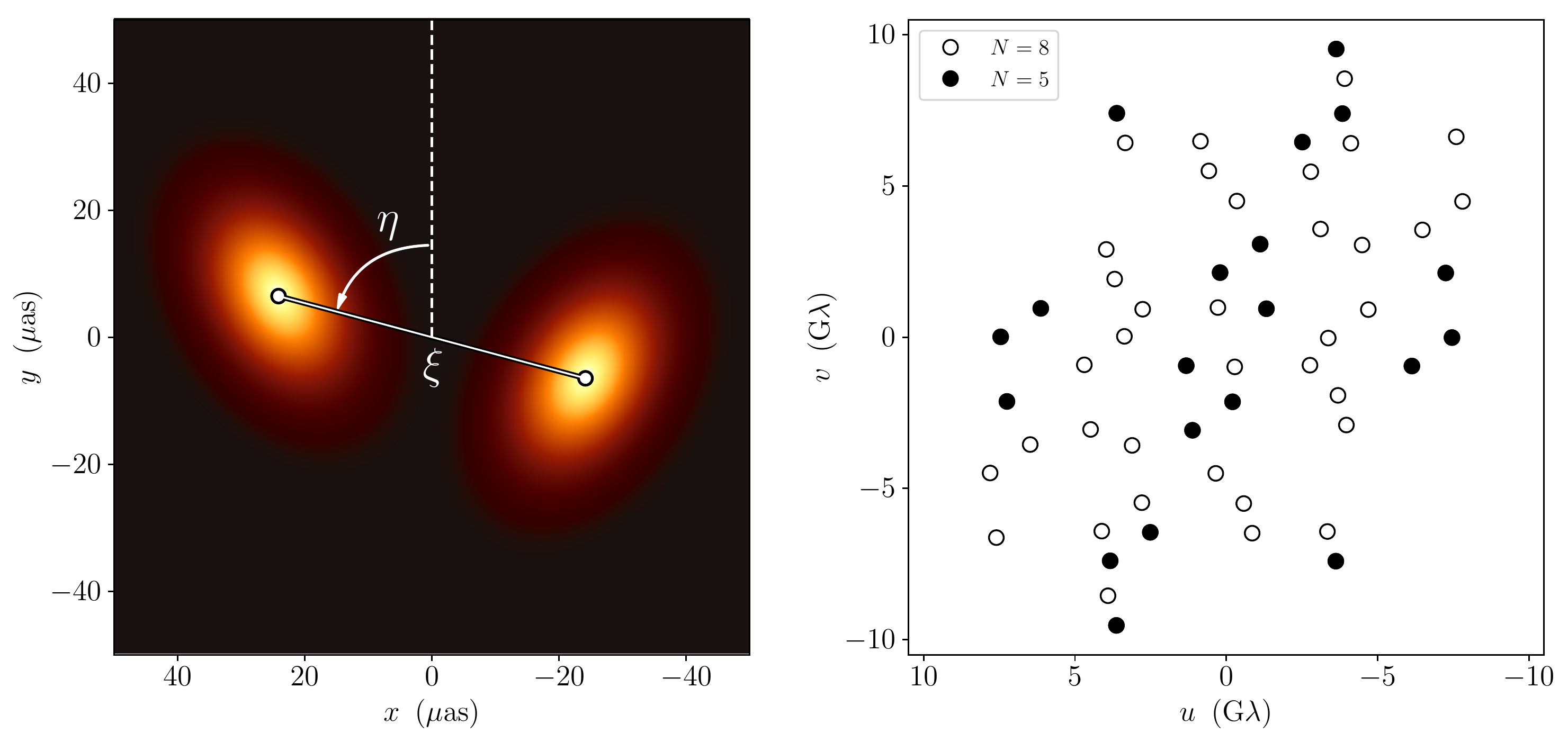}
    \caption{Truth image (left panel) and $(u,v)$ coverage (right panel) for the model considered in \autoref{sec:MonteCarlo}. The truth image contains two elliptical Gaussian components with an arbitrary but specific separation and relative orientation angle; all defining parameters are fixed during model fitting except for the component separation $\xi$ and position angle $\eta$. The $(u,v)$ coverage represents that from a single simultaneous observation with mutual visibility to all stations; we consider both an array with $N=8$ stations of equal sensitivity as well as a subset containing only $N = 5$. The resulting baseline S/N measurements span a factor of $\sim$20.
    }
    \label{fig:TruthImage}
\end{figure*}

We consider the simple geometric truth image shown in the left panel of \autoref{fig:TruthImage}.  This image is constructed from the sum of two elliptical Gaussian components that are symmetrically positioned about the origin with a mutual separation of $\xi = 50$\,$\mu$as and a position angle of $\eta = 75$\,degrees east of north.  Both components have major and minor axis Gaussian $\sigma$-values of $9$\,$\mu$as and $6$\,$\mu$as, respectively, and each has a flux density of 0.5\,Jy.  The major axis of the eastern component is oriented at 30\,degrees east of north, while the western component has a $-30$\,degree orientation.  These specific choices of parameter values are largely arbitrary, and they serve primarily to give the image sufficient asymmetry to produce nontrivial visibility phases and sufficient compactness to produce nonzero visibility amplitudes.

To produce synthetic visibility data, we sample the Fourier transform of the truth image at discrete locations in $(u,v)$ space.  We consider two sets of $(u,v)$ coverage, corresponding to (1) a single snapshot from an $N = 8$ station array with mutual visibility to all stations and (2) an $N = 5$ station subset of that array.  Both sets of coverage are shown in the right panel of \autoref{fig:TruthImage}.  Visibility amplitudes and phases are given by the magnitude and argument of the complex visibilities from each $(u,v)$ point.  The visibilities are then multiplied by their associated station gains, which are simulated as complex Gaussian-distributed random variables with unit mean and standard deviation of 0.1 along each dimension.  We add a single realization of Gaussian thermal noise to the visibilities corresponding to a median S/N of 10.2 and spanning an S/N range from 2.3 to 44.9. Closure phases are constructed from the visibility phases using \autoref{eqn:ClosurePhaseDefinition}, and log-closure amplitudes are constructed from the visibility amplitudes using \autoref{eqn:LogCampDefinition}.

We model the data as the sum of two elliptical Gaussians, with all parameters except for $\xi$ and $\eta$ held fixed at their corresponding truth values.  By restricting the model to this two-dimensional subspace of its natural 12-dimensional parameter space, we simplify the fitting process while retaining enough model complexity to provide nontrivial parameter correlations.  We perform parameter estimation using Gaussian likelihoods analogous to \autoref{eqn:VPhaseLikelihood} for all data products.  Unless otherwise specified, we apply uniform priors on the range $[40,60]$\,$\mu$as for $\xi$ and $[0,180]$\,degrees for $\eta$; when fitting gains, our ``maximally uninformative'' priors are log-uniform on the range $[10^{-5},10^5]$ for all gain amplitudes and uniform priors on the range $[0,360)$\,degrees for all gain phases.  We use the Python nested sampling code \texttt{dynesty}\footnote{\url{https://github.com/joshspeagle/dynesty}} \citep{Speagle_2019} to produce parameter posteriors for all model fits.

\Needspace*{4\baselineskip}
\subsection{Phase and amplitude modeling} \label{sec:PhaseModeling}

\begin{figure*}
    \centering
    \includegraphics[width=1.00\textwidth]{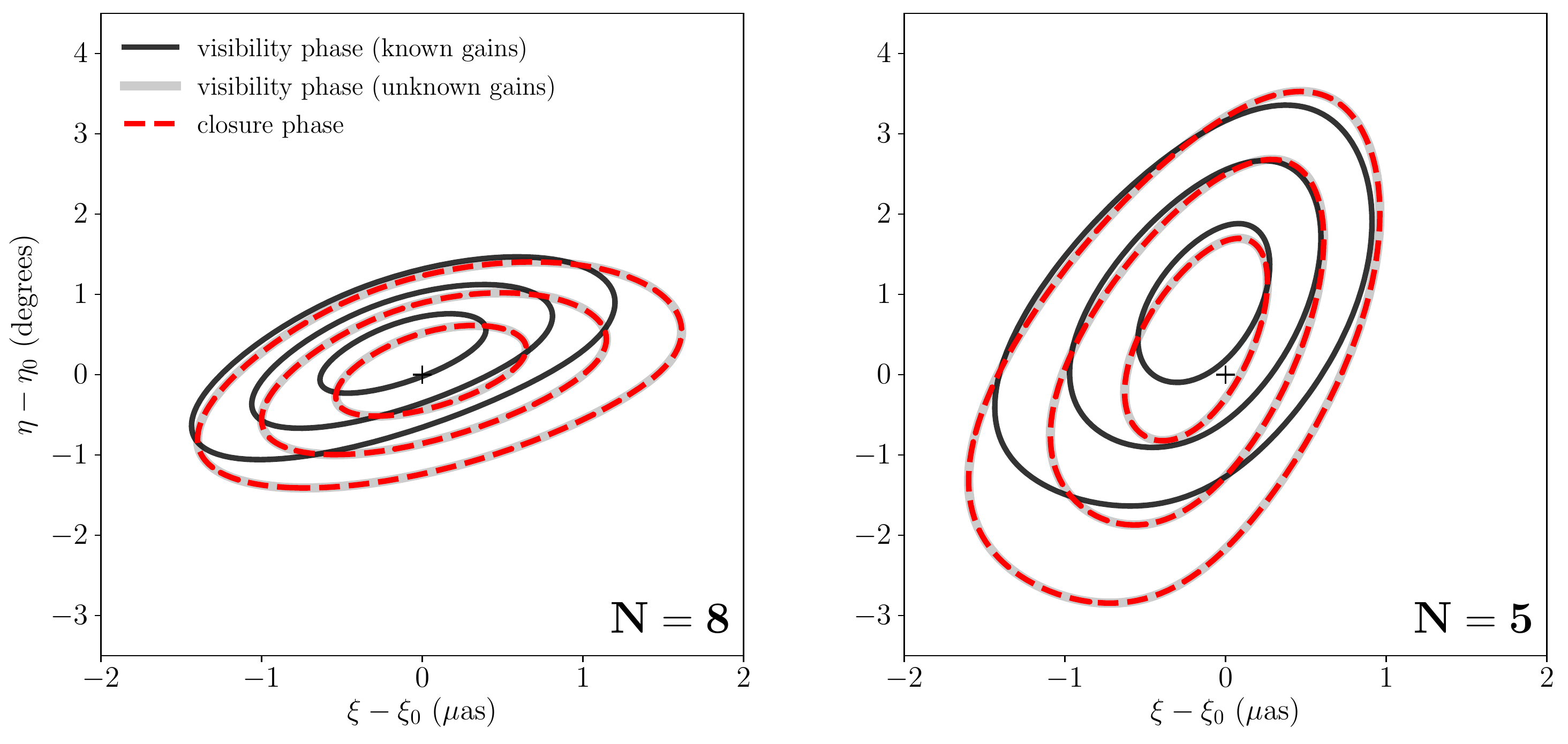}
    \caption{Joint posterior distributions for residual separation ($\xi - \xi_0$) and position angle ($\eta - \eta_0$) when fitting the model described in \autoref{sec:MonteCarlo} to visibility phases with perfectly known gain phases (black contours), visibility phases with completely unknown gain phases (gray contours), and closure phases with covariant structure accounted for (red dashed contours). We also show the results from numerically marginalizing over the gains (thin black contours), which match the covariant treatment as expected (see \autoref{sec:gainmarginalization}). The model is fitted to the eight-station array data on the left and to the five-station array data on the right; we can see that the relative loss of information when going from perfectly calibrated phases to closure phases increases for smaller arrays.  In both cases, the closure phase fits accurately recover the posteriors derived from visibility phase fits, within sampling uncertainties. Contours enclose 50\%, 90\%, and 99\% of the posterior probability.}
    \label{fig:posterior_cphase}
\end{figure*}

\begin{figure*}
    \centering
    \includegraphics[width=1.00\textwidth]{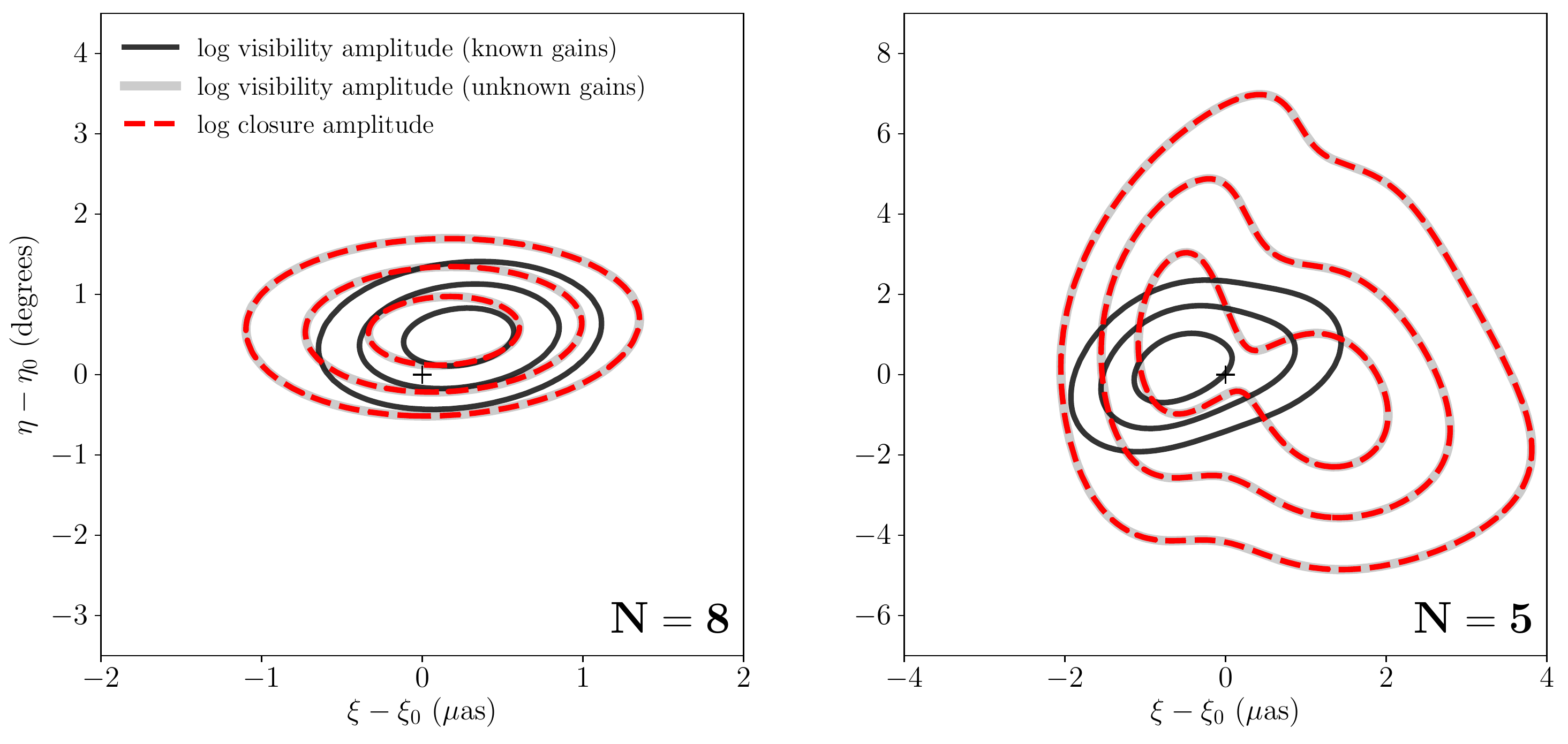}
    \caption{Same as \autoref{fig:posterior_cphase}, but for visibility amplitudes and log-closure amplitudes rather than visibility phases and closure phases. Contours enclose 50\%, 90\%, and 99\% of the posterior probability.}
    \label{fig:posterior_logcamp}
\end{figure*}

We fit the model to our single realization of synthetic data represented in a variety of ways, starting with visibility phases under the assumption that the gain phases are perfectly known (or equivalently, that they are perfectly calibrated).  The likelihood function for this representation is given by \autoref{eqn:VPhaseLikelihood}, and because the gain phases are known, the visibility phase covariance matrix is diagonal.  \autoref{fig:posterior_cphase} shows the two-dimensional $(\xi,\eta)$ posteriors for such fits to the $N=8$ array and $N=5$ array data in black contours.

We also fit the model to visibility phase data without assuming any \textit{a priori} knowledge of the gain phases. The likelihood function remains \autoref{eqn:VPhaseLikelihood}, but in this case, the covariance matrix is no longer diagonal. The gray contours in \autoref{fig:posterior_cphase} show the corresponding joint posteriors for $(\xi,\eta)$, which exhibit the expected loss of constraining power compared to the posteriors derived from calibrated visibility phases.  We can see that this loss becomes less severe as the number of stations increases, a consequence of the fact that the fraction of the visibility phase information required to constrain the gain phases decreases as $2/N$.

The other phase data representations we consider are closure phases.  For a minimal subset of closure phases described by covariance matrix \cpcov (see \autoref{sec:CphaseDesign}), the likelihood function is given by
\autoref{eqn:cphase_likelihood}.
Posteriors derived from this likelihood are shown as red dashed contours in \autoref{fig:posterior_cphase}.  Within the $\sim$1\% numerical sampling uncertainties of our posterior contours, the closure phases provide parameter constraints that are identical to those imposed by the uncalibrated visibility phases.


We perform a corresponding set of model fits
to visibility amplitudes and log-closure amplitudes rather than visibility phases and closure phases.  The black contours in \autoref{fig:posterior_logcamp} show the $(\xi,\eta)$ posteriors for fits to the $N=8$ and $N=5$ array visibility amplitude data, using a likelihood analogous to \autoref{eqn:VPhaseLikelihood} under the assumption of perfectly calibrated gain amplitudes.  The gray contours show fits to visibility amplitudes using the same likelihood but assuming no knowledge of the gain amplitudes.  We again see the relative loss of information increasing as the number of stations decreases, becoming particularly severe for the case of $N=5$ (in which there are only three degrees of freedom remaining in the data to constrain the model, compared to eight degrees of freedom when the gain amplitudes are calibrated).

We compare the visibility amplitude results to those obtained from fitting to log-closure amplitudes.  For a minimal subset of log-closure amplitudes described by a covariance matrix \lcacov (see \autoref{sec:CampDesign}), the likelihood function is given by
\autoref{eqn:camp_likelihood}.
The posteriors derived using this likelihood expression are plotted in \autoref{fig:posterior_logcamp} as red dashed contours.  As with the closure phases, we find that the log-closure amplitudes provide constraints that are identical to those provided by the uncalibrated visibility amplitudes.

\begin{figure*}
    \centering
    \includegraphics[width=1.00\textwidth]{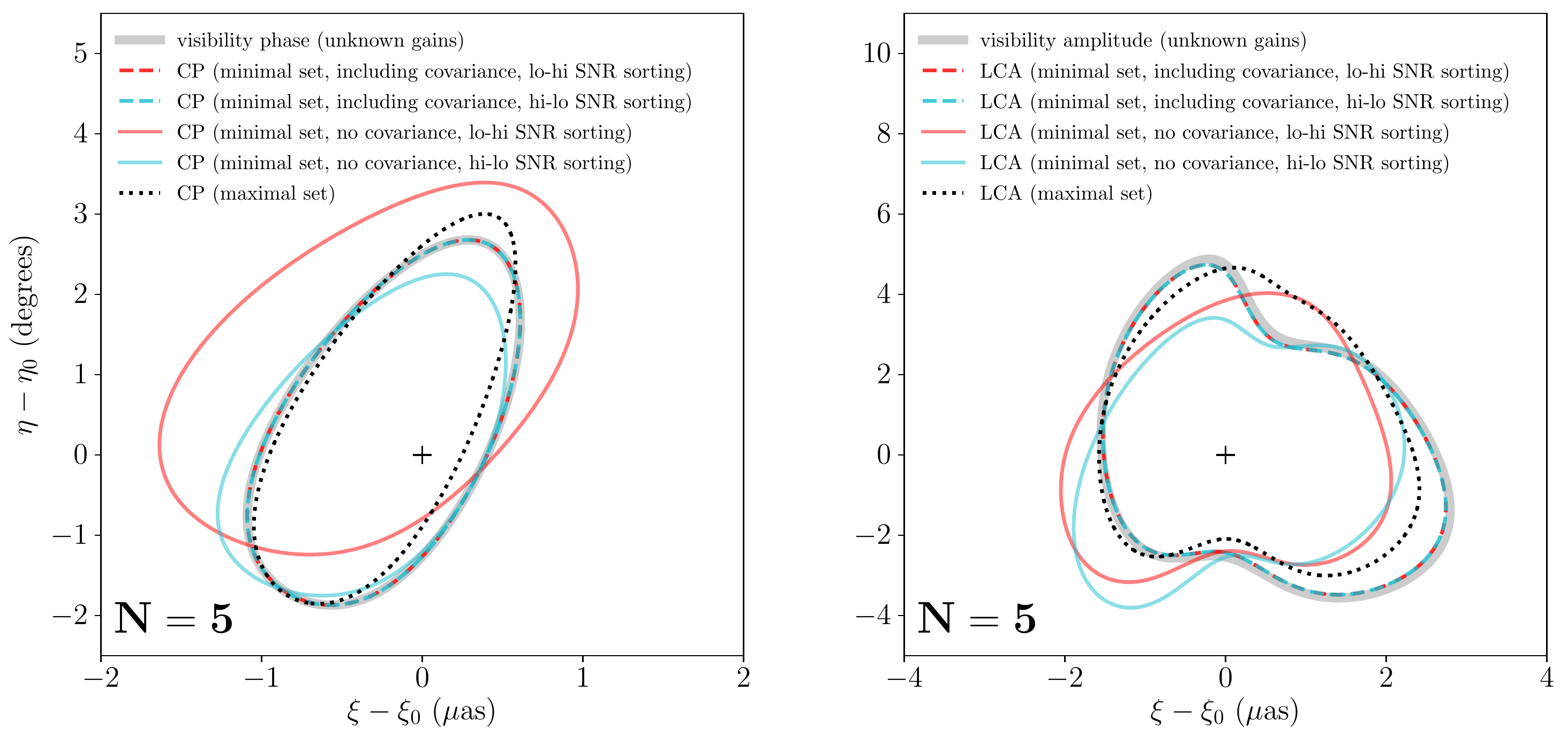}
    \caption{\textit{Left panel}: same as the left panel of \autoref{fig:posterior_cphase}, but showing additional posteriors as calculated for different closure phase (CP) likelihood constructions. The posteriors obtained when including closure phase covariance (dashed contours) demonstrate the most consistency with the uncalibrated visibility phase posterior (yellow contour). The solid blue and red contours show posteriors constructed from fitting a minimal closure phase set but ignoring covariance, while the dotted black contour shows the posterior constructed from a maximal (redundant) set of closure phases that has been corrected for redundancy factor (see \autoref{eqn:ClosurePhaseRedundancy}). These three contours that do not account for covariance show artificial distortions in the confidence regions.  \textit{Right}: analogous to the left panel but for the log-closure amplitudes (LCA) rather than closure phases; the redundancy factor for the maximal set is given by \autoref{eqn:ClosureAmpRedundancy}.  The contours in both panels enclose 90\% of the posterior probability.
    }
    \label{fig:posterior_multi}
\end{figure*}

We also consider two alternative treatments of the closure phases and log-closure amplitudes that attempt to avoid accounting for covariances, and we show here that these efforts fail.  In the first such treatment, we use a minimal closure phase subset but assume all measurements are independent.  This assumption amounts to using only the diagonal elements of \cpcov (i.e., all off-diagonal elements are set to zero), and the likelihood function remains \autoref{eqn:cphase_likelihood}.  The red and blue contours in the left panel of \autoref{fig:posterior_multi} show posteriors derived under this assumption, for two different choices of minimal closure phase subset constructed by ordering the stations from lowest to highest (red) and highest to lowest (light blue) mean baseline S/N. We can see that these contours systematically deviate from the visibility phase contour.  In the second treatment, we use the maximal (redundant) set of closure phases (see \autoref{sec:CphaseDesign}), but we retain the assumption that all measurements are independent.  The likelihood is then simply the product of the individual measurement likelihoods taken over all closure phases in the maximal set.  The dotted black contour in the left panel of \autoref{fig:posterior_multi} represents the resulting posterior after scaling the individual measurement variances by

\begin{equation}
R_{\psi} = \binom{N}{3} \Big/ \binom{N-1}{2} = \frac{N}{3} , \label{eqn:ClosurePhaseRedundancy}
\end{equation}

\noindent which is a redundancy factor that accounts for the fact that the maximal set contains an increased number of measurements without a corresponding increase in the number of degrees of freedom.  Even after accounting for this redundancy, however, we see a similar systematic discrepancy in the posterior relative to those derived from the visibility phases.  Note that for the unusual case of equal S/N on all baselines, this redundancy factor scaling does produce the correct likelihood (see \autoref{app:WorkedExamples}).

In the right panel of \autoref{fig:posterior_multi}, we again compare the posteriors obtained using (1) a minimal set of log-closure amplitudes without accounting for covariance, and (2) a maximal set of log-closure amplitudes. In both cases, the likelihood function is the product of the individual measurement likelihoods, where the product is taken over all log-closure amplitudes in the minimal or maximal set, as appropriate. We again consider two choices of minimal subset, constructed via the same station-ordering scheme used for phases. For the posteriors derived from the maximal set, shown using a dotted black contour in the right panel of \autoref{fig:posterior_multi}, we have scaled the measurement variances by the redundancy factor

\begin{equation}
R_{A} = 3 \binom{N}{4} \Big/ \left( \frac{N (N-3)}{2} \right) = \frac{(N-1)(N-2)}{4} . \label{eqn:ClosureAmpRedundancy}
\end{equation}

\noindent Regardless of the redundancy correction, we find in both cases that the posterior distributions do not match those expected from fits to the visibility amplitudes.

\Needspace*{4\baselineskip}
\subsection{Gain uncertainty modeling} \label{sec:PartialGains}

\begin{figure*}
    \centering
    \includegraphics[width=1.00\textwidth]{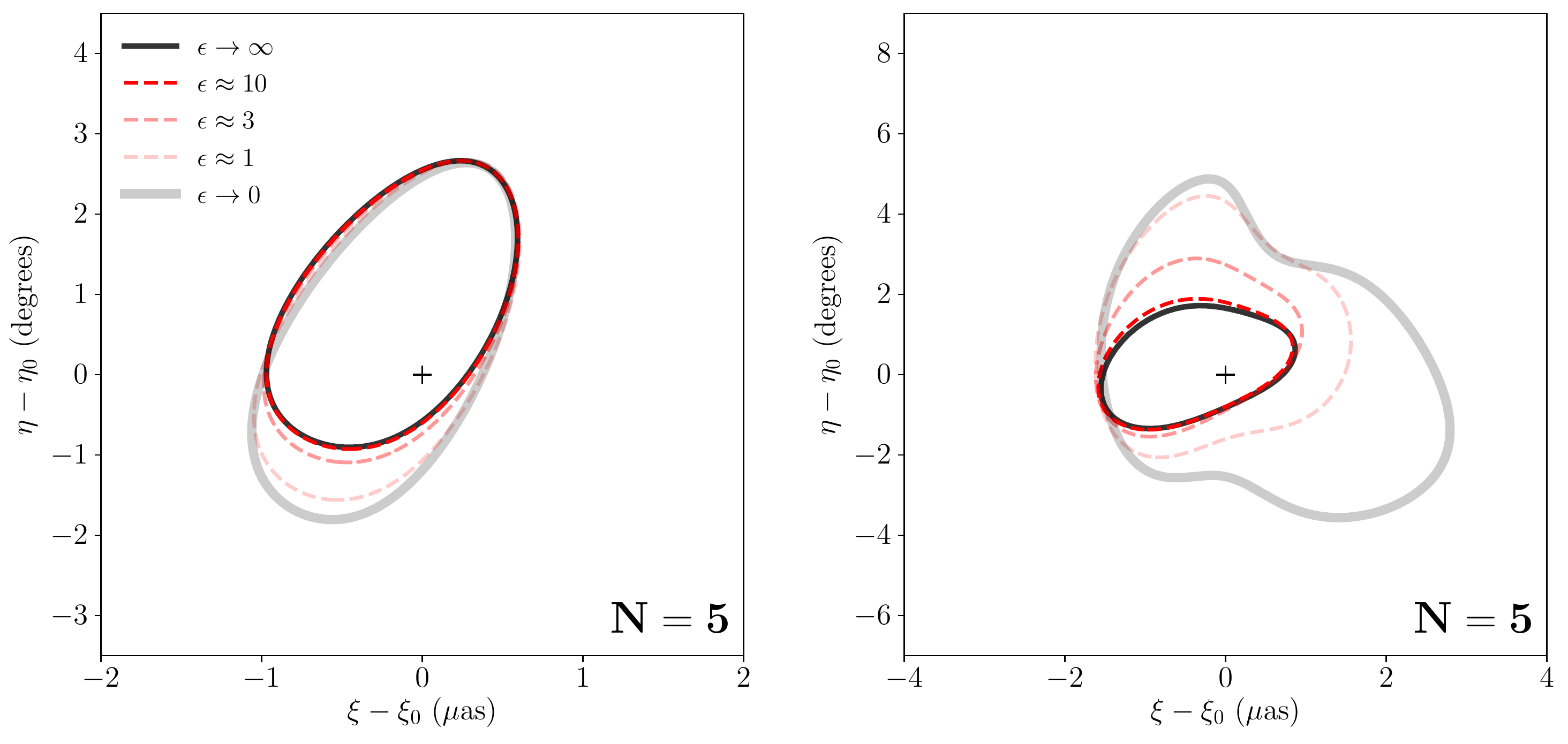}
    \caption{\textit{Left panel}: posterior contours for fits to visibility phases with varying degrees of prior gain phase knowledge assumed.  The gray contour matches the posterior recovered when fitting to closure phases (see \autoref{fig:posterior_cphase}).  \textit{Right panel}: same as the left panel, but fitting to the visibility amplitudes rather than visibility phases.  The contours in both panels enclose 90\% of the posterior probability.
    } 
    \label{fig:partial_gains}
\end{figure*}

We have shown that our level of knowledge about the gains dictates how much source information the closure quantities contain relative to the visibility quantities.  For perfectly known (or, equivalently, perfectly calibrated) gains, the visibility quantities provide more information about the source than the closure quantities; for small arrays, this difference may be quite large (see, e.g., \autoref{fig:posterior_logcamp}).  When the gains are completely unknown (or, equivalently, when the gains must be fully determined along with the source information), both the visibility and closure quantities contain identical source information.  We now explore the case of partially known gains.

We quantify how well we know the gains by comparing our gain uncertainty, $\sigma_i$, to the uncertainties in the data (i.e., in the visibilities), $\sigma_{ij}$, using

\begin{equation}
\varepsilon \equiv \frac{\left\langle \sigma_{ij} \right\rangle}{\left\langle \sigma_{i} \right\rangle} ,
\end{equation}

\noindent where $\left\langle \right\rangle$ denotes a sample average; the averages are taken over all stations and all baselines for the gain uncertainties and visibility uncertainties, respectively.  The quantity $\varepsilon$ tracks our knowledge of the gains; $\varepsilon \rightarrow 0$ when we have no information about the gains, and $\varepsilon \rightarrow \infty$ when the gains are perfectly calibrated.  Note that both $\sigma_i$ and $\sigma_{ij}$ refer to logarithmic uncertainties when considering amplitude data products, meaning that $\varepsilon$ can also be thought of as the ratio of the ``gain S/N'' to the data S/N.

Within the context of our model-fitting procedure, the assumed level of gain knowledge can be straightforwardly incorporated using Gaussian priors on the gain parameters. \autoref{fig:partial_gains} shows the results of fitting to visibility quantities while varying the value of $\varepsilon$.  We find that noticeable improvements in the posterior constraints start to occur for $\varepsilon \gtrsim 1$, and that for $\varepsilon \gtrsim 10$ the posteriors better approximate the perfect-knowledge case (black contours in \autoref{fig:partial_gains}) than they do the no-knowledge case (gray contours). This matches our expectation that knowledge of gains begins to inform an overconstrained model as soon as its precision approaches that of the thermal uncertainties. For an underconstrained problem such as imaging using a sparse array, typical regularization imposes much weaker relationships across points in the $(u,v)$ domain, and partial gain calibration can matter much earlier by providing unique information not sampled by the closure quantities.

While the demonstrations presented here are all done using simulated observations, the recent parameterization of the horizon-scale emission and shadow of the supermassive black hole in M87 by the \citet{PaperVI} utilized cross-validation of results across several techniques to handle gain uncertainty. These included explicit semi-analytic marginalization of amplitude gains by Laplace approximation \citep[via \textsc{Themis};][]{themis}, minimization of closure phase covariance through selection of a highly sensitive reference antenna (ALMA), and use of diagonalized closure phases and log-closure amplitudes by accounting for covariance (via \texttt{dynesty}, as described in this work). The multiple approaches resulted in a high degree of consistency as reflected by their posterior parameter distributions. A detailed study of the effects of covariant interferometric errors on imaging and on parameter reconstruction is forthcoming (Pesce, D. W., et al., {\em in preparation}).

\section{Summary}
\label{sec:summary}

We have explored in detail the statistics of closure phase and closure amplitude for S/N $\gtrsim$ 1, characteristic of high-frequency radio interferometry where both phase and amplitude calibration have significant uncertainties, and where phase coherence timescales are short relative to the length of a continuous observation. The analysis unifies and clarifies several concepts that have been previously discussed in the literature regarding the independence of closure quantities, the nature and number of statistical degrees of freedom, best practices for constructing and fitting to closure quantities, and the relationship of closure quantities to self calibration and marginalization over unknown gains.
Due to the large number of topics covered, we delineate the main statements and findings from this work across three primary topics.

\vspace{1em}

\Needspace*{4\baselineskip}
\noindent 1) Formation of closure quantities and non-Gaussian errors:

\begin{itemize}
\setlength\itemsep{0em}

\item{Non-Gaussian errors
become significant for S/N below $\sim$2--5 for phase, amplitude, and log amplitude. Reciprocal amplitude is unstable below S/N $\sim$5, which provides motivation to use log amplitude instead when there is a chance for low-S/N amplitudes to appear in the denominator of an amplitude ratio. (\autoref{app:distributions})}

\item{The ensemble distribution of measured log amplitude for known S/N is fully characterized in terms of moments, from which expected distributions for log-closure amplitude can be derived. (\autoref{app:distributions})}

\item{In practice, a noisy estimate of S/N prevents a reliable characterization of phase and amplitude errors, particularly for weak signals, and can lead to significant bias from self-selection of data. (\autoref{sec:ignorance}})

\end{itemize}

\noindent 2) The covariance structure for closure quantities -- fitting those quantities to a model and characterizing their fundamental degrees of freedom:

\begin{itemize}
\setlength\itemsep{0em}

\item{Closure quantities formed from a common set of baseline visibilities are covariant due to shared thermal noise. This must be included for a particular realization to recover both proper $\chi^2$ statistics and a correct likelihood. (\autoref{sec:ClosureCovariance})}

\item{When covariant errors are included, both the $\chi^2$ and the likelihood (in the Gaussian limit) are independent of the specific minimal set of closure quantities used for the calculation. (\autoref{sec:ClosureCovariance})}

\item{In the limit of equal S/N on all baselines, the $\chi^2$ for a specific minimal set reduces to an evaluation over all closure quantities weighted equally, scaled to the appropriate degrees of freedom. (\autoref{app:WorkedExamples})}

\item{If closure quantities are assumed to be independent and the covariance structure is ignored, results do depend on the specific choice of minimal set. Certain selections can be chosen to minimize off-diagonal terms in the covariance matrix, but this choice depends on the specific arrangement of baseline S/N. (\autoref{sec:construction}, \autoref{fig:posterior_multi} of \autoref{sec:MonteCarlo})}

\item{Two different direct constructions for selecting nonredundant sets of closure amplitudes are given. They verify explicitly the expected ${N(N-3)/2}$ independent degrees of freedom contained in the closure amplitudes. (\autoref{sec:construction} and \autoref{sec:CampDesign})}


\item{A unified matrix construction for creating visibilities and closure quantities is given, which systematically builds up design matrices for increasing station number. These are used to derive the covariance and other relationships across different quantities. (\autoref{app:AllDesign})}

\end{itemize}

\noindent 3) Relationship of closure information to station gain information, and behavior in the limit of completely known gains, partially known gains, and completely unknown gains:

\begin{itemize}
\setlength\itemsep{0em}

\item{Under a model for systematic station-based gain error, residual visibility phase and log amplitude also assume a covariance structure with nonzero off-diagonal elements due to gain model error. (\autoref{sec:visibilitycovariance})}

\item{Using this covariance structure for visibilities is equivalent to explicitly marginalizing over additional free gain parameters under a Gaussian prior. We note however that wide log-amplitude gain priors will often be a poor characterization of expected telescope performance, which is bounded. A standard Bayesian approach of direct numerical marginalization over nuisance gain parameters would be needed to take full advantage of more realistic priors. (\autoref{sec:gainmarginalization})}

\item{In the limit of small thermal error compared to gain error, the $\chi^2$ derived from visibility measurements reduces to the $\chi^2$ derived from only closure quantities, after accounting for covariance. Thus, the closure quantities contain all non-station-based information. (\autoref{sec:generalN})}

\item{We apply the likelihood constructions introduced in this paper toward direct sampling of the posterior distribution of a simple source model and simulated observation. We confirm that the inferred parameter posterior derived using closure quantities matches that derived using baseline visibilities in the limit of unknown gains and that the uncertainties are larger than those derived under known gain calibration, reflecting the relative loss of information. (\autoref{sec:MonteCarlo})}

\item{Under modeling of partially known gains, with systematic station gain uncertainty comparable to that from baseline thermal noise, we see that the model posterior distribution transitions smoothly between the case of perfectly calibrated, corresponding to zero gain error, and completely unknown calibration, using only closure quantities. (\autoref{sec:PartialGains})}

\end{itemize}


\section*{acknowledgements}
The authors would like to thank Jim Moran, Geoff Bower, Ramesh Narayan, Kazunori Akiyama, Katie Bouman, Christiaan Brinkerink, Avery Broderick, Andre Young, Josh Speagle, and our anonymous referee for helpful discussions, comments, and ideas.
We thank the National Science Foundation (AST-1440254, AST-1614868) and the Gordon and Betty Moore Foundation (GBMF-5278) for financial support of this work. This work was supported in part by the Black Hole Initiative at Harvard University, which is supported by a grant from the John Templeton Foundation and the Gordon and Betty Moore Foundation.

\onecolumngrid
\vspace{2em}


\bibliographystyle{yahapj2}
\bibliography{main}

\clearpage
\appendix

\restartappendixnumbering

\section{Distributions due to thermal noise}
\label{app:distributions}

\begin{figure}
\centering
\includegraphics[width=2.7in]{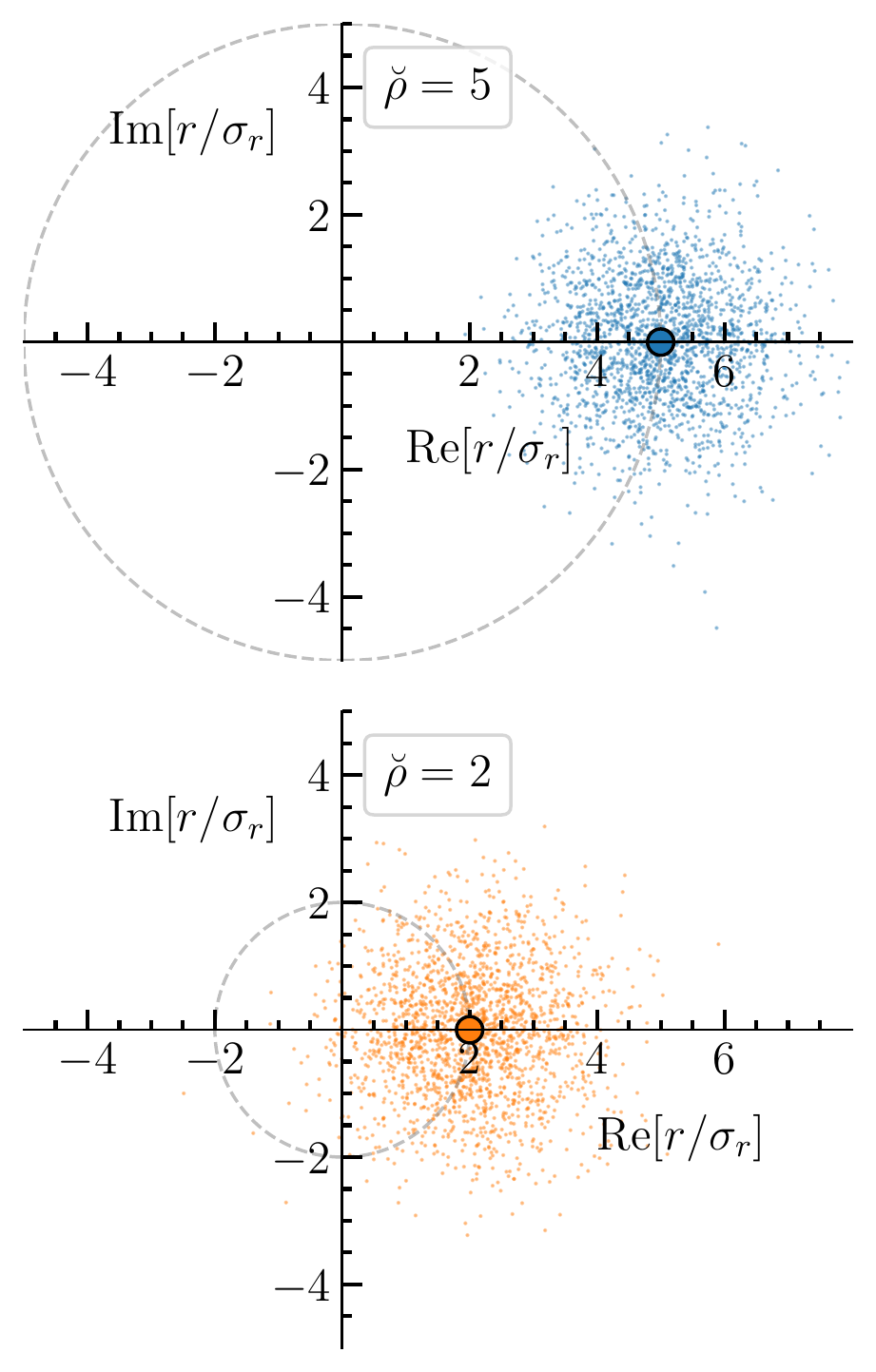}
\caption{Two thousand random realizations of the measured complex correlation coefficient $r/\sigma_r$ given intrinsic signal-to-noise $\breve{\rho}$ = 2 and~5.
}
\label{fig:scatter}
\end{figure}

Here, we discuss the statistical distributions of measured amplitude and phase quantities used for model fitting, quality of the normal distribution approximations, and influence of the estimation of an intrinsic signal-to-noise parameter $\breve{\rho}$. In the thermal-noise-dominated regime, the fundamental measured quantity, complex correlation coefficient $r$, follows a circularly-symmetric complex normal distribution with mean $\breve{\rho}\,\sigma_r$. Without loss of generality, we choose our coordinates such that the mean of the complex correlation distribution is real. An associated standard deviation of both real and imaginary components $\sigma_r$ can be computed from first principles \citep{tms}. Hence, it is useful to work with the normalized complex random variable $r/\sigma_r$, with unit standard deviation (\autoref{fig:scatter}). Probability densities for closure quantities, as shown in \autoref{fig:closure_phase_distr} and \autoref{fig:closure_ampl_distr}, can then be derived from the ones presented here with elementary operations such as convolution.

\subsection{Phase and amplitude distributions}

\begin{figure*}
    \centering
    \includegraphics[width=6.5in]{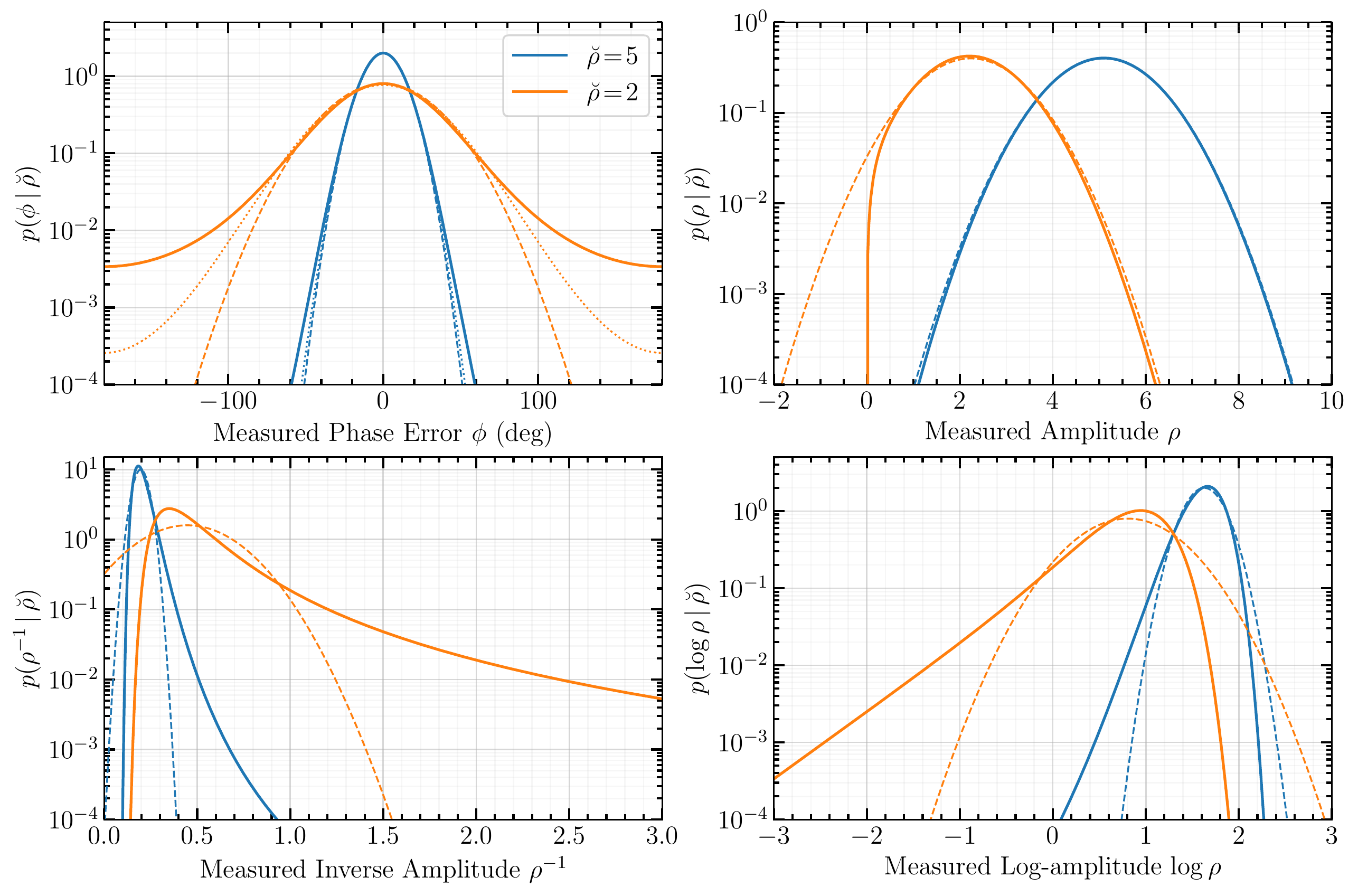}
    \caption{Analytic distributions of phase and amplitude quantities (continuous lines). Normal distribution approximations, exact in the $\breve{\rho} \rightarrow \infty$ limit, are shown with dashed lines. For the visibility phase (top left panel), the von Mises distribution approximation is shown with a dotted line. All presented approximations assume knowledge of a hidden parameter $\breve{\rho}$, which in general must be estimated from noisy measurements.}
    \label{fig:distributions_measured}
\end{figure*}

The correlation coefficient phase $\phi$ is the argument of a circular complex normal variable and, as such, obeys the following circular distribution \citep{tms}:
\begin{multline}
p(\phi| \breve{\rho}) = \frac{1}{2 \pi} \exp \left( {-\frac{\breve{\rho}^2}{2}} \right) \Bigg\{ 1 + \sqrt{\frac{\pi}{2}}\breve{\rho}\, \cos \phi \,\exp \left(\frac{\breve{\rho}^2 \, \cos^2 \phi}{2} \right)  \\
\times \left[1 + {\rm Erf} \left(\frac{\breve{\rho} \, \cos \phi }{\sqrt{2}} \right) \right] \Bigg\}.
\end{multline}
We choose the coordinates in such a way that the true visibility phase is zero, and we denote the error function with ${\rm Erf}$. Examples showing the probability density $p(\phi | \breve{\rho})$ for different values of $\breve{\rho}$ are shown in the top left panel of \autoref{fig:distributions_measured}. This somewhat complicated distribution can be approximated either by a normal distribution (dashed lines in \autoref{fig:distributions_measured}),
\begin{align}
&p_{\rm N}(\phi | \breve{\rho} )=     \frac{\breve{\rho}}{\sqrt{2 \pi}} \exp \left( \frac{- \phi^2 \breve{\rho}^2}{2} \right), 
&\sigma_{\rm N}^2 = \frac{1}{\breve{\rho}^2},
\end{align}
or by the von Mises distribution \citep[dotted lines in \autoref{fig:distributions_measured};][]{Christian_2019},
\begin{align}
& p_{\rm M} (\phi | \breve{\rho})   = \frac{\breve{\rho}\, \exp \left(\breve{\rho}^2 \, \cos \phi \right)}{2 \pi \, I_0(\breve{\rho}^2)},
& \sigma_{\rm M}^2 = 1 -  \frac{I_1(\breve{\rho}^2/2)}{I_0(\breve{\rho}^2/2)}.
\end{align}
which outperforms the normal distribution for low S/N.

For a given model $\breve{\rho}$, the measured normalized correlation coefficient amplitude ${\rho = |r/\sigma_r|} \geq 0$ follows a Rice distribution,
\begin{equation}
  p(\rho | \breve{\rho})= \rho \, \exp\left(-\frac{\rho^2+\breve{\rho}^2}
{2}\right)I_0\left(\rho \breve{\rho}\right) \, ,
\label{eq:rice}
\end{equation}
where $I_0$ is a modified Bessel function of the first kind with order zero.
Distributions of visibility amplitude are shown in Figure \ref{fig:distributions_measured} (top right panel). The dashed lines represent a normal distribution approximation with mean
$m^2 = (\breve{\rho}^2 + 1)$
and unit standard deviation, which is accurate in the limit $\breve{\rho} \rightarrow \infty$.
The normal approximation can not properly handle strictly nonnegative random variables, which becomes a problem at low S/N. The mean of the correlation amplitude is also positively biased with respect to $\breve{\rho}$ due to its noise contribution, and we find $\Eop{\rho} = 2.272$ for $\breve{\rho} = 2$ and $\Eop{\rho} = 5.101$ for $\breve{\rho} = 5$, which illustrates why debiasing is important for incoherent averaging over many realizations and for estimating $\breve{\rho}$ from low-S/N data.

When working with closure amplitudes, we need to  utilize the reciprocal amplitudes $ y = 1/\rho$, distributed according to
\begin{equation}
  p(y | \breve{\rho})= \frac{1}{y^3} \exp\left(-\frac{1/y^2+\breve{\rho}^2}
{2}\right)I_0\left(\frac{\breve{\rho}}{y}\right) \, .
\label{eq:inv_rice}
\end{equation}
Although this distribution can be approximated at high S/N as a normal distribution with mean
$m^2 = (\breve{\rho}^2 + 1)^{-1}$ and standard deviation
$\breve{\rho}^{-2}$ (\autoref{fig:distributions_measured} bottom left panel), the probability distribution exhibits heavy tails at the low S/N, related to inversion of potentially arbitrarily small amplitude. The fact that amplitude is always positive is one indication that log amplitude
might be a more natural space in which to characterize the distribution. Another benefit of using log amplitude is that amplitude and
squared-amplitude (a more natural quantity for incoherent sums of Gaussian
components) are simply related. Logarithms of the correlation amplitude $z = \log \rho$ ($\log$ denotes a natural logarithm) obey the following log--Rice distribution
 \begin{equation}
  p(z | \breve{\rho})=  \exp\left(2z -\frac{\breve{\rho}^2}
{2} - \frac{\exp 2z}{2} \right)I_0\left( \breve{\rho} \exp z\right).
\end{equation}
The distributions of the logarithm of amplitude for different $\breve{\rho}$ are shown in Figure \ref{fig:distributions_measured}, bottom right panel. Moments of the log--Rice distribution are treatable analytically, and the distribution can be approximated with a normal distribution of mean 
$m = 0.5 \log (\breve{\rho}^2 + 1)$
and standard deviation $1/\breve{\rho}$. A more general exact treatment of incoherent averages of $M$ amplitude measurements follows.
 
\subsection{Log-amplitude ensemble distribution}

Amplitude gain is a property of antenna efficiency and system noise and is generally quite stable when compared to variations in atmospheric phase. This leads to the concept of incoherent averaging for a series of amplitude measurements \citep{rogers1995,johnson2015}.
Consider a set of $M$ independent complex visibility measurements $v_i$ where
each complex component has thermal noise of 1. Thus, $v_i = \breve{\rho}_i + n_i$ where
$\breve{\rho}_i$ is some expected signal-to-noise ratio for each measurement, and $n_i$
is a Gaussian complex random variable with $\sigma=1$ for each component. The
sum-squared amplitudes follow a $\chi^2$ distribution with $2M$ degrees of
freedom. This will be a noncentral $\chi^2$ distribution
if it includes a nonzero expected source contribution.
\begin{equation}
x = \sum_i |v_i|^2 \qquad f(x) = \chi^2_{2M,\lambda}
\end{equation}
where $\lambda$ is the non-centrality parameter,
\begin{equation}
\lambda = \sum_i |\breve{\rho}_i|^2
\end{equation}
The expectation value of $\log{x}$ is
\begin{equation}
    \Eop{\log{x}} = g_M(\lambda)
\end{equation}
and $g(\cdot)$ is the function \citep{lapidoth2003capacity}
\begin{equation}
g(\lambda) = \begin{dcases}
\log{\lambda} - \Ei\left[-\frac{\lambda}{2\sigma^2}\right]
\kern5em (\text{if }\lambda > 0) \\
\quad + \sum\limits_{j=1}^{M-1} (-1)^j \left[
  e^{-\lambda}(j-1)! - \frac{(M-1)!}{j(M-1-j)!} \right]\left(\frac{1}{\lambda}\right)^j \\
  \log{2\sigma^2} -\gamma \; + \sum\limits_{j=1}^{M-1}\frac{1}{j}
\kern5em (\text{if }\lambda = 0)
\end{dcases}
\end{equation}
where $\gamma \approx 0.577$ is the Euler--Mascheroni constant and $\Ei$ is the
exponential integral. We have introduced $\sigma$ for the case where amplitudes
are uniformly scaled away from $\sigma=1$.
From this, the expectation
value $\mathrm{E}\big[\log{\sqrt{x}}\big] = \mathrm{E}\big[\log{x}\big]/2$ is easy to calculate, for example, in
the case of a single Rice-distributed complex visibility (where ``measured'' $\rho = |v|$)
\begin{equation}
    \Eop{\log{\rho}} = \log{\breve{\rho}} - \Ei\left[-\frac{\breve{\rho}^2}{2}\right].
\label{eqn:logricebias}
\end{equation}

The log-closure amplitude $c$ is formed from the linear combination of four
log amplitudes $A,B,C,D$,
\begin{equation}
c = A+B-C-D
\end{equation}
so that the expectation value (from which a bias is derived) is trivial,
\begin{equation}
\Eop{c}\,=\,\Eop{A}\,+\,\Eop{B}\,-\,\Eop{C}\,-\,\Eop{D}
\end{equation}

To characterize the distribution of measured closure amplitudes, we require
additional moments beyond the first moment (bias). For a multivariate Gaussian
approximation suitable for a least-squares fitting of log-closure quantities with
known covariance, we need to estimate the second moment of each log amplitude.
High-order moments of the log-noncentral $\chi^2$ distribution can be
derived as Poisson-weighted infinite series of polygamma functions $\psi^{(m)}(z)$ \citep{pav2015},
\begin{equation}
\label{eqn:noncentralmoments}
\Eop{x^k} = \sum_{j=0}^\infty \frac{e^{-\lambda/2}\left(\lambda/2\right)^j}{j!} \mu'_{k_{2M+2j}}
\end{equation}
where $\mu'_{k_{2M+2j}}$ is the $k^\mathrm{th}$ moment (not central moment) of a
log chi-square ($\lambda=0$) distribution with $2M+2j$ degrees of freedom,
\begin{gather}
\mu'_n = \kappa_n + \sum_{m=1}^{n-1}\binom{n-1}{m-1}\kappa_m\mu'_{n-m} \\
\kappa_n = \begin{dcases}
\log{2} + \psi(M+j) & n = 1 \\
\psi^{(n-1)}(M+j) & n > 1
\end{dcases}
\end{gather}
in terms of cumulants $\kappa_n$. Note that the second and third cumulants are
equal to the corresponding central moments. The cumulants in terms of noncentral moments are
\begin{equation}
\kappa_n = \mu'_n - \sum_{m=1}^{n-1}\binom{n-1}{m-1}\kappa_m\mu'_{n-m}
\end{equation}
For a single log-central $\chi^2$
distribution of 2 degrees of freedom (i.e. an exponential distribution with
mean value~2), the first and second cumulants are particularly simple,
\begin{gather}
\Eop{\log{\chi^2_2}} = \log{2} + \psi(1) = \log{2} - \gamma \\
\mathrm{Var}\left[\log{\chi^2_2}\right] = \psi^1(1) = \frac{\pi^2}{6}
\end{gather}
which are the same as the cumulants for a log-Rayleigh distribution scaled
appropriately by a factor of two.

Recurrence relationships for polygamma can be used to quickly derive cumulants,
including higher-order cumulants, of the log-central $\chi^2$ distribution of
$2M$ degrees of freedom. Aside from $\kappa_1$, the higher-order cumulants
approach zero as $M\to\infty$. We see that calculating cumulants at increasing
number of degrees of freedom is simply adding one more term to the series.
\begin{gather}
\kappa_1 = \log{2} - \gamma + \sum_{k=1}^{M-1} \frac{1}{k} \\
\kappa_2 = \frac{\pi^2}{6} - \sum_{k=1}^{M-1} \frac{1}{k^2} \\
\kappa_3 = -2\,\zeta(3) + \sum_{k=1}^{M-1} \frac{2}{k^3} \\
\kappa_4 = \frac{\pi^4}{15} - \sum_{k=1}^{M-1} \frac{6}{k^4}
\end{gather}
or, more generally,
\begin{equation}
\frac{\psi^{(n-1)}(M)}{(-1)^{n}(n-1)!} = \zeta(n) -
\sum_{k=1}^{M-1}\frac{1}{k^{n}} = \sum_{k=M}^{\infty} \frac{1}{k^{n}}
\end{equation}

\begin{figure*}
\centering
\includegraphics[width=6.45in]{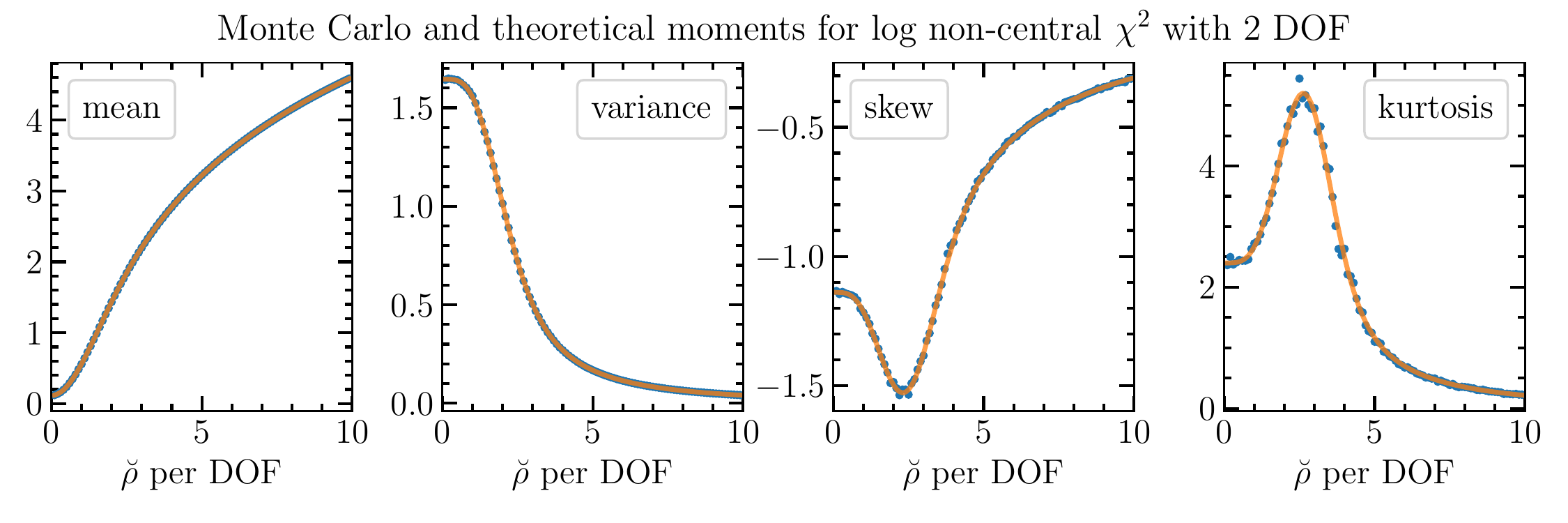}
\caption{Moments of the log-noncentral $\chi^2$ distribution (two degrees of freedom) as a function
of signal-to-noise. Blue dots correspond to a Monte Carlo estimation, while orange lines correspond to the moment expansion (\autoref{eqn:noncentralmoments}) over a finite number of Poisson terms. The noncentral $\chi^2$ distribution itself is not well captured by a small number of moments (due to the tail), but for log-closure amplitude, the propagated moments (\ref{eqn:moments1}--\ref{eqn:moments4}) can be used to fit to good approximations such as an exponentially modified Gaussian distribution.}
\label{fig:chisqmoments}
\end{figure*}

These relatively simple expressions are for the cumulants of a log {\em
central} $\chi^2$ distribution. They must be converted into moments of a {\em
noncentral} distribution by summing over the appropriate Poisson mixture
(\autoref{eqn:noncentralmoments}), which depends on the non-centrality
parameter.  The noncentral moments can then be converted back into cumulants
of the noncentral distribution to build cumulants of the log-closure amplitude
distribution. For a log-closure amplitude $c = A+B-C-D$, the cumulants of $c$ are
formed as
\begin{align}
\kappa_{c,1} &= \kappa_{A,1} + \kappa_{B,1} - \kappa_{C,1} - \kappa_{D,1} & \text{(mean)} \\
\label{eqn:moments1}
\kappa_{c,2} &= \kappa_{A,2} + \kappa_{B,2} + \kappa_{C,2} + \kappa_{D,2} & \text{(variance)} \\
\kappa_{c,3} &= \kappa_{A,3} + \kappa_{B,3} - \kappa_{C,3} - \kappa_{D,3} & (\text{skew} \times \kappa_2^{3/2})\\
\kappa_{c,4} &= \kappa_{A,4} + \kappa_{B,4} + \kappa_{C,4} + \kappa_{D,4} & (\text{ex. kurtosis} \times \kappa_2^2)
\label{eqn:moments4}
\end{align}
and so on. \autoref{fig:chisqmoments} shows the first four moments calculated this way using a finite number of nonzero terms from the Poisson mixture, and compared to a Monte Carlo estimation.

\begin{figure*}
    \centering
    \includegraphics[width=3.35in,trim=8 8 0 8, clip]{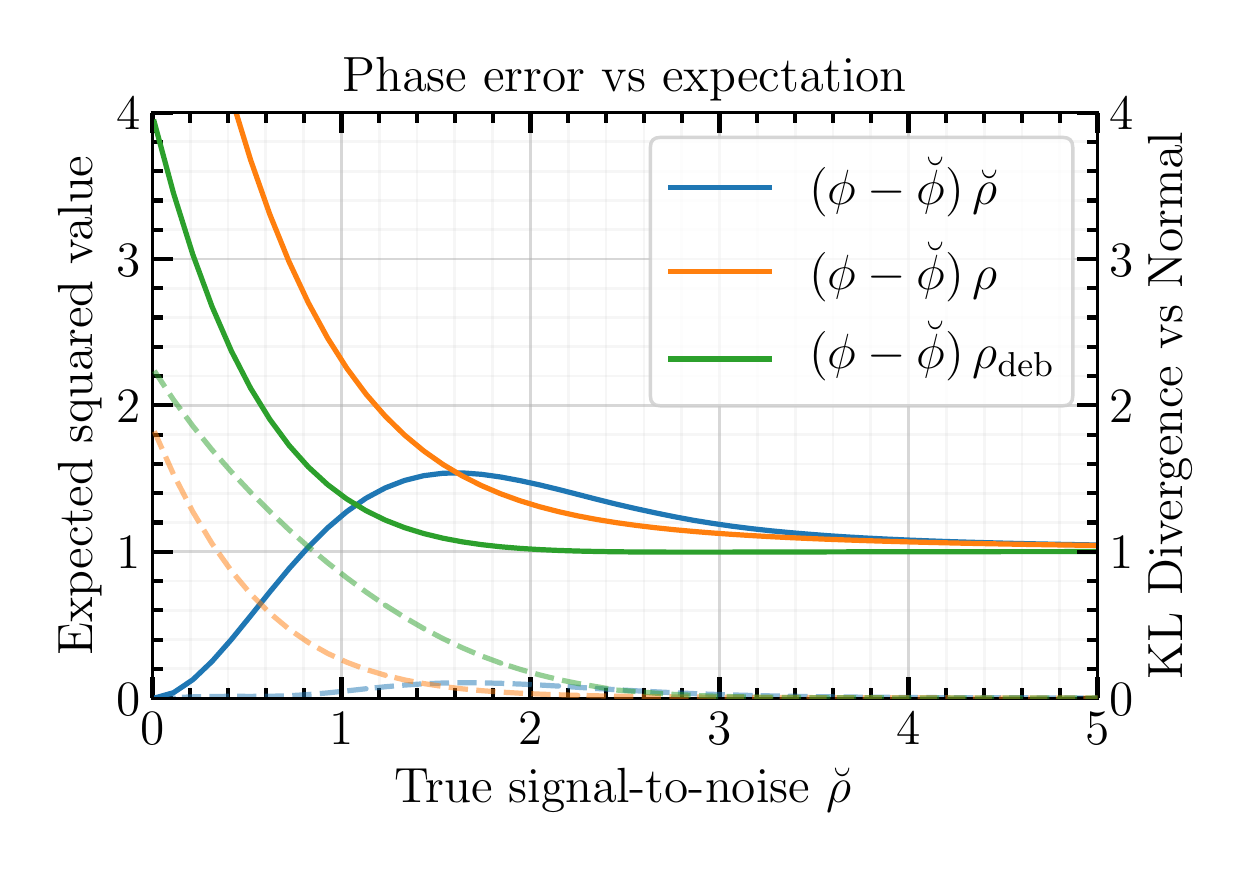}
    \includegraphics[width=3.35in,trim=0 8 8 8, clip]{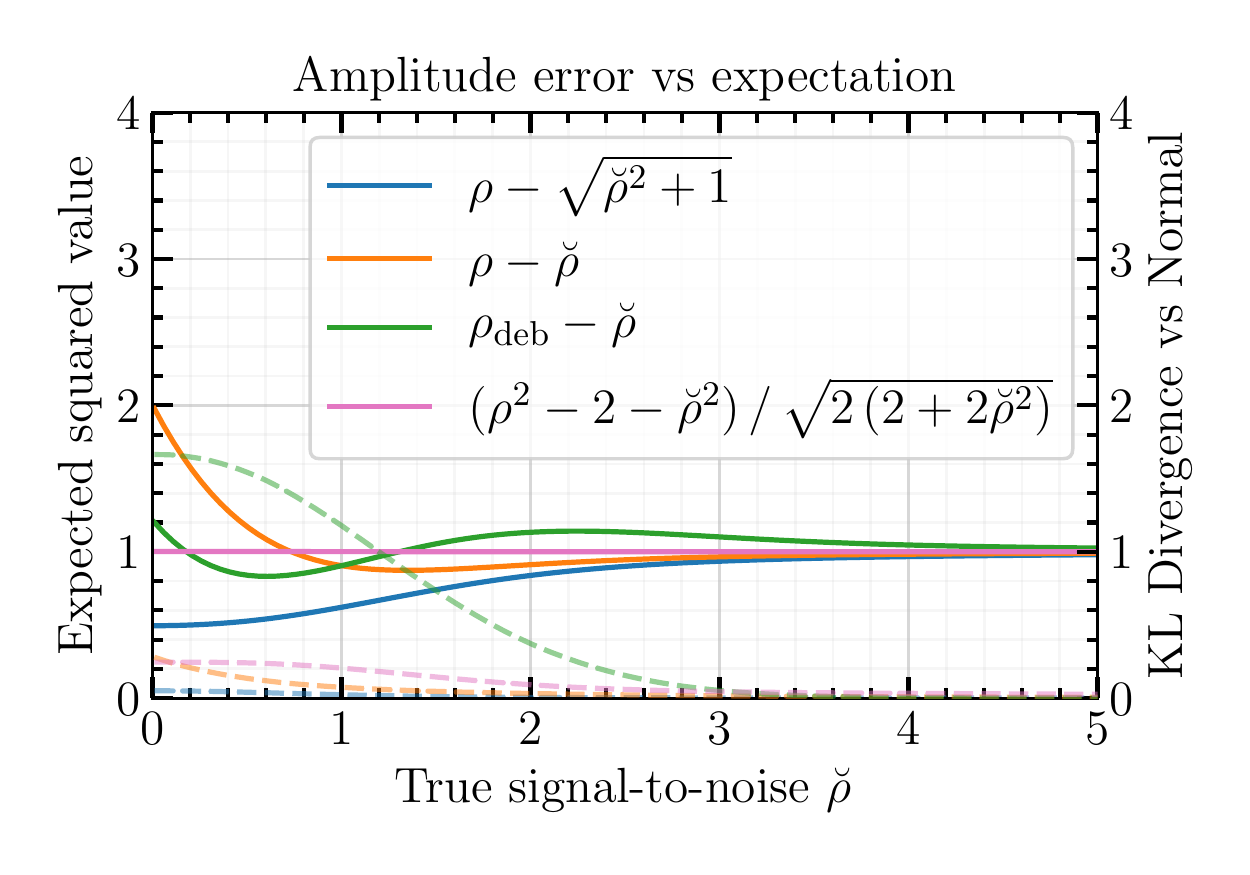} \\
    \includegraphics[width=3.35in,trim=8 8 0 4, clip]{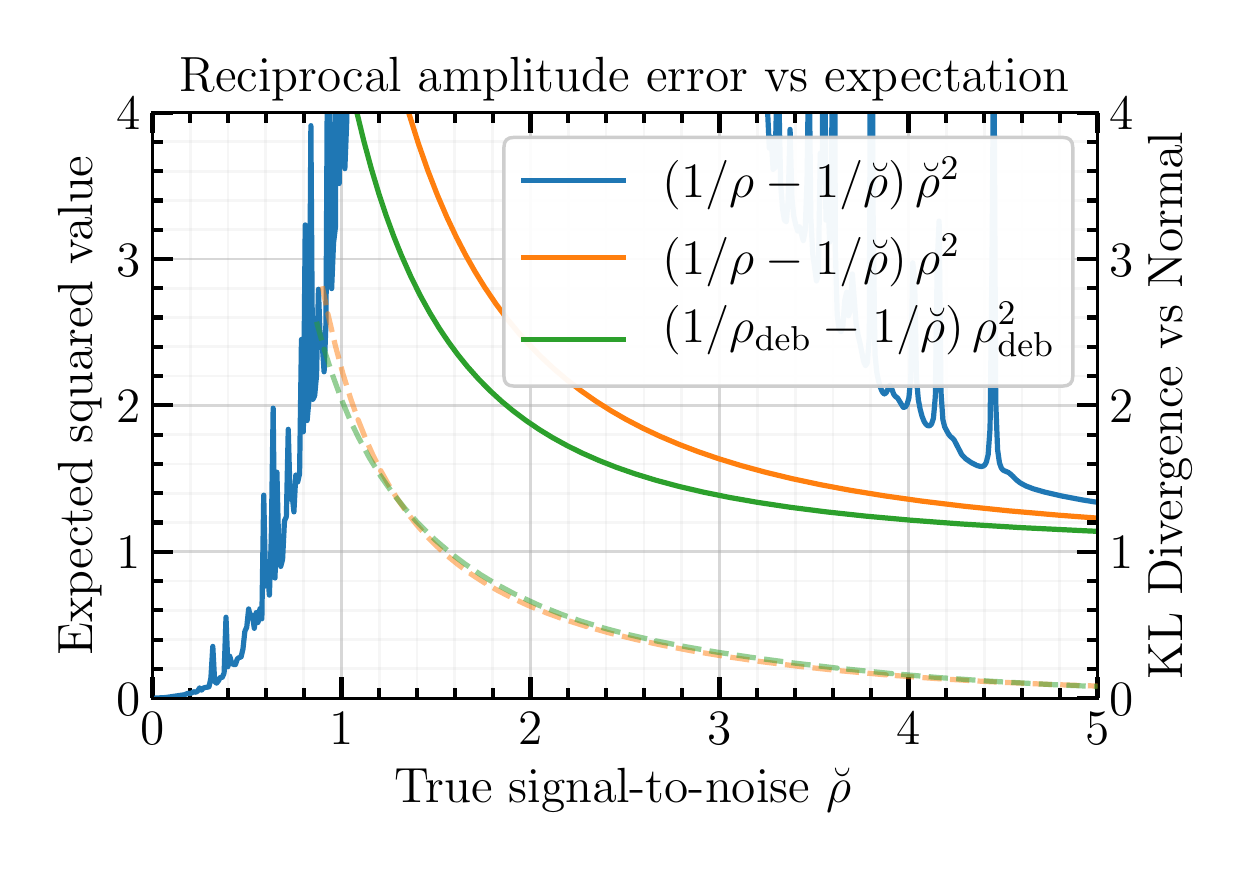}
    \includegraphics[width=3.35in,trim=0 8 8 4, clip]{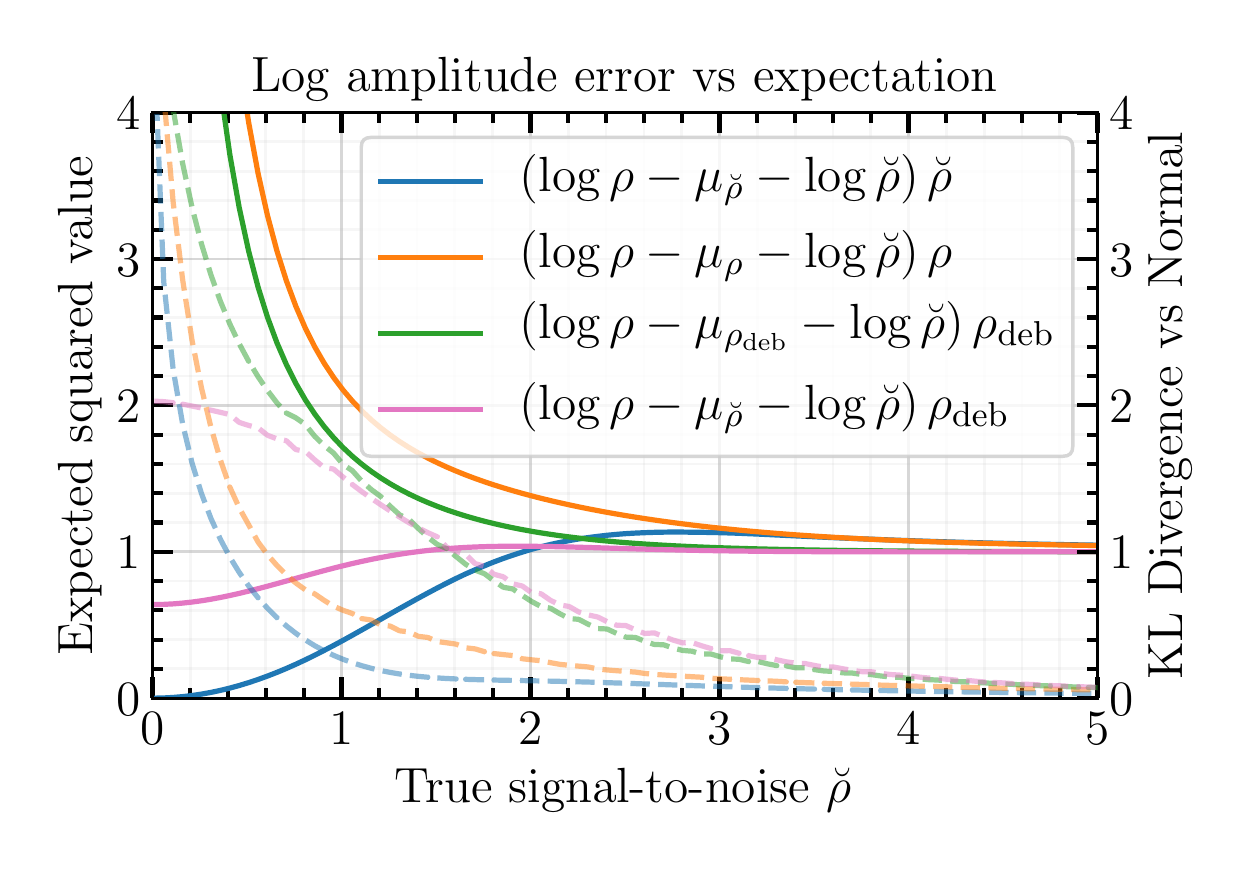}
    \caption{The four panels show the quality of the Normal approximation for different phase and amplitude distributions as a function of model S/N. The solid lines show the expected squared value of the normalized residual quantity (expected reduced $\chi^2$) from the legend, while the dashed lines show the relative entropy (Kullback--Leibler divergence) between the true distribution of each quantity and a standard Normal distribution. For example, the values at $\breve{\rho}=2$ reflect an ensemble of complex visibilities with intrinsic $\breve{\rho} = 2$ and measured $\rho=|r/\sigma_r|$ for each random realization (see \autoref{fig:scatter}). The orange line in the top left panel thus corresponds to an expected squared deviation in measured phase $\phi=\Arg{\rho}$ away from the truth value $\breve{\phi}$, where the deviation is normalized by an empirical error estimate $\sigma_\phi = 1/\rho$. Other curves show different error estimates based on the model $\breve{\rho}$ (which is typically not known in a real observation), or a noise-debiased estimate $\rho_\mathrm{deb} = \sqrt{\rho^2 -1}$. For log amplitude, $\mu$ corresponds to the small expected bias from \autoref{eqn:logricebias}.}
    \label{fig:chisq_invlogamp}
\end{figure*}

\subsection{Quality of distribution approximations}

The true underlying value of $\breve{\rho}$ remains generally unknown, and our ability to estimate $\breve{\rho}$ will influence the quality of our derived distribution for the measured value. This contributes a source of error in addition to a possible mismatch due to any approximations used. In \autoref{fig:chisq_invlogamp} we evaluate the influence of both these effects using a $\chi^2$ test and also by calculating the Kullback--Leibler divergence between the ground truth and a normal distribution characterized by the two approximated moments. At low S/N, uncertainties are typically underestimated leading to large $\chi^2$ values. In the context of inferred model parameters, this leads to erroneously narrow derived posteriors.

Given knowledge of true $\breve{\rho}$, it is possible in principle to achieve perfect statistics due to full knowledge of the distribution, rather than the high-S/N approximations used in the figure. However, this does not extend to the empirical (realistic) estimators. Furthermore, we see that the estimator with $\chi^2_r$ closest to one is not always the estimator with the best Gaussianity according to the KL divergence. Lastly, although reciprocal amplitude is very difficult to characterize due to values near zero, motivating the use of log amplitude, visibility amplitude itself is comparatively well behaved and easy to approximate, even to low S/N.

\section{Design and covariance matrix construction}
\label{app:AllDesign}

\subsection{Baseline phase and amplitude matrices}
\label{app:VisDesign}

A pair of complex visibilities may share a station, so station-based gain effects result in covariances between visibility measurements.  The covariance between two visibility phase measurements \vps[ij] and \vps[k\ell] can be expressed as

\begin{eqnarray}
\text{Cov} (\vps[ij],\vps[k\ell]) & = & \vpvar[ij] \left( \delta_{ik} \delta_{j\ell} - \delta_{i \ell} \delta_{j k} \right) \nonumber \\
& + & \gpvar[i] \left( \delta_{ik} - \delta_{i\ell} \right) - \gpvar[j] \left( \delta_{jk} - \delta_{j\ell} \right) , \label{eqn:VisPhasePairwiseCovariance}
\end{eqnarray}

\noindent where \vpvar[ij] is the thermal variance of the visibility phase measurement \vps[ij], \gpvar[i] is the gain phase variance for station $i$, and $\delta_{ij}$ is the Kronecker delta.  A similar expression holds for the covariance between two log visibility amplitude measurements \lvas[ij] and \lvas[k\ell],

\begin{eqnarray}
\text{Cov} (\lvas[ij],\lvas[k\ell]) & = & \lvavar[ij] \left( \delta_{ik} \delta_{j\ell} + \delta_{i \ell} \delta_{j k} \right) \nonumber \\
& + & \lgavar[i] \left( \delta_{ik} + \delta_{i\ell} \right) + \lgavar[j] \left( \delta_{jk} + \delta_{j\ell} \right) , \label{eqn:LogVisAmpPairwiseCovariance}
\end{eqnarray}

\noindent where \lvavar[ij] is the thermal variance of the log visibility amplitude measurement \lvas[ij] and \lgavar[i] is the log gain amplitude variance for station $i$.

We can see from \autoref{eqn:VisPhasePairwiseCovariance} and \autoref{eqn:LogVisAmpPairwiseCovariance} that the visibility measurement covariances separate into baseline-based and station-based terms.  We can thus write the visibility phase covariance matrix \vpcov as

\begin{equation}
\vpcov = \vpd \gpcov \vpd^{\transpose} + \; \textbf{S}_{\boldsymbol{\phi}} .
\label{eqn:VisibilityCovarianceMatrix}
\end{equation}

\noindent The \gpcov and $\textbf{S}_{\boldsymbol{\phi}}$ matrices are $\gpn \times \gpn$ and $\vpn \times \vpn$ diagonal matrices constructed from the individual station gain phase variances and visibility phase variances, respectively.  The visibility phase ``design matrix'' \vpd is rectangular in general, with \vpn rows and \gpn columns, and provides a mapping from the station-based representation to the baseline-based representation.  Each row of \vpd contains only two nonzero entries, the first being a $1$ and the second being a $-1$.  There are \vpn different ways of writing a length-\gpn row in this fashion, and these constitute the \vpn rows of the matrix.  The ordering of these rows depends on the chosen baseline ordering scheme.  In this section, we assume a ``second station first'' baseline ordering scheme, which increments the visibility phases via a nested loop method.  The ``inner loop'' iterates through the second station in increasing order, and the ``outer loop'' iterates through the first station in increasing order.  An example such ordering would be (\vps[12], \vps[13], \vps[14], \ldots, \vps[1\gpn], \vps[23], \vps[24], \ldots, \vps[2\gpn], \vps[34], \ldots, \vps[\gpn-1,\gpn]), where the indices here correspond to the two stations forming each baseline.

For a general \gpn-station array with $\gpn > 2$, we present the following recursive relationship for the visibility phase design matrix:

\begin{equation}
\vpd_{\gpn} = \begin{pmatrix}
\boldsymbol{1} & - \textbf{I}_{\gpn-1} \\
\boldsymbol{0} & \vpd_{\gpn-1}
\end{pmatrix} , \label{eqn:VisPhaseDesignMatrix}
\end{equation}

\noindent where $\boldsymbol{1}$ is an $(\gpn-1) \times 1$ vector containing only 1s, $\boldsymbol{0}$ is an $\binom{\gpn-1}{2} \times 1$ vector containing only 0s, $\textbf{I}_{\gpn-1}$ is the identity matrix of rank $\gpn-1$, and $\vpd_{\gpn-1}$ is the visibility phase design matrix for an array with $\gpn-1$ stations.  The rank of $\vpd_{\gpn}$ is equal to $\gpn - 1$.  \autoref{tab:VisibilityPhaseTable} lists examples of \vpd and the corresponding \vpcov matrices.

The log visibility amplitude design matrix \lvad shares the same structure as the visibility phase design matrix, with the only difference being that the negative elements of \vpd become positive for \lvad.  As a result, for $\lgan > 2$ the rank of $\lvad_{\lgan}$ for the log visibility amplitudes is equal to \lgan.  \autoref{tab:VisibilityAmpTable} lists examples of log visibility amplitude design and covariance matrices.

\begin{table*}
\centering{}
\begin{tabular}{lccc}
\hline \hline
 & &  \multicolumn{2}{c}{\textbf{Number of stations (\gpn)}} \\ \cline{3-4}
\textbf{Matrix} & \textbf{Shape}   &   $\gpn=2$     &   $\gpn=3$ \\
\hline
\gpcov & $\gpn \times \gpn$ & $\begin{pmatrix}
\gpvar[1] & 0 \\
0 & \gpvar[2]
\end{pmatrix}$ & $\begin{pmatrix}
\gpvar[1] & 0 & 0 \\
0 & \gpvar[2] & 0 \\
0 & 0 & \gpvar[3]
\end{pmatrix}$ \\
\hline
$\textbf{S}_{\boldsymbol{\phi}}$ & $\vpn \times \vpn$ & $\begin{pmatrix}
\vpvar[12]
\end{pmatrix}$ & $\begin{pmatrix}
\vpvar[12] & 0 & 0 \\
0 & \vpvar[13] & 0 \\
0 & 0 & \vpvar[23]
\end{pmatrix}$ \\
\hline
\vpd & $\vpn \times \gpn$ & $\begin{pmatrix}
1 & -1
\end{pmatrix}$ & $\begin{pmatrix}
1 & -1 & 0 \\
1 & 0 & -1 \\
0 & 1 & -1
\end{pmatrix}$ \\
\hline
\vpcov & $\vpn \times \vpn$ & $\begin{pmatrix}
\vpvar[12] + \gpvar[1] + \gpvar[2]
\end{pmatrix}$ & $\begin{pmatrix}
\vpvar[12] + \gpvar[1] + \gpvar[2] & \gpvar[1] & - \gpvar[2] \\
\gpvar[1] & \vpvar[13] + \gpvar[1] + \gpvar[3] & \gpvar[3] \\
- \gpvar[2] & \gpvar[3] & \vpvar[23] + \gpvar[2] + \gpvar[3]
\end{pmatrix}$ \\
\hline
\end{tabular}
\caption{Visibility phase design and covariance matrices for two- and three-element arrays, along with matrices relevant for their construction.  Here, $\vpn = \binom{\gpn}{2}$ is the number of baselines.}
\label{tab:VisibilityPhaseTable}
\end{table*}

\begin{table*}
\centering{}
\begin{tabular}{lccc}
\hline \hline
 & &  \multicolumn{2}{c}{\textbf{Number of stations (\lgan)}} \\ \cline{3-4}
\textbf{Matrix} & \textbf{Shape}   &   $\lgan=2$     &   $\lgan=3$ \\
\hline
\lgacov & $\lgan \times \lgan$ & $\begin{pmatrix}
\lgavar[1] & 0 \\
0 & \lgavar[2]
\end{pmatrix}$ & $\begin{pmatrix}
\lgavar[1] & 0 & 0 \\
0 & \lgavar[2] & 0 \\
0 & 0 & \lgavar[3]
\end{pmatrix}$ \\
\hline
$\textbf{S}_{\textbf{\textit{a}}}$ & $\lvan \times \lvan$ & $\begin{pmatrix}
\lvavar[12]
\end{pmatrix}$ & $\begin{pmatrix}
\lvavar[12] & 0 & 0 \\
0 & \lvavar[13] & 0 \\
0 & 0 & \lvavar[23]
\end{pmatrix}$ \\
\hline
\lvad & $\lvan \times \lgan$ & $\begin{pmatrix}
1 & 1
\end{pmatrix}$ & $\begin{pmatrix}
1 & 1 & 0 \\
1 & 0 & 1 \\
0 & 1 & 1
\end{pmatrix}$ \\
\hline
\lvacov & $\lvan \times \lvan$ & $\begin{pmatrix}
\lvavar[12] + \lgavar[1] + \lgavar[2]
\end{pmatrix}$ & $\begin{pmatrix}
\lvavar[12] + \lgavar[1] + \lgavar[2] & \lgavar[1] & \lgavar[2] \\
\lgavar[1] & \lvavar[13] + \lgavar[1] + \lgavar[3] & \lgavar[3] \\
\lgavar[2] & \lgavar[3] & \lvavar[23] + \lgavar[2] + \lgavar[3]
\end{pmatrix}$ \\
\hline
\end{tabular}
\caption{Log visibility amplitude design and covariance matrices for two- and three-station arrays, along with matrices relevant for their construction.  Here, $\lvan = \binom{\lgan}{2}$ is the number of baselines.}
\label{tab:VisibilityAmpTable}
\end{table*}

\subsection{Closure phase matrices}
\label{sec:CphaseDesign}

It is possible for two closure triangles to have a baseline in common, so in general, two closure phase measurements may be covariant.  The covariance between two closure phase measurements \cps[ijk] and \cps[\ell m n] can be expressed as

\begin{eqnarray}
\text{Cov} (\cps[ijk],\cps[\ell m n]) & = & \vpvar[ij] \left( \delta_{i \ell} \delta_{j m} + \delta_{i m} \delta_{j n} + \delta_{i n} \delta_{j \ell} \right) \nonumber \\
& - & \vpvar[ij] \left( \delta_{i \ell} \delta_{j n} + \delta_{i m} \delta_{j \ell} + \delta_{i n} \delta_{j m} \right) \nonumber \\
& + & \vpvar[jk] \left( \delta_{j \ell} \delta_{k m} + \delta_{j m} \delta_{k n} + \delta_{j n} \delta_{k \ell} \right) \nonumber \\
& - & \vpvar[jk] \left( \delta_{j \ell} \delta_{k n} + \delta_{j m} \delta_{k \ell} + \delta_{j n} \delta_{k m} \right) \nonumber \\
& + & \vpvar[ik] \left( \delta_{i \ell} \delta_{k n} + \delta_{i m} \delta_{k \ell} + \delta_{i n} \delta_{k m} \right) \nonumber \\
& - & \vpvar[ik] \left( \delta_{i \ell} \delta_{k m} + \delta_{i m} \delta_{k n} + \delta_{i n} \delta_{k \ell} \right) .
\end{eqnarray}

\noindent This lengthy expression encodes two symmetries of closure phases.  The first symmetry is a ``cycling invariance'',

\begin{equation}
\cps[ijk] = \cps[jki] = \cps[kij] , \label{eqn:CPhaseCycling}
\end{equation}

\noindent which indicates that the choice of starting baseline does not affect the value of the closure phase.  The second symmetry is a sign flip imparted upon reversing the direction of the sequence,

\begin{equation}
\cps[ijk] = - \cps[jik] . \label{eqn:CPhaseReversal}
\end{equation}

\begin{figure*}
    \centering
    \includegraphics[width=0.2\textwidth]{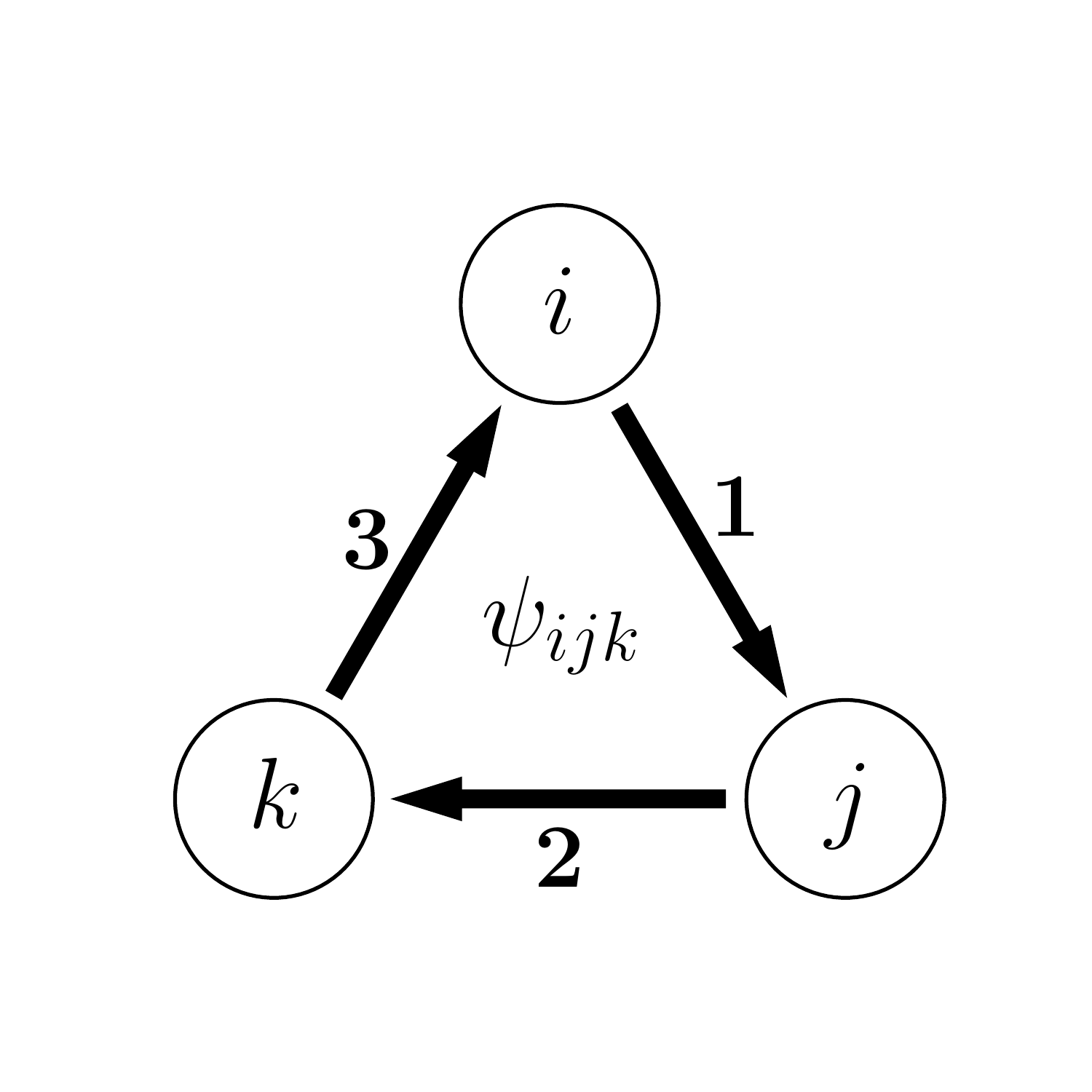}
    \hspace{2in}
    \includegraphics[width=0.2\textwidth]{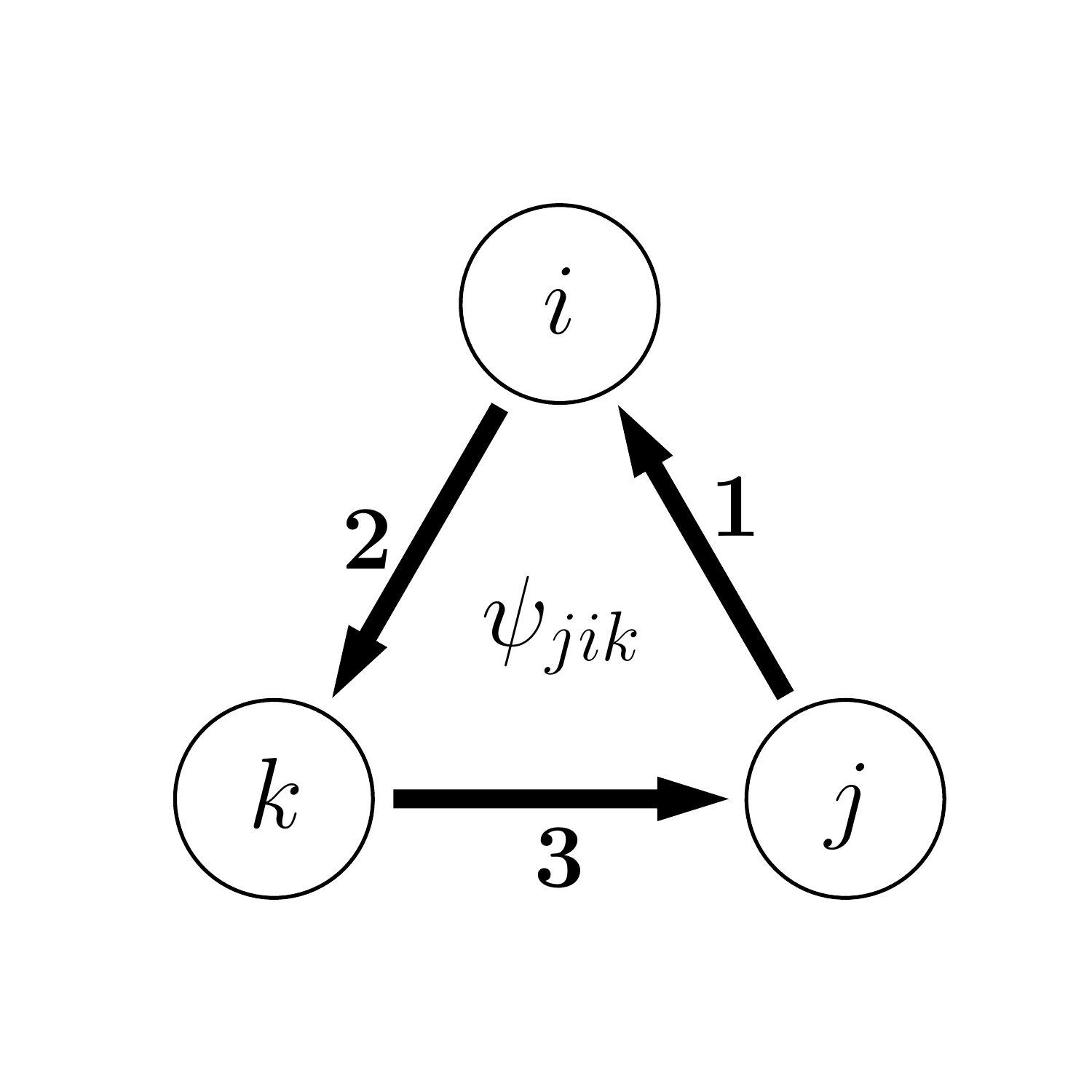}\\
    \includegraphics[width=0.2\textwidth]{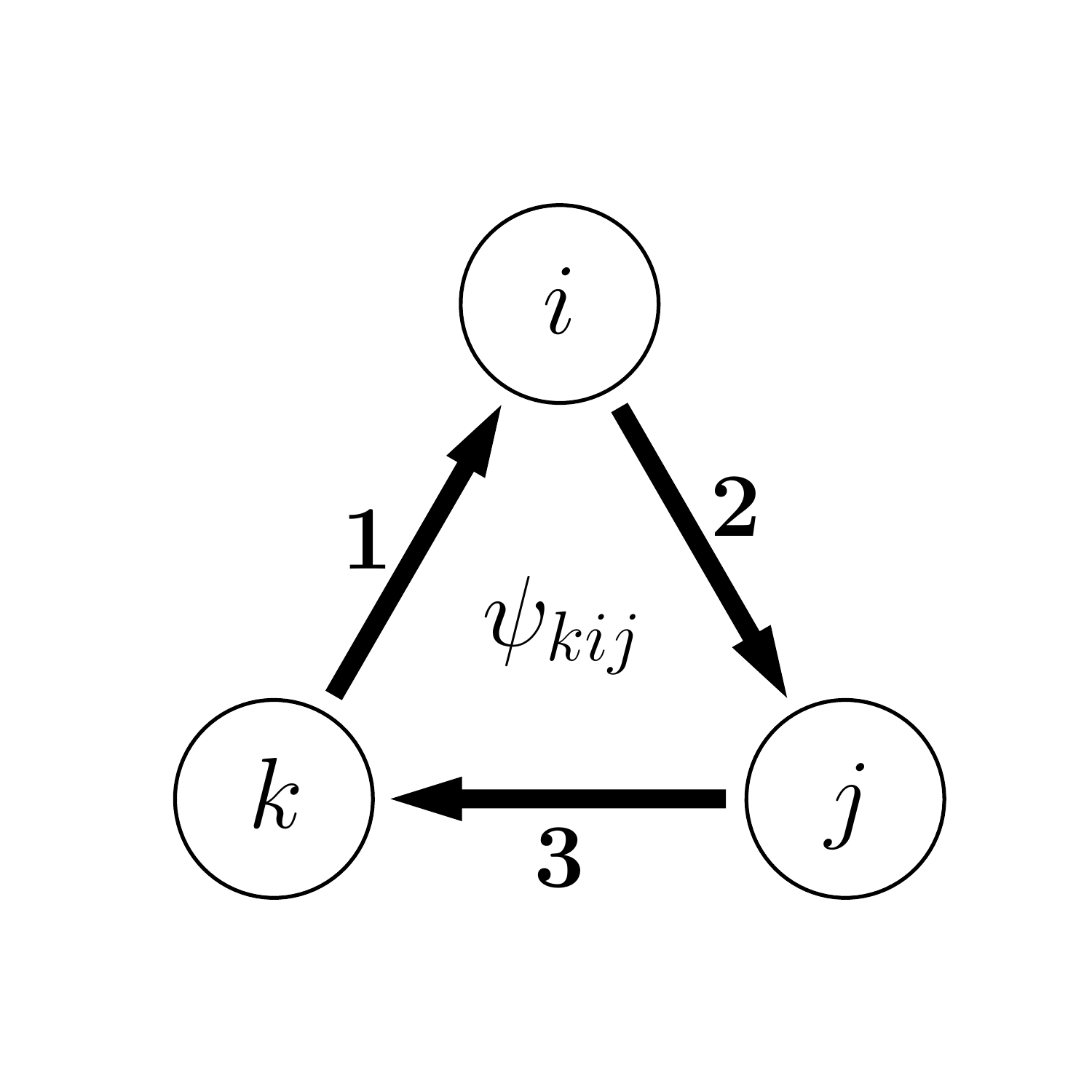}
    \includegraphics[width=0.2\textwidth]{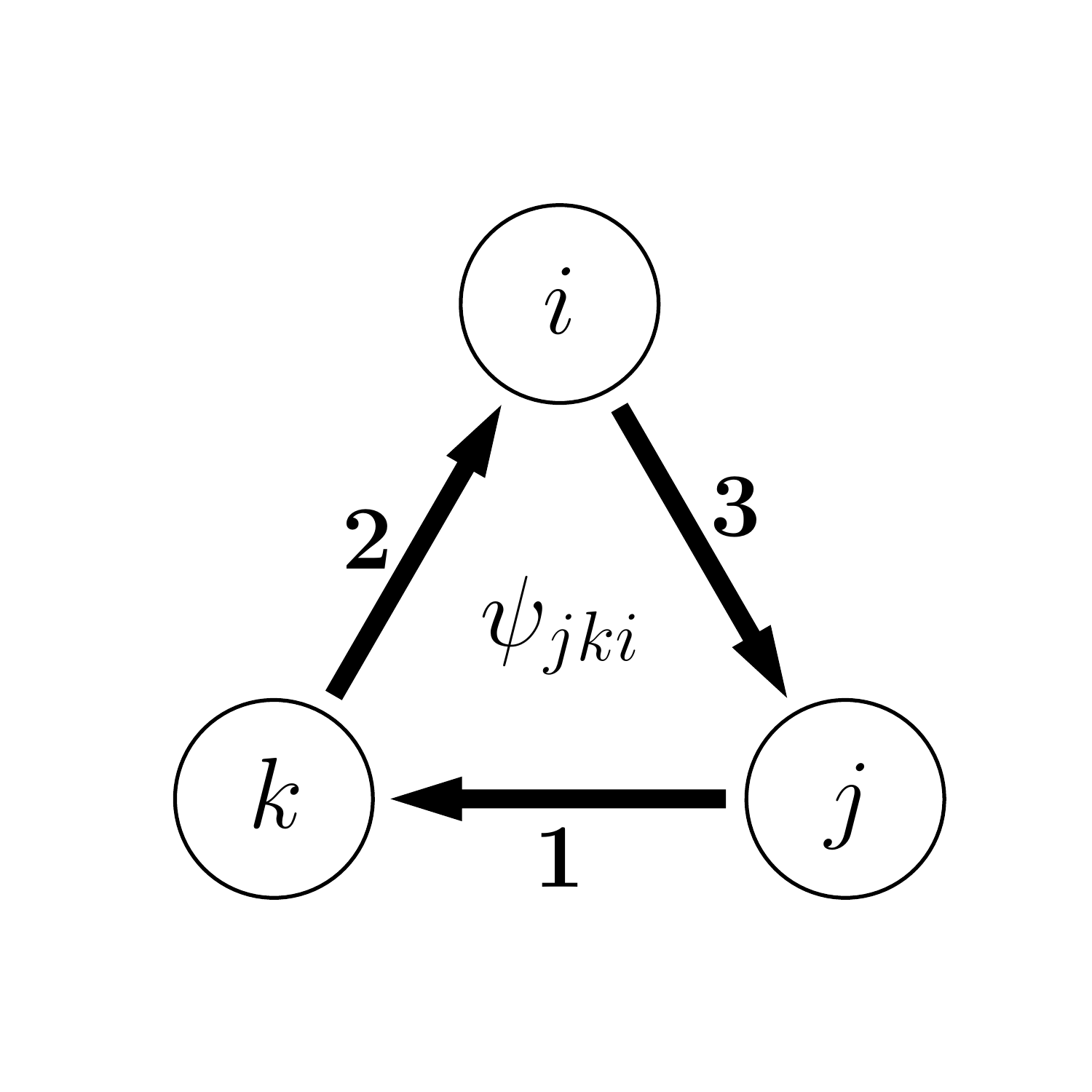}
    \hspace{0.5in}
    \includegraphics[width=0.2\textwidth]{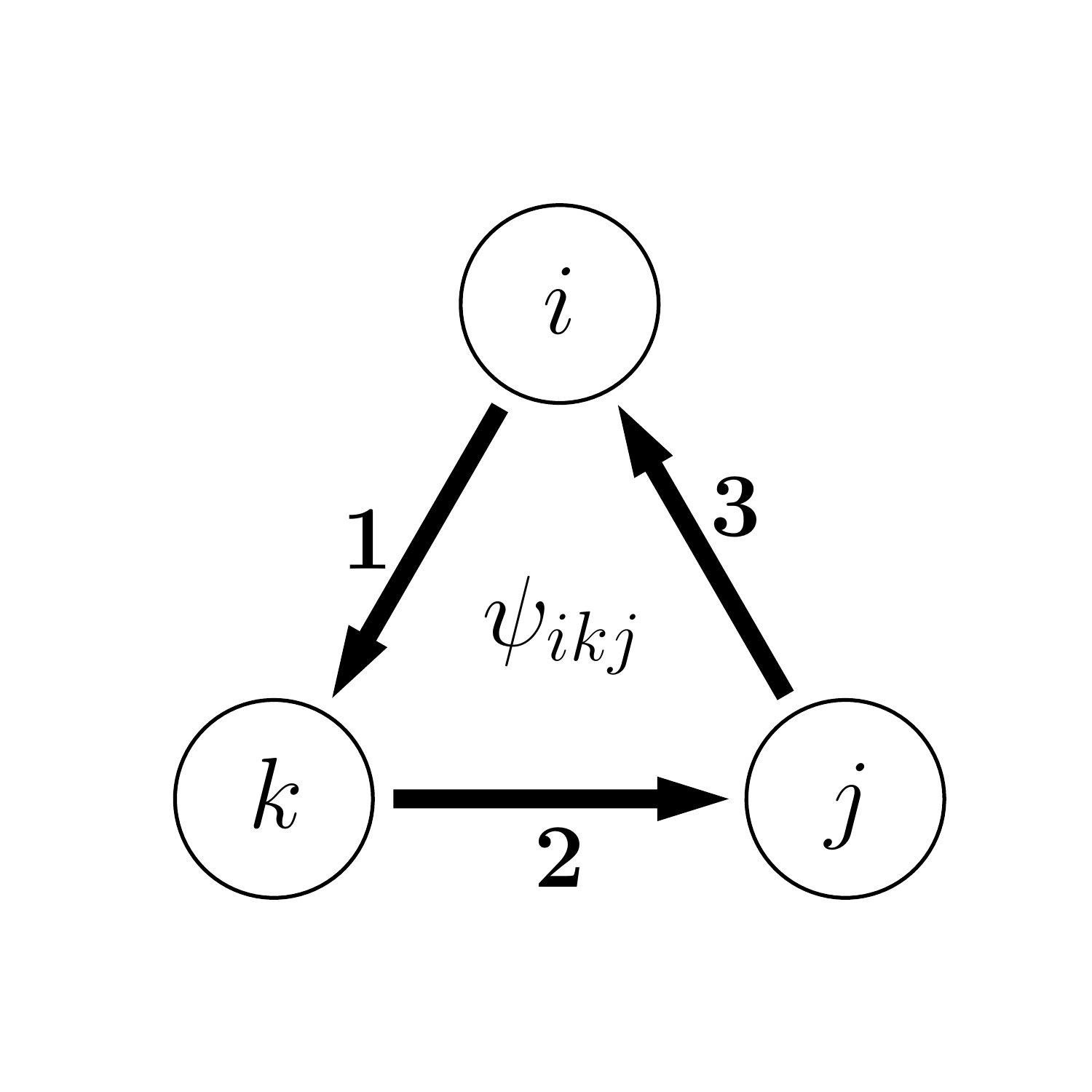}
    \includegraphics[width=0.2\textwidth]{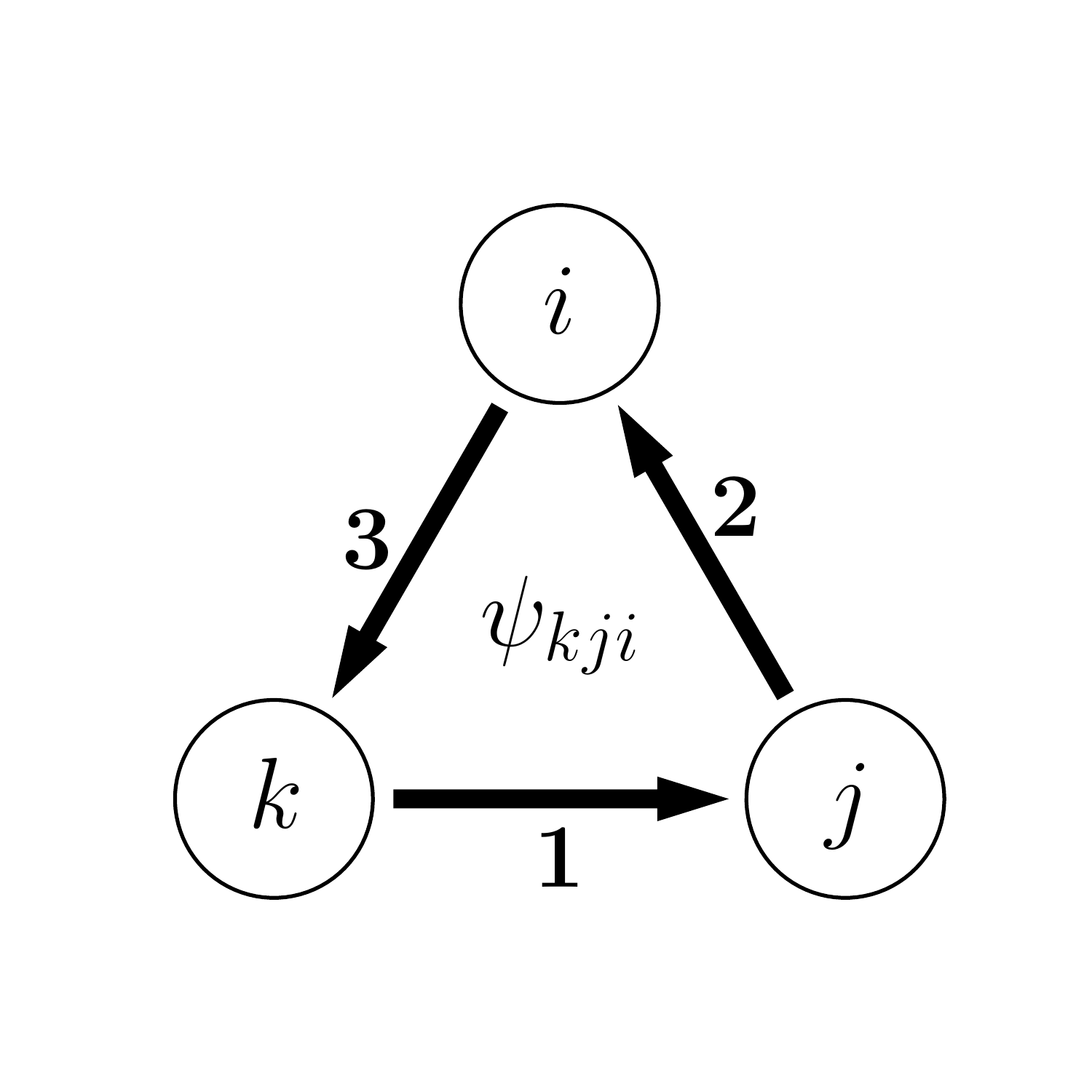}
    \caption{Diagrams of closure phase symmetries for a single triangle containing stations $i$, $j$, and $k$, with baselines numbered in the sequence used to construct the closure phase.  All closure phases in the left block of diagrams have the same value, and all closure phases in the right block of diagrams have the same value; however, the values corresponding to the two blocks of diagrams differ in sign.  The left three diagrams illustrate closure phases constructed in a clockwise manner using a different starting baseline each time; the value of the closure phase is invariant to the choice of starting baseline (\autoref{eqn:CPhaseCycling}).  The left and right blocks of diagrams differ by a reversal in the direction of closure phase construction; the value of the closure phase changes sign upon direction reversal (\autoref{eqn:CPhaseReversal}).}
    \label{fig:cphase_symmetries}
\end{figure*}

\noindent These symmetries are illustrated in \autoref{fig:cphase_symmetries}.


As with the visibilities (see \autoref{app:VisDesign}), we can construct a design matrix \cpd that maps from the visibility phase space to the closure phase space,

\begin{equation}
\cpv = \cpd \vpv ,
\end{equation}

\noindent where \vpv and \cpv are vectors of visibility phases and closure phases, respectively.  This design matrix allows us to express the closure phase covariance matrix \cpcov in terms of the visibility phase covariance matrix \vpcov,

\begin{equation}
\cpcov = \cpd \, \vpcov \cpd^{\transpose} . \label{eqn:CPCovariance}
\end{equation}

\noindent By construction, the closure phases use combinations of visibility phases for which the gain contributions cancel, a property referred to as ``phase aberration annihilation'' by \cite{Lannes_1991}.  This cancellation manifests in the design matrices as well: the product of closure phase and visibility phase design matrices evaluates to the zero matrix,

\begin{equation}
\cpd \vpd = \boldsymbol{0} . \label{eqn:AberrationAnnihilation}
\end{equation}

\noindent We can thus express the closure phase covariance matrix more simply in terms of the diagonal matrix containing only visibility phase thermal variances,

\begin{equation}
\cpcov = \cpd \, \textbf{S}_{\boldsymbol{\phi}} \cpd^{\transpose} .
\end{equation}

For a general \gpn-station array with $\gpn > 3$, we present the following recursive relationship for constructing the design matrix corresponding to a maximal set of closure phases:

\begin{equation}
\cpd_{\gpn,\text{max}} = \begin{pmatrix}
\vpd_{\gpn-1} & \textbf{I}_{\binom{\gpn-1}{2}} \\
\boldsymbol{0} & \cpd_{\gpn-1,\text{max}}
\end{pmatrix} ,
\end{equation}

\noindent where $\vpd_{\gpn-1}$ is the visibility phase design matrix for an array with $\gpn-1$ stations (see \autoref{eqn:VisPhaseDesignMatrix}), $\boldsymbol{0}$ is an $\binom{\gpn-1}{3} \times (\gpn-1)$ matrix containing only 0s, $\textbf{I}_{\binom{\gpn-1}{2}}$ is the identity matrix of rank $\binom{\gpn-1}{2}$, and $\cpd_{\gpn-1,\text{max}}$ is the maximal closure phase design matrix for an array with $\gpn-1$ stations. \\

To obtain a minimal (nonredundant) set of closure phases for an $\gpn$-station array, we can use a modified design matrix:

\begin{equation}
\cpd_{\gpn} = \begin{pmatrix}
\vpd_{\gpn-1} & \textbf{I}_{\binom{\gpn-1}{2}}
\end{pmatrix} . \label{eqn:MinimalCphaseDesign}
\end{equation}

\noindent \autoref{tab:ClosurePhaseTable} lists example closure phase design and covariance matrices.

\begin{table*}
\centering{}
\begin{tabular}{lccc}
\hline \hline
 & &  \multicolumn{2}{c}{\textbf{Number of stations (\gpn)}} \\ \cline{3-4}
\textbf{Matrix} & \textbf{Shape}   &   $\gpn=3$  & $\gpn=4$ \\
\hline
$\vpv^{\transpose}$ & $1 \times B$ & $\begin{pmatrix}
\vps[12] & \vps[13] & \vps[23]
\end{pmatrix}$ & $\begin{pmatrix}
\vps[12] & \vps[13] & \vps[14] & \vps[23] & \vps[24] & \vps[34]
\end{pmatrix}$ \\ \hline
$\cpd_{\text{max}}$ & $T \times B$ & $\begin{pmatrix}
1 & -1 & 1
\end{pmatrix}$ & $\begin{pmatrix}
1 & -1 & 0 & 1 & 0 & 0 \\
1 & 0 & -1 & 0 & 1 & 0 \\
0 & 1 & -1 & 0 & 0 & 1 \\
0 & 0 & 0 & 1 & -1 & 1
\end{pmatrix}$ \\
\hline
\cpd & $t \times B$  & $\begin{pmatrix}
1 & -1 & 1
\end{pmatrix}$ & $\begin{pmatrix}
1 & -1 & 0 & 1 & 0 & 0 \\
1 & 0 & -1 & 0 & 1 & 0 \\
0 & 1 & -1 & 0 & 0 & 1
\end{pmatrix}$ \\
\hline
$\boldsymbol{\Sigma}_{\boldsymbol{\psi},\text{max}}$ & $T \times T$ & $\begin{pmatrix}
\vpvar[12] + \vpvar[13] + \vpvar[23]
\end{pmatrix}$ & $\begin{pmatrix}
\vpvar[12] + \vpvar[13] + \vpvar[23] & \vpvar[12] & - \vpvar[13] & \vpvar[23] \\
\vpvar[12] & \vpvar[12] + \vpvar[14] + \vpvar[24] & \vpvar[14] & - \vpvar[24] \\
- \vpvar[13] & \vpvar[14] & \vpvar[13] + \vpvar[14] + \vpvar[34] & \vpvar[34] \\
\vpvar[23] & - \vpvar[24] & \vpvar[34] & \vpvar[23] + \vpvar[24] + \vpvar[34]
\end{pmatrix}$ \\
\hline
\cpcov & $t \times t$ & $\begin{pmatrix}
\vpvar[12] + \vpvar[13] + \vpvar[23]
\end{pmatrix}$ & $\begin{pmatrix}
\vpvar[12] + \vpvar[13] + \vpvar[23] & \vpvar[12] & - \vpvar[13] \\
\vpvar[12] & \vpvar[12] + \vpvar[14] + \vpvar[24] & \vpvar[14] \\
- \vpvar[13] & \vpvar[14] & \vpvar[13] + \vpvar[14] + \vpvar[34]
\end{pmatrix}$ \\
\hline
\end{tabular}
\caption{Closure phase design and covariance matrices for three- and four-element arrays.  Here, $\cpn = \binom{\gpn}{3}$ is the number of triangles in a maximal set, $t = \binom{\gpn-1}{2}$ is the number of triangles in a minimal set, and $\vpn = \binom{\gpn}{2}$ is the number of baselines.}
\label{tab:ClosurePhaseTable}
\end{table*}

\subsection{Log-closure amplitude matrices}
\label{sec:CampDesign}

A pair of closure quandrangles can have up to two baselines in common, meaning that, in general, log-closure amplitudes will be covariant.  The covariance between log-closure amplitude measurements \lcas[ijk\ell] and \lcas[mnpq] can be expressed as

\begin{eqnarray}
\text{Cov} \left( \lcas[ijk\ell] , \lcas[mnpq] \right) & = & \lcavar[ij] \left( \delta_{im} \delta_{jn} + \delta_{in} \delta_{jm} + \delta_{ip} \delta_{jq} + \delta_{iq} \delta_{jp} \right) \nonumber \\
& - & \lcavar[ij] \left( \delta_{im} \delta_{jp} + \delta_{in} \delta_{jq} + \delta_{ip} \delta_{jm} + \delta_{iq} \delta_{jn} \right) \nonumber \\
& + & \lcavar[k\ell] \left( \delta_{km} \delta_{\ell n} + \delta_{kn} \delta_{\ell m} + \delta_{kp} \delta_{\ell q} + \delta_{kq} \delta_{\ell p} \right) \nonumber \\
& - & \lcavar[k\ell] \left( \delta_{km} \delta_{\ell p} + \delta_{kn} \delta_{\ell q} + \delta_{kp} \delta_{\ell m} + \delta_{kq} \delta_{\ell n} \right) \nonumber \\
& + & \lcavar[ik] \left( \delta_{im} \delta_{kp} + \delta_{in} \delta_{kq} + \delta_{ip} \delta_{km} + \delta_{iq} \delta_{kn} \right) \nonumber \\
& - & \lcavar[ik] \left( \delta_{im} \delta_{kn} + \delta_{in} \delta_{km} + \delta_{ip} \delta_{kq} + \delta_{iq} \delta_{kp} \right) \nonumber \\
& + & \lcavar[j\ell] \left( \delta_{jm} \delta_{\ell p} + \delta_{jn} \delta_{\ell q} + \delta_{jp} \delta_{\ell m} + \delta_{jq} \delta_{\ell n} \right) \nonumber \\
& - & \lcavar[j\ell] \left( \delta_{jm} \delta_{\ell n} + \delta_{jn} \delta_{\ell m} + \delta_{jp} \delta_{\ell q} + \delta_{jq} \delta_{\ell p} \right) , \nonumber \\
\end{eqnarray}

\noindent where \lcavar[ij] is the variance in the log visibility amplitude measurement \lvas[ij].  There are three symmetries encoded in the above expression.  The first of these is a cycling invariance,

\begin{equation}
\lcas[ijk\ell] = \lcas[\ell k j i] , \label{eqn:CAmpCycling}
\end{equation}

\noindent indicating that, as for the closure phases, the log-closure amplitude value does not change with choice of starting baseline.  The second symmetry is a direction invariance,

\begin{equation}
\lcas[ijk\ell] = \lcas[k \ell i j] , \label{eqn:CAmpReversal}
\end{equation}

\noindent showing that, unlike for closure phases, the log-closure amplitude value does not change when the sequence of baselines is reversed.  The third symmetry is a sign flip imparted on the value of the log-closure amplitude upon swapping the numerator and denominator,

\begin{equation}
\lcas[ijk\ell] = - \lcas[ikj\ell] . \label{eqn:CAmpSwap}
\end{equation}

\begin{figure*}
    \centering
    \includegraphics[width=0.2\textwidth]{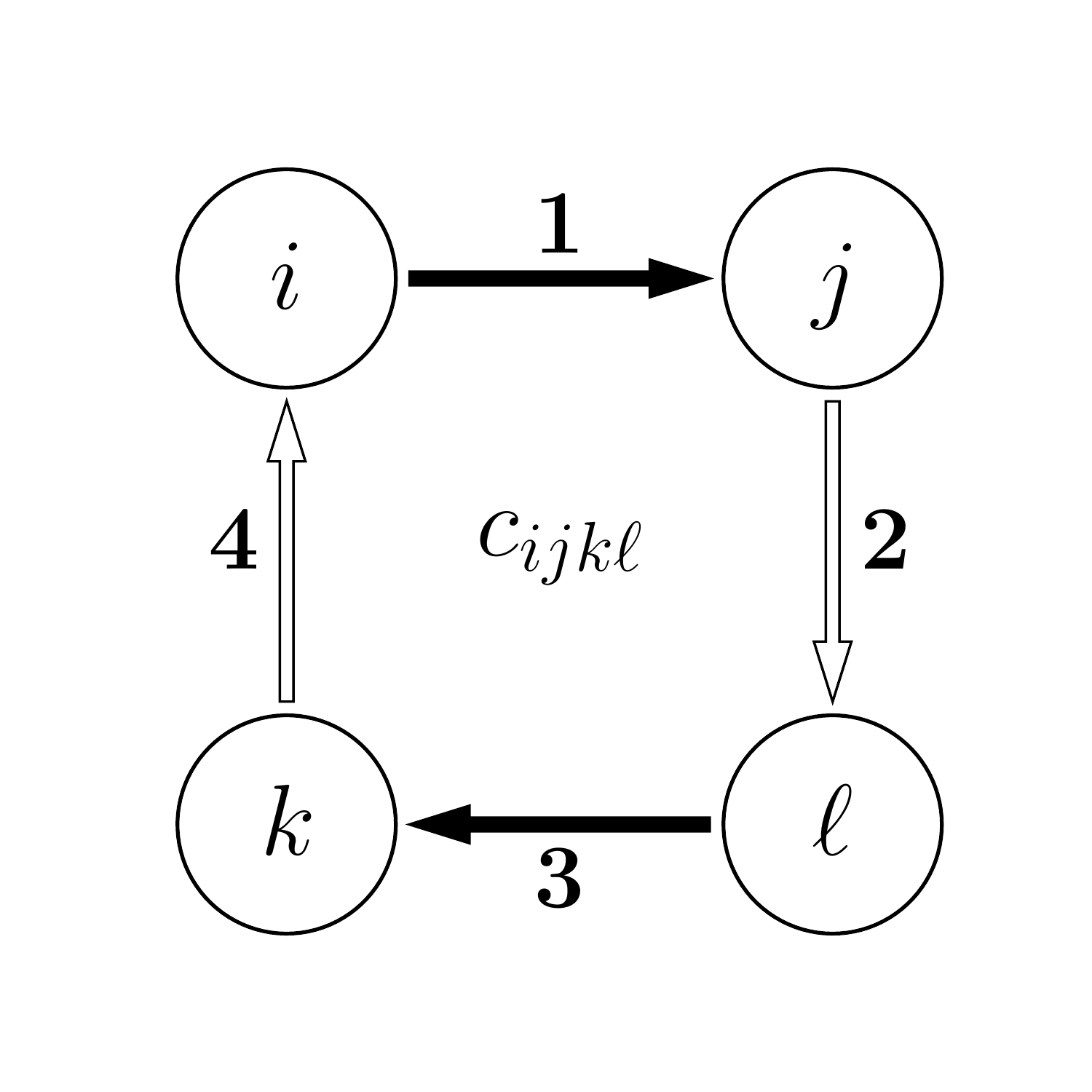}
    \includegraphics[width=0.2\textwidth]{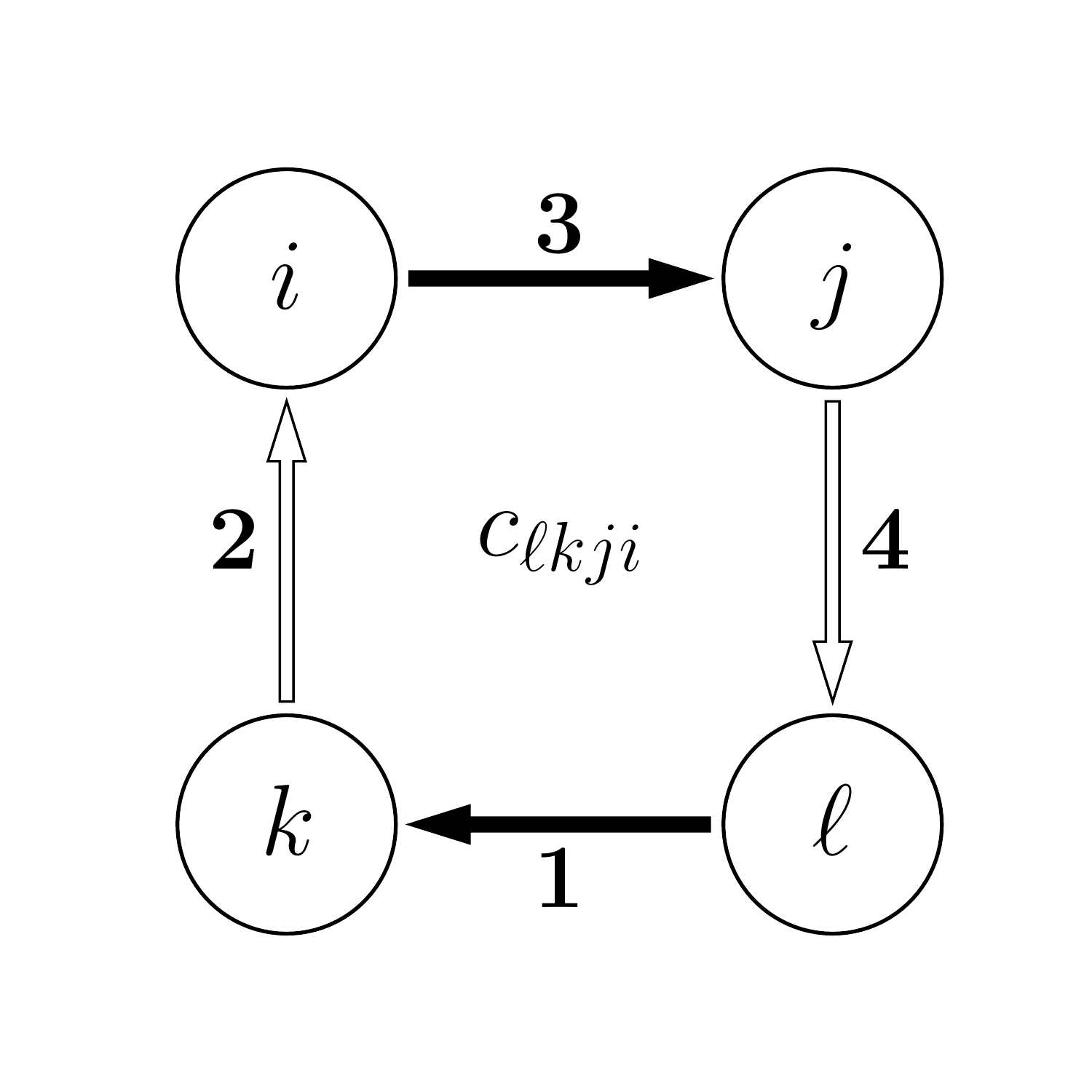}
    \hspace{0.5in}
    \includegraphics[width=0.2\textwidth]{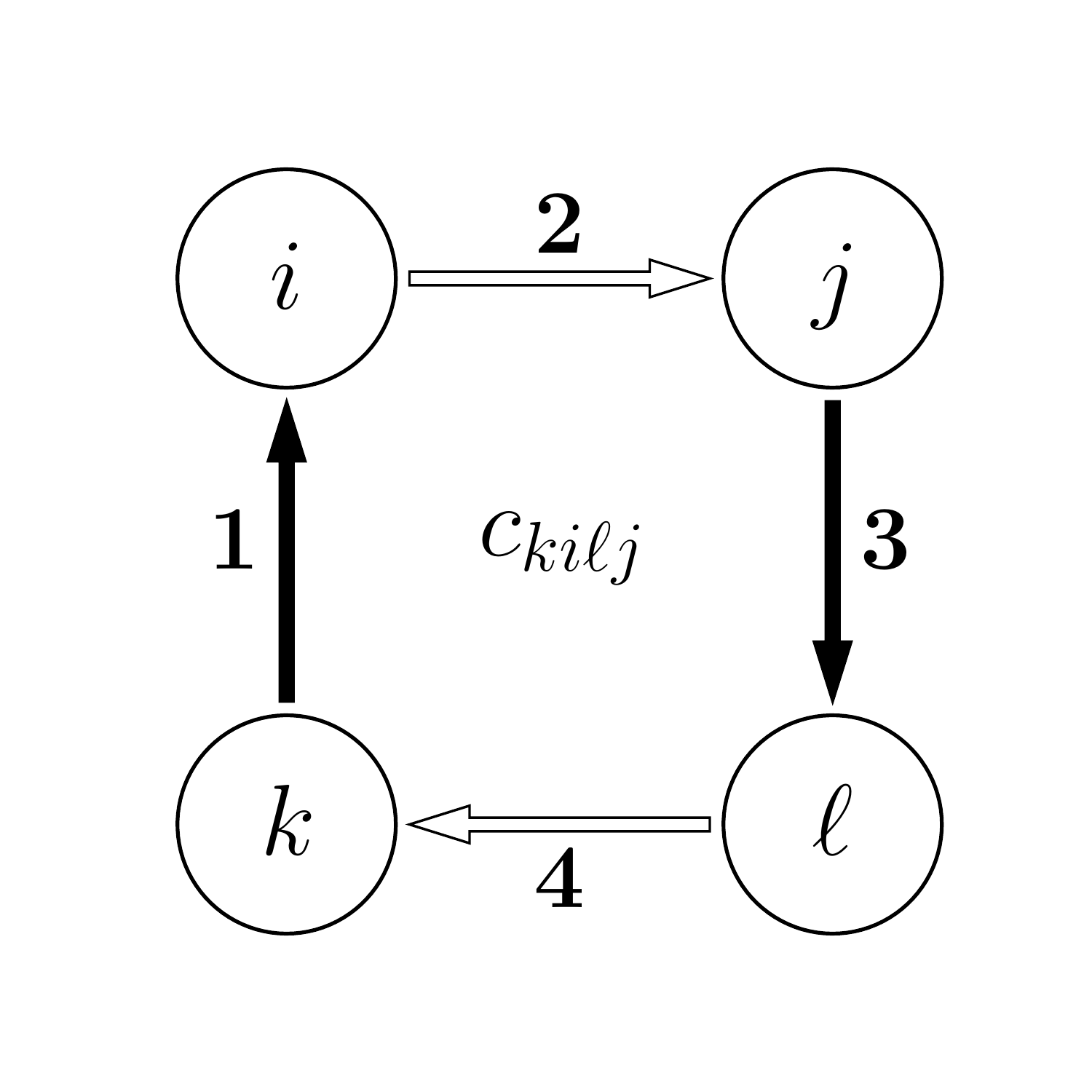}
    \includegraphics[width=0.2\textwidth]{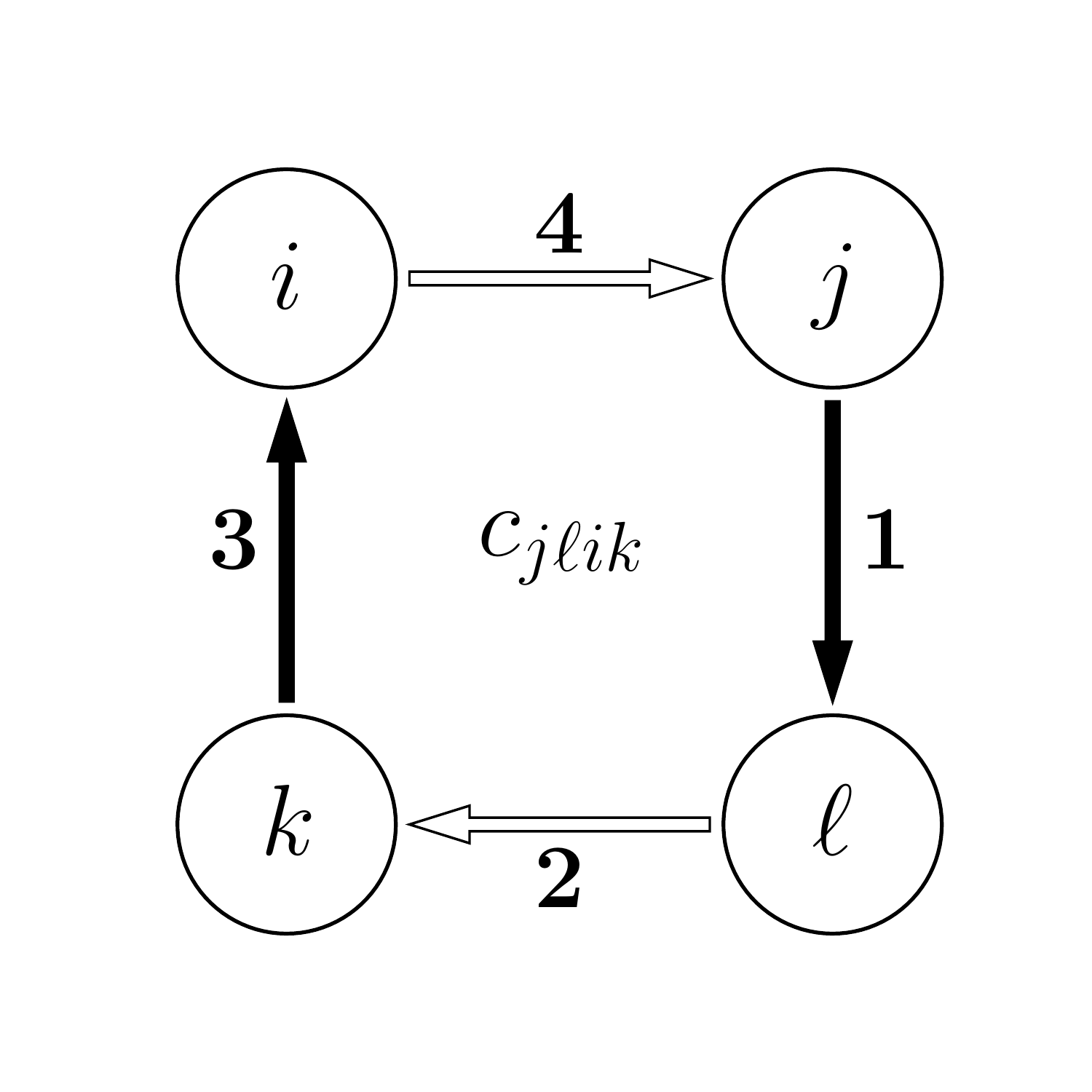}\\
    \includegraphics[width=0.2\textwidth]{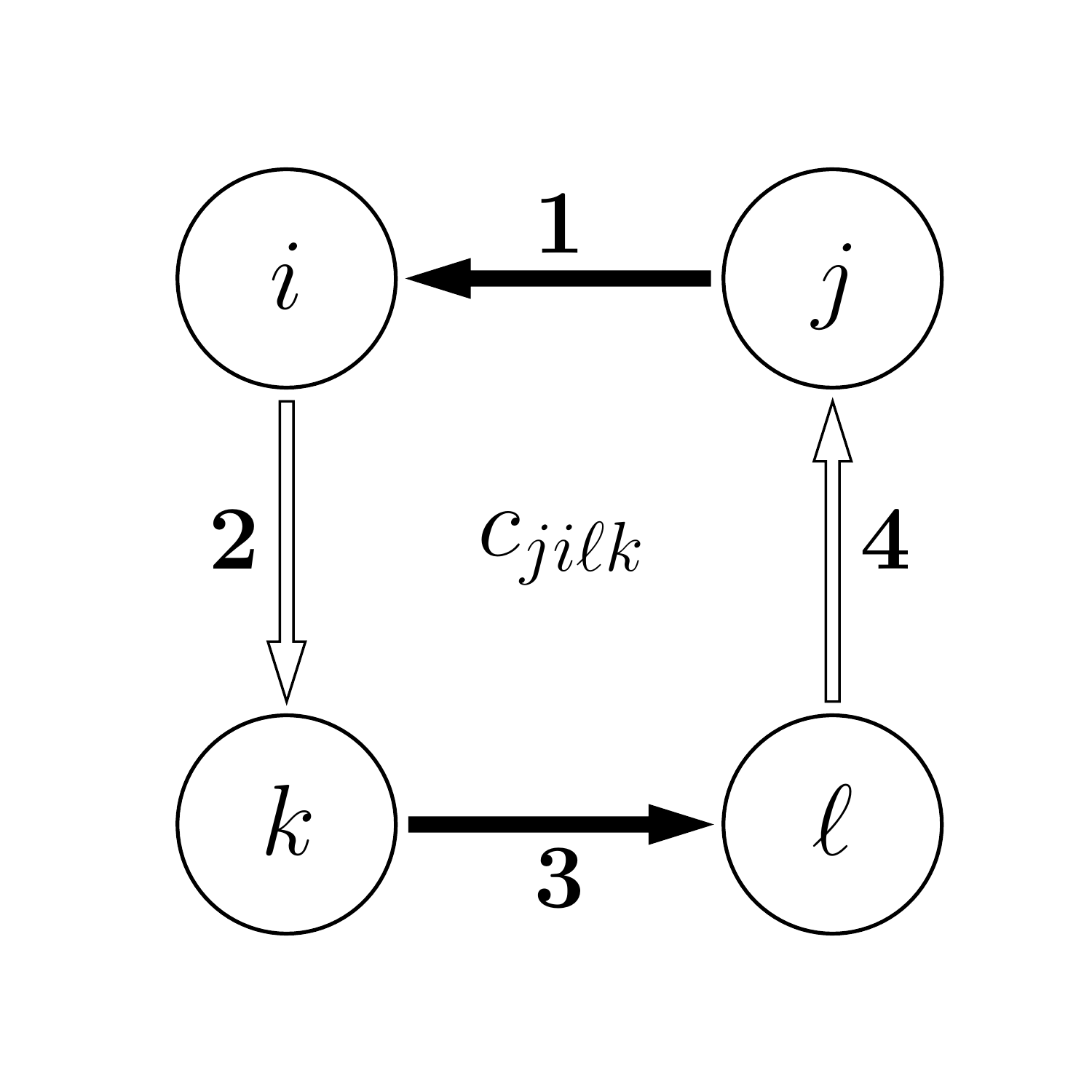}
    \includegraphics[width=0.2\textwidth]{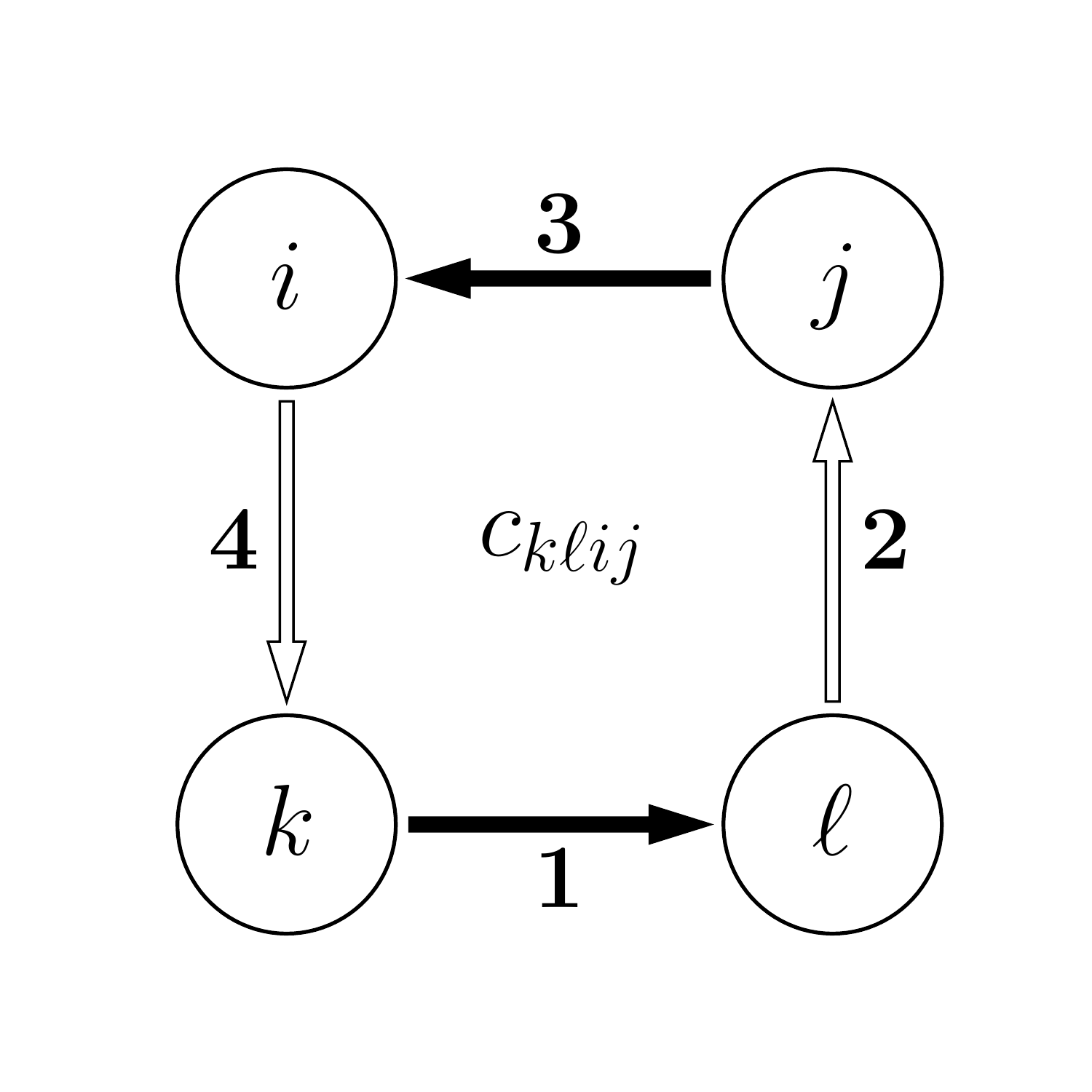}
    \hspace{0.5in}
    \includegraphics[width=0.2\textwidth]{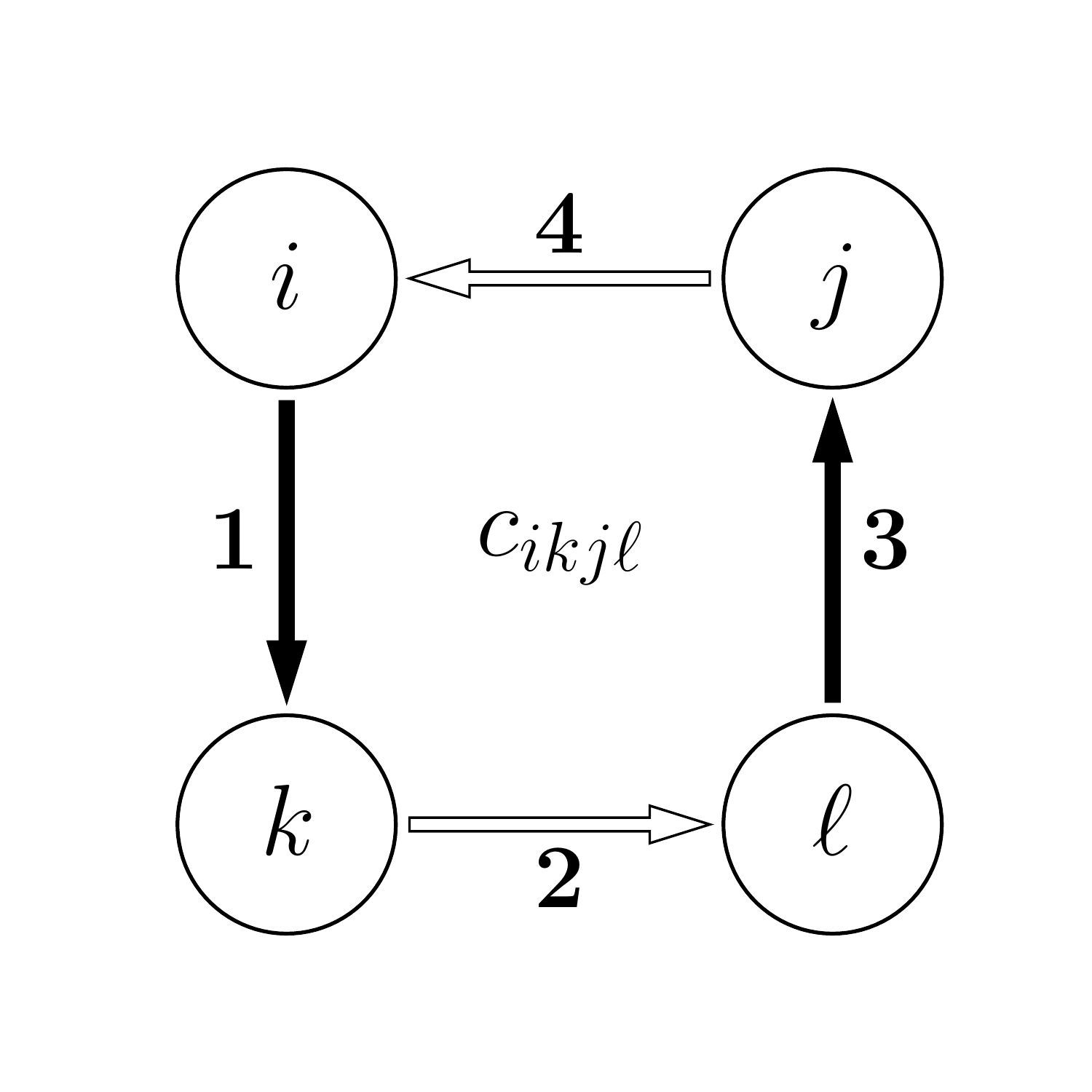}
    \includegraphics[width=0.2\textwidth]{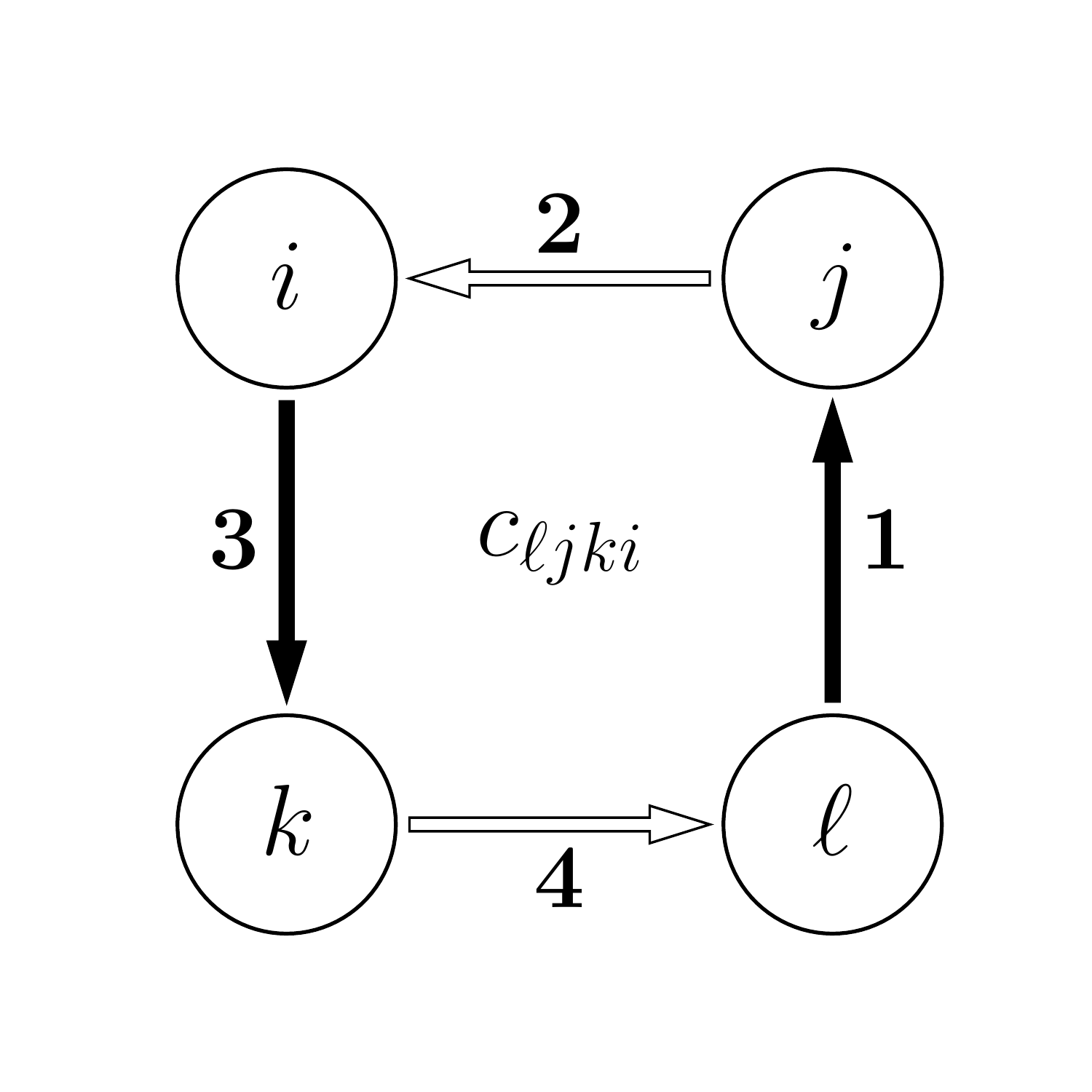}
    \caption{Diagrams of log-closure amplitude symmetries for a single quadrangle containing stations $i$, $j$, $k$, and $\ell$, with baselines numbered in the sequence used to construct the log-closure amplitude; baselines in the numerator of the closure amplitude are filled in black, while those in the denominator are filled in white.  All log-closure amplitudes in the left block of diagrams have the same value, and all log-closure amplitudes in the right block of diagrams have the same value, but the values corresponding to the two blocks of diagrams differ in sign.  Within a single block of diagrams, each row illustrates log-closure amplitudes constructed in the same cycle direction but using a different starting baseline; the value of the log-closure amplitude is invariant to the choice of starting baseline (\autoref{eqn:CAmpCycling}).  Within a single block of diagrams, each column illustrates a reversal in the cycle direction of log-closure amplitude construction; the value of the log-closure amplitude is invariant upon direction reversal (\autoref{eqn:CAmpReversal}).  The left and right blocks of diagrams differ by a swap of numerator and denominator; the value of the log-closure amplitude changes sign upon swapping numerator and denominator (\autoref{eqn:CAmpSwap}).
    }
    \label{fig:camp_symmetries}
\end{figure*}

\noindent These symmetries are illustrated in \autoref{fig:camp_symmetries}.

We construct a minimal design matrix \lcad that maps from the log visibility amplitude space to the log-closure amplitude space,

\begin{equation}
\lcav = \lcad \, \lvav ,
\end{equation}

\noindent where \lvav and \lcav are vectors of log visibility amplitudes and log-closure amplitudes, respectively.  This log-closure amplitude design matrix is equivalent to the ``amplitude closure operator'' of \cite{Lannes_1990a} and the ``alternate amplitude compilation operator'' of \cite{Lannes_1991}.  We express the log-closure amplitude covariance matrix \lcacov in terms of this design matrix \lcad and the log visibility amplitude covariance matrix \lvacov,

\begin{equation}
\lcacov = \lcad \, \lvacov \lcad^{\transpose} = \lcad \, \textbf{S}_{\textbf{\textit{a}}} \lcad^{\transpose} , \label{eqn:LCACovariance}
\end{equation}

\noindent where, as with \autoref{eqn:AberrationAnnihilation}, we have used the fact that ${\lcad \lvad = \boldsymbol{0}}$ to simplify the construction. This cancellation is referred to as ``amplitude aberration annihilation'' by \cite{Lannes_1991}.

For an \lgan-station array with $\lgan > 4$, the design matrix for a minimal set of log-closure amplitudes can be constructed using

\begin{equation}
\lcad_{\lgan} = \begin{pmatrix}
\textbf{X}_{\lgan} & \textbf{Y}_{\lgan} \\
\boldsymbol{0} & \lcad_{\lgan-1}
\end{pmatrix} , \label{eqn:LogClosureAmpDesign}
\end{equation}

\noindent where $\lcad_{\lgan-1}$ is the design matrix for an array with $\lgan-1$ stations, $\boldsymbol{0}$ is an $\left( \frac{(\lgan-1) (\lgan-4)}{2} \right) \times \left( \lgan - 1 \right)$ matrix of all zeros,

\begin{equation}
\textbf{X}_{\lgan} = \begin{pmatrix}
\textbf{I}_{\lgan-2} & - \boldsymbol{1}
\end{pmatrix} ,
\end{equation}

\noindent and $\textbf{Y}_{\lgan}$ is an $(\lgan-2) \times \binom{\lgan-1}{2}$ matrix constructed by ``cycling'' through pairs of baselines that do not contain the first station:

\begin{equation}
\resizebox{0.95\columnwidth}{!}{$\textbf{Y}_{\lgan} = \begin{pmatrix}
0 & 0 & 0 & \ldots & 0 & -1 & 0 & \ldots & 0 & 0 & 0 & 0 & 0 & 0 & 0 & 0 & 0 & 1 \\
-1 & 0 & 0 & \ldots & 0 & 0 & 1 & \ldots & \cdot & \cdot & \cdot & \cdot & \cdot & \cdot & \cdot & \cdot & \cdot & 0 \\
\vdots & & & \ddots & & & & \ddots & & & & & & & & & & \vdots \\
\cdot & \cdot & \cdot & \ldots & \cdot & \cdot & \cdot & \ldots & -1 & 0 & 0 & 1 & \cdot & \cdot & \cdot & \cdot & \cdot & 0 \\
\cdot & \cdot & \cdot & \ldots & \cdot & \cdot & \cdot & \ldots & \cdot & \cdot & \cdot & \cdot & -1 & 0 & 1 & \cdot & \cdot & 0 \\
\cdot & \cdot & \cdot & \ldots & \cdot & \cdot & \cdot & \ldots & \cdot & \cdot & \cdot & \cdot & \cdot & \cdot & \cdot & -1 & 1 & 0
\end{pmatrix}$} .
\end{equation}

\noindent Here, matrix elements represented by a dot indicate zero-valued entries.  \autoref{tab:LogClosureAmpTable} lists example log-closure amplitude design and covariance matrices.

\begin{table}
\vspace{1em}
\centering{}
\begin{tabular}{lcc}
\hline \hline
 & &  \textbf{Number of stations (\lgan)} \\ \cline{3-3}
\textbf{Matrix} & \textbf{Shape}   &   $\lgan=4$ \\ 
\hline
$\lvav^{\transpose}$ & $1 \times B$ & $\begin{pmatrix}
\lvas[12] & \lvas[13] & \lvas[14] & \lvas[23] & \lvas[24] & \lvas[34]
\end{pmatrix}$ \\ \hline
\lcad & $q \times \lvan$  & $\begin{pmatrix}
0 & 1 & -1 & -1 & 1 & 0 \\
1 & 0 & -1 & -1 & 0 & 1
\end{pmatrix}$ \\
\hline
\lcacov & $q \times q$ & $\begin{pmatrix}
\lvavar[13] + \lvavar[14] + \lvavar[23] + \lvavar[24] & \lvavar[14] + \lvavar[23] \\
\lvavar[14] + \lvavar[23] & \lvavar[12] + \lvavar[14] + \lvavar[23] + \lvavar[34]
\end{pmatrix}$ \\
\hline
\end{tabular}
\caption{Minimal log-closure amplitude design and covariance matrices for four-element array.  Here, $q = \frac{\lgan (\lgan-3)}{2}$ is the number of quadrangles in a minimal set and $\lvan = \binom{\lgan}{2}$ is the number of baselines.}
\label{tab:LogClosureAmpTable}
\end{table}

\section{Worked examples of information content}
\label{app:WorkedExamples}

For an array of \gpn stations, an accounting of the nonredundant closure phases reveals that they differ by an amount $\gpn-1$ from the number of visibility phases.  This offset, which is equal to the number of unique gain phases in the array, suggests that closure phases contain all of the source phase information and that the additional degrees of freedom afforded by the visibility phases only describe the gains.  A similar situation holds for the closure amplitudes, where the number of nonredundant quadrangles differs from the number of visibility amplitudes by an amount equal to the number of gain amplitudes, \gpn.  In the limit where we have no \textit{a priori} information about the gains, then, the information content in the visibility quantities should be identical to that contained within the closure quantities.  In this section, we demonstrate the reality of this equality and its lack of dependence on the specific choice of nonredundant closure subset for some selected test cases.

\subsection{Closure phase for $\gpn=3$ stations}
\label{sec:CphaseN3}

We consider a three-station interferometer with measured visibility phases (\vps[12], \vps[13], \vps[23]) and model visibility phases (\vpms[12], \vpms[13], \vpms[23]) related by the station gain phases (\gpms[1], \gpms[2], \gpms[3])~as

\begin{equation}
\vps[ij] = \vpms[ij] + \gpms[i] - \gpms[j] . \label{eqn:VisibilityPhases}
\end{equation}

\noindent The information contained in the measured visibilities is captured by their joint likelihood distribution, $\mathcal{L}$.  If the visibility phases have Gaussian thermal variances (\vpvar[12], \vpvar[13], \vpvar[23]), and if we assume that the gain contributions are also Gaussian distributed with variances (\gpvar[1], \gpvar[2], \gpvar[3]), then the likelihood of the measured visibility phases can be expressed as a multivariate Gaussian,

\begin{equation}
\mathcal{L} = \frac{1}{\sqrt{(2 \pi)^3 \det( \vpcov )}} \exp\left[ - \frac{1}{2} \vprv^{\transpose} \vpcov^{-1} \vprv \right] , \label{eqn:VPhaseLikelihood_app}
\end{equation}

\noindent where

\begin{equation}
\vprv = \begin{pmatrix}
\vps[12] - ( \vpms[12] + \gpms[1] - \gpms[2] ) \\
\vps[13] - ( \vpms[13] + \gpms[1] - \gpms[3] ) \\
\vps[23] - ( \vpms[23] + \gpms[2] - \gpms[3] )
\end{pmatrix} \equiv \begin{pmatrix}
\vprs[12] \\
\vprs[13] \\
\vprs[23]
\end{pmatrix}
\end{equation}

\noindent is the vector of visibility phase residuals, and \vpcov is the visibility phase covariance matrix (see \autoref{app:VisDesign}).  Because the likelihood is Gaussian and the variances are constant-valued, the quantity

\begin{equation}
\vpchisq = \vprv^{\transpose} \vpcov^{-1} \vprv \label{eqn:VisPhaseChisquared}
\end{equation}

\noindent contains the same information as $\mathcal{L}$ in a more compact form; we will thus proceed through the use of \vpchisq rather than $\mathcal{L}$.

Our expectation is that in the high-S/N limit (i.e., when the thermal noise is negligible compared to gain variations), the information content in the visibility phases will be identical to that in the closure phases; equivalently, \vpchisq for the visibility phases should equal \cpchisq for the closure phases.  To simplify the mathematics and notation, let us now suppose that the array is perfectly homogeneous such that we can denote $\gpvar[1] = \gpvar[2] = \gpvar[3] \equiv \sigma^2$ and $\vpvar[12] = \vpvar[13] = \vpvar[23] \equiv \varepsilon^2 \sigma^2$.  The high-S/N limit thus corresponds to $\varepsilon^2 \ll 1$.  To leading order in $\varepsilon^2$, the inverse of the covariance matrix is

\begin{equation}
\vpcov^{-1} = \frac{1}{3 \, \varepsilon^2 \sigma^2} \begin{pmatrix}
1 & - 1 & 1 \\
- 1 & 1 & - 1 \\
1 & - 1 & 1
\end{pmatrix} ,
\end{equation}

\noindent which corresponds to a \vpchisq in the same limit of

\begin{equation}
\vpchisq = \frac{\left( \vprs[12] - \vprs[13] + \vprs[23] \right)^2}{3 \, \varepsilon^2 \sigma^2} . \label{eqn:VisibilityPhaseChisqN3}
\end{equation}

Because the array contains only $\gpn=3$ stations, $\binom{\gpn}{3} = \binom{\gpn-1}{2}$ and the complete set of closure phases is equal to the nonredundant set, both of which contain only a single element.  We can write the model closure phase as $\cpms[123] = \vpms[12] - \vpms[13] + \vpms[23]$ and the measured closure phase as $\cps[123] = \vps[12] - \vps[13] + \vps[23]$, with corresponding thermal noise given by $\cpvar[123] = \vpvar[12] + \vpvar[13] + \vpvar[23] = 3 \epsilon \sigma^2$.  The value of \cpchisq is then written simply as

\begin{eqnarray}
\cpchisq & = & \frac{\left( \cps[123] - \cpms[123] \right)^2}{\cpvar[123]} \nonumber \\
& = & \frac{\left( \vprs[12] - \vprs[13] + \vprs[23] \right)^2}{3\, \varepsilon^2 \sigma^2} ,
\end{eqnarray}

\noindent which we can see is identical to \autoref{eqn:VisibilityPhaseChisqN3}.

\subsection{Closure phase for $\gpn=4$ stations}

We consider now a four-station interferometer with measured visibility phases (\vps[12], \vps[13], \vps[14], \vps[23], \vps[24], \vps[34]) and model visibility phases (\vpms[12], \vpms[13], \vpms[14], \vpms[23], \vpms[24], \vpms[34]) related by the station gain phases (\gpms[1], \gpms[2], \gpms[3], \gpms[4]) as specified in \autoref{eqn:VisibilityPhases}.  Following the same procedure as in the previous section, the inverse of the covariance matrix in the high-S/N limit is

\begin{equation}
\vpcov^{-1} = \frac{1}{4\, \varepsilon^2 \sigma^2} \begin{pmatrix}
2 & -1 & -1 & 1 & 1 & 0 \\
-1 & 2 & -1 & -1 & 0 & 1 \\
-1 & -1 & 2 & 0 & -1 & -1 \\
1 & -1 & 0 & 2 & -1 & 1 \\
1 & 0 & -1 & -1 & 2 & -1 \\
0 & 1 & -1 & 1 & -1 & 2
\end{pmatrix} ,
\end{equation}

\noindent corresponding to

\begin{eqnarray}
\vpchisq & = & \frac{\left( \vprs[12] - \vprs[13] + \vprs[23] \right)^2}{4 \, \varepsilon^2 \sigma^2} + \frac{\left( \vprs[12] - \vprs[14] + \vprs[24] \right)^2}{4 \, \varepsilon^2 \sigma^2} \nonumber \\
& + & \frac{\left( \vprs[13] - \vprs[14] + \vprs[34] \right)^2}{4 \, \varepsilon^2 \sigma^2} + \frac{\left( \vprs[23] - \vprs[24] + \vprs[34] \right)^2}{4 \, \varepsilon^2 \sigma^2} . \nonumber \\
\label{eqn:VisPhaseChisqN4}
\end{eqnarray}

The four-station array has four closure phases in total, of which three are nonredundant.  We specify a measured closure phase \cps[ijk] as

\begin{equation}
\cps[ijk] = \vps[ij] - \vps[ik] + \vps[jk] , \label{eqn:ClosurePhaseDefinition}
\end{equation}

\noindent with an analogous specification for the corresponding model closure phase \cpms[ijk].  For a particular choice of nonredundant closure phase subset, the value of \cpchisq will depend on the covariance matrix \cpcov for the closure phases (see \autoref{sec:CphaseDesign} and \autoref{eqn:CPCovariance}) and on the vector \cprv of closure phase residuals,

\begin{equation}
\cprv = \begin{pmatrix}
\cps[123] - \cpms[123] \\
\cps[124] - \cpms[124] \\
\cps[134] - \cpms[134]
\end{pmatrix} \equiv \begin{pmatrix}
\cprs[123] \\
\cprs[124] \\
\cprs[134]
\end{pmatrix} .
\end{equation}

\noindent After computing the inverse of $\boldsymbol{\Sigma}_{\psi}$,

\begin{equation}
\cpcov^{-1} = \frac{1}{4 \, \varepsilon^2 \sigma^2} \begin{pmatrix}
2  & -1 & 1 \\
-1 & 2 & -1 \\
1 & -1 & 2
\end{pmatrix} ,
\end{equation}

\noindent it is a tedious but straightforward algebraic exercise to obtain

\begin{eqnarray}
\cpchisq & = & \cprv^{\transpose} \cpcov^{-1} \cprv \nonumber \\
& = & \frac{\cprs[123]^2 + \cprs[124]^2 + \cprs[134]^2 + \cprs[234]^2}{4 \, \varepsilon^2 \sigma^2} . \label{eqn:CphaseChisqN4}
\end{eqnarray}

\noindent Because closure phases are constructed purely from sums and differences of visibility phases, $\cprs[ijk] = \vprs[ij] - \vprs[ik] + \vprs[jk]$ and thus, \autoref{eqn:CphaseChisqN4} is equivalent to \autoref{eqn:VisPhaseChisqN4}.  Furthermore, \autoref{eqn:CphaseChisqN4} no longer shows any signature of the original nonredundant closure phase subset choice; rather, each element of the full redundant set of four closure phases is represented equally, and the $\cpchisq$ includes a 3/4 redundancy correction factor (see \autoref{eqn:ClosurePhaseRedundancy}) corresponding to the ratio of linearly independent to total closure phases (note that $\sigma_{ijk}^2 = 3\, \varepsilon^2 \sigma^2)$.

\subsection{Closure amplitude for $\gpn=4$ stations}

We consider again a four-station interferometer with measured log visibility amplitudes (\lvas[12], \lvas[13], \lvas[14], \lvas[23], \lvas[24], \lvas[34]) and model log visibility amplitudes (\lvams[12], \lvams[13], \lvams[14], \lvams[23], \lvams[24], \lvams[34]) related by the log station gain amplitudes (\lgams[1], \lgams[2], \lgams[3], \lgams[4]) as

\begin{equation}
\Eop{\lvas[ij]} = \lvams[ij] + \lgams[i] + \lgams[j] . \label{eqn:LVAsingle}
\end{equation}

\noindent As in \autoref{sec:CphaseN3}, if the measured log visibility amplitudes have Gaussian thermal variances (\lvavar[12], \lvavar[13], \lvavar[14], \lvavar[23], \lvavar[24], \lvavar[34]) and the log gain amplitude contributions are also Gaussian distributed with variances (\lgavar[1], \lgavar[2], \lgavar[3], \lgavar[4]), then the joint distribution of the measured log visibility amplitudes can be expressed as a multivariate Gaussian.  The covariance matrix for this distribution can be constructed using the procedure described in \autoref{app:VisDesign}.

If we once again treat the array as perfectly homogeneous and take the high-S/N limit, then to leading order in $\varepsilon^2$ we find

\begin{equation}
\lvacov^{-1} = \frac{1}{6 \, \varepsilon^2 \sigma^2} \begin{pmatrix}
2 & -1 & -1 & -1 & -1 & 2 \\
-1 & 2 & -1 & -1 & 2 & -1 \\
-1 & -1 & 2 & 2 & -1 & -1 \\
-1 & -1 & 2 & 2 & -1 & -1 \\
-1 & 2 & -1 & -1 & 2 & -1 \\
2 & -1 & -1 & -1 & -1 & 2
\end{pmatrix} .
\end{equation}

\noindent The corresponding $\lvachisq = \lvarv^{\transpose} \lvacov^{-1} \lvarv$ can then be written

\begin{eqnarray}
\lvachisq & = & \frac{\left( \lvars[12] + \lvars[34] - \lvars[13] - \lvars[24] \right)^2}{6 \, \varepsilon^2 \sigma^2} \nonumber \\
& + & \frac{\left( \lvars[12] + \lvars[34]  - \lvars[14] - \lvars[23] \right)^2}{6 \, \varepsilon^2 \sigma^2} \nonumber \\
& + & \frac{\left( \lvars[13] + \lvars[24] - \lvars[14] - \lvars[23] \right)^2}{6 \, \varepsilon^2 \sigma^2} . \label{eqn:LogVisAmpChisqN4}
\end{eqnarray}

The four-station array has three closure amplitudes in total, of which two are nonredundant.  We specify a measured log-closure amplitude \lcas[ijk\ell] as

\begin{equation}
\lcas[ijk\ell] = \lvas[ij] + \lvas[k\ell] - \lvas[ik] - \lvas[j\ell] , \label{eqn:LogCampDefinition}
\end{equation}

\noindent with an analogous specification for the corresponding model log-closure amplitude \lcams[ijk\ell].  For a particular choice of nonredundant closure amplitude subset, the value of \lcachisq will depend on the covariance matrix \lcacov for the log-closure amplitudes (see \autoref{sec:CampDesign} and \autoref{eqn:LCACovariance}) and on the vector \lcarv of log-closure amplitude residuals,

\begin{equation}
\lcarv = \begin{pmatrix}
\lcas[1234] - \lcams[1234] \\
\lcas[1243] - \lcams[1243]
\end{pmatrix} \equiv \begin{pmatrix}
\lcars[1234] \\
\lcars[1243]
\end{pmatrix} .
\end{equation}

\noindent Written out more explicitly, the covariance matrix is given by

\begin{equation}
\lcacov = 2 \, \varepsilon^2 \sigma^2 \begin{pmatrix}
2 & 1 \\
1 & 2
\end{pmatrix} ,
\end{equation}

\noindent with corresponding inverse

\begin{equation}
\lcacov^{-1} = \frac{1}{6 \, \varepsilon^2 \sigma^2} \begin{pmatrix}
2 & -1 \\
-1 & 2
\end{pmatrix} .
\end{equation}

\noindent We thus obtain

\begin{eqnarray}
\lcachisq & = & \lcarv^{\transpose} \lcacov^{-1} \lcarv \nonumber \\
& = & \frac{\lcars[1234]^2 + \lcars[1243]^2 + \lcars[1342]^2}{6 \, \varepsilon^2 \sigma^2} ,
\end{eqnarray}

\noindent which is equal to \autoref{eqn:LogVisAmpChisqN4}.  As with the closure phases, we see that the initial choice of minimal log-closure amplitude subset has no bearing on the value of \lcachisq, and accounts for the redundancy factor of total versus linearly independent closure amplitudes.

\subsection{Closure quantities for arbitrary $N$}
\label{sec:generalN}

We introduce the notion of ``mixed phases'' that retain all of the information contained in the visibility phases, but we separate it into two components: one component that is captured by the closure phases and a second component that captures the remaining station-based effects.  For an array with $N$ stations, the mixed phase design matrix operates on the $\vpn$ baseline phases and is given by

\begin{eqnarray}
\label{eqn:MixedPhaseDesign}
\mpd & = & \begin{pmatrix}
\textbf{I}_{\gpn-1,\vpn} \\
\cpd_\gpn
\end{pmatrix} \nonumber \\
& = & \begin{pmatrix}
\textbf{I}_{\gpn-1} & \boldsymbol{0} \\
\vpd_{\gpn-1} & \textbf{I}_{\binom{\gpn-1}{2}}
\end{pmatrix} .
\end{eqnarray}

\noindent $\textbf{I}_{\gpn-1,\vpn}$ is an $(\gpn-1) \times \vpn$ ``rectangular identity matrix'' that extracts the first $\gpn-1$ baseline phases by combining $\textbf{I}_{\gpn-1}$, a standard square identity matrix of rank $\gpn-1$, with $\boldsymbol{0}$, an $(\gpn-1) \times {(\vpn-\gpn+1)}$ matrix of all zeros. $\cpd_\gpn$ is the minimal closure phase design matrix for $\gpn$ stations (see \autoref{eqn:MinimalCphaseDesign}), which can be expanded into the visibility phase design matrix $\vpd_{\gpn-1}$ for $\gpn-1$ stations (see \autoref{eqn:VisPhaseDesignMatrix}) and a standard square identity matrix of rank $\binom{N-1}{2}$. This design matrix maps from the visibility phase space to the mixed phase space,

\begin{equation}
\mpv = \mpd \vpv .
\end{equation}

\noindent For example, the mixed phase design matrix for an array with $N=4$ stations is given by

\begin{equation}
\boldsymbol{\Psi}_{+} = 
\begin{pmatrix}
1 & 0 & 0 & 0 & 0 & 0\\
0 & 1 & 0 & 0 & 0 & 0 \\
0 & 0 & 1 & 0 & 0 & 0 \\
1 & -1 & 0 & 1 & 0 & 0 \\
1 & 0 & -1 & 0 & 1 & 0 \\
0 & 1 & -1 & 0 & 0 & 1
\end{pmatrix} ,
\end{equation}

\noindent and the corresponding mixed phase vector is

\begin{equation}
\boldsymbol{\psi}_+ =  \begin{pmatrix}
\vps[12] \\
\vps[13] \\
\vps[14] \\
\vps[12] - \vps[13] + \vps[23] \\
\vps[12] - \vps[14] + \vps[24] \\
\vps[13] - \vps[14] + \vps[34]
\end{pmatrix}.
\end{equation}

The mixed phase covariance matrix is given by

\begin{equation}
\mpcov = \mpd \vpcov \mpd^\transpose . \label{eqn:MixedPhaseCovariance}
\end{equation}

\noindent Using the inverse of the mixed phase design matrix,

\begin{equation}
\mpd^{-1} = \begin{pmatrix}
\textbf{I}_{\gpn-1} & \boldsymbol{0} \\
- \vpd_{\gpn-1} & \textbf{I}_{\binom{\gpn-1}{2}}
\end{pmatrix} ,
\end{equation}

\noindent we can invert \autoref{eqn:MixedPhaseCovariance} to obtain an expression for the visibility phase covariance matrix in terms of the mixed phase covariance matrix,

\begin{equation}
\vpcov = \mpd^{-1} \mpcov {\mpd^\transpose}^{-1} ,
\end{equation}

\noindent where we note that the inverse transpose is equal to the transposed inverse for the mixed phase design matrix.  We can use the above to substitute for \vpcov in our expression for the visibility phase $\chi^2$ (see \autoref{eqn:VisPhaseChisquared}),

\begin{eqnarray}
\vpchisq & = & \vprv^{\transpose} \vpcov^{-1} \vprv \nonumber \\
& = & \vprv^{\transpose} \left( \mpd^{-1} \mpcov {\mpd^{\transpose}}^{-1} \right)^{-1} \vprv \nonumber \\
& = & \vprv^{\transpose} \left( \mpd^{\transpose} \mpcov^{-1} \mpd \right) \vprv \nonumber \\
& = & \left( \mpd \vprv \right)^{\transpose} \mpcov^{-1} \left( \mpd \vprv \right) \nonumber \\
& = & \mprv^{\transpose} \mpcov^{-1} \mprv ,
\end{eqnarray}

\noindent revealing that the $\chi^2$ constructed from mixed phases is equal to that constructed from visibility phases, when all covariances are taken into account. This is due to the fact that the mixed phases are generated through a non-singular linear transformation of the visibility phases.

To see how the mixed phases reduce to purely closure phases in the high-S/N limit, it is convenient to consider the following decomposition of the mixed phase covariance matrix:

\begin{equation}
\mpcov = \begin{pmatrix}
\vpcov' & \textbf{W}^{\transpose} \\
\textbf{W} & \cpcov
\end{pmatrix} , \label{eqn:MixedPhaseCovarianceBlock}
\end{equation}

\noindent where $\vpcov'$ is the first $(N-1) \times (N-1)$ upper left subset of the full visibility phase covariance matrix $\vpcov$, $\cpcov$ is the closure phase covariance matrix, and
$\textbf{W} = \vpd_{N-1}\,\textbf{S}'_{\boldsymbol{\phi}}$ is the covariance between the closure phases and the first $\gpn-1$ visibility phases. Since the closure phases are independent of station gain, both $\textbf{W}$ and $\cpcov$ include only baseline thermal noise.

Using a strategy analogous to that employed in the previous sections, where parameter $\varepsilon^2 \sim \mathcal{O}(\vpvar[ij]/\gpvar[i])$ relates statistical error in visibility phase to that from gain uncertainty, we examine the behavior of $\mpcov^{-1}$ as $\varepsilon \rightarrow 0$.  The sub-matrices of $\mpcov$ scale with $\varepsilon$ as

\begin{eqnarray*}
\vpcov' & \propto & 1 \\
\textbf{W} & \propto & \varepsilon^2 \\
\cpcov & \propto & \varepsilon^2 .
\end{eqnarray*}

\noindent From the block matrix form of \autoref{eqn:MixedPhaseCovarianceBlock}, we can write the inverse mixed phase covariance matrix as

\begin{multline}
\mpcov^{-1} = \left( \begin{matrix}
\left( \vpcov' - \textbf{W}^{\transpose} \cpcov^{-1} \textbf{W} \right)^{-1} \\
- \cpcov^{-1} \textbf{W} \left( \vpcov' - \textbf{W}^{\transpose} \cpcov^{-1} \textbf{W} \right)^{-1}
\end{matrix} \right. \; , \\
\left. \begin{matrix}
- \left( \vpcov' - \textbf{W}^{\transpose} \cpcov^{-1} \textbf{W} \right)^{-1} \textbf{W}^{\transpose} \cpcov^{-1} \\
\cpcov^{-1} + \cpcov^{-1} \textbf{W} \left( \vpcov' - \textbf{W}^{\transpose} \cpcov^{-1} \textbf{W} \right)^{-1} \textbf{W}^{\transpose} \cpcov^{-1}
\end{matrix} \right).
\end{multline}

\noindent These four sub-matrices scale with $\varepsilon$ as

\begin{eqnarray*}
\left( \vpcov' - \textbf{W}^{\transpose} \cpcov^{-1} \textbf{W} \right)^{-1} & \propto & 1 \\
\left( \vpcov' - \textbf{W}^{\transpose} \cpcov^{-1} \textbf{W} \right)^{-1} \textbf{W}^{\transpose} \cpcov^{-1} & \propto & 1 \\
\cpcov^{-1} \textbf{W} \left( \vpcov' - \textbf{W}^{\transpose} \cpcov^{-1} \textbf{W} \right)^{-1} & \propto & 1 \\
\cpcov^{-1} + \cpcov^{-1} \textbf{W} \left( \vpcov' - \textbf{W}^{\transpose} \cpcov^{-1} \textbf{W} \right)^{-1} \textbf{W}^{\transpose} \cpcov^{-1} & \propto & \frac{1}{\varepsilon^2} ,
\end{eqnarray*}

\noindent and we can see that the $\cpcov^{-1}$ term in the lower right sub-matrix dominates $\mpcov^{-1}$.  The product of $\mpcov^{-1}$ with $\mprv$ in this limit will therefore serve to isolate the last $t$ terms of $\mprv$ (which are just the closure phases \cprv), and multiply them by $\cpcov^{-1}$:

\begin{eqnarray}
\label{eqn:MixedPhaseQED}
\lim_{\varepsilon \rightarrow 0}\left( \vpchisq \right) & = & \cprv^{\transpose} \cpcov^{-1} \cprv \nonumber \\
& = & \cpchisq .
\end{eqnarray}

\noindent The final result is that the visibility phase $\chi^2$, in the limit where the uncertainty in the baseline-based quantities is much smaller than the uncertainty in the station-based quantities, is equal to the closure phase $\chi^2$.

The equivalence between $\lvachisq$ derived from a complete set of log visibility amplitudes in the $\varepsilon \rightarrow 0$ limit and $\lcachisq$ derived from log-closure amplitudes is demonstrated the same way. Here, the corresponding design matrix and mixed log amplitudes are,
\begin{equation}
    \label{eqn:MixedAmplitudeDesign}
\lmad = \begin{pmatrix}
\textbf{I}_{\gan,\van} \\
\lcad_\gan
\end{pmatrix} \qquad
\lmav = \lmad \, \lvav,
\end{equation}
which draws from the first $\gan$ visibility log amplitudes followed by a minimal set of log-closure amplitudes. The transformation to mixed quantities is non-singular, as long as the first $\gan$ baselines drawn do not form any closed quadrangles (or that the first $\gpn-1$ baselines do not form any closed triangles, in the case of mixed phases). This condition is met by the baseline ordering convention used in this paper. The reduction in Equations \ref{eqn:MixedPhaseCovarianceBlock}--\ref{eqn:MixedPhaseQED} then follows under substitution of phase with log-amplitude quantities.

\subsection{Explicit gain marginalization}
\label{sec:gainmarginalization}

In the previous Sections \ref{sec:CphaseN3}--\ref{sec:generalN}, we have shown that the information content in closure quantities is equivalent to that in the baseline visibilities for the limit of completely unconstrained gains, so long as the covariance structure of the corresponding observables is taken into account. We now relate the use of covariance in the residual visibility likelihood construction (\autoref{eqn:VPhaseLikelihood_app}) to explicit analytic marginalization over Gaussian uncertainties in phase or log-amplitude station gain. Thus, in the limit of completely unconstrained gains, use of closure quantities should give identical results to explicit numerical marginalization over all possible gains; and, for the case of finite Gaussian uncertainties in phase or log-amplitude station gain, use of the residual visibility covariance should give identical results to explicit numerical marginalization over Gaussian priors for the gains.

For an array with $N$ stations under modeled gain corrections, we can write \autoref{eqn:LVAsingle} for all baselines using

\begin{equation}
\Eop{\lvarv} = \Eop{\lvav} - \lvamv = \lvad (\lgav - \lgamv) = \lvad \lgarv,
\end{equation}

\noindent where $\lvad \lgarv$ is a residual vector of log visibility amplitude correction factors. For example, a three-station array would have

\begin{equation}
\lvad \lgarv = \begin{pmatrix}
1 & 1 & 0 \\
1 & 0 & 1 \\
0 & 1 & 1
\end{pmatrix} \begin{pmatrix}
\lgars[1] \\
\lgars[2] \\
\lgars[3]
\end{pmatrix} = \begin{pmatrix}
\lgars[1] + \lgars[2] \\
\lgars[1] + \lgars[3] \\
\lgars[2] + \lgars[3]
\end{pmatrix} .
\end{equation}

\noindent The Gaussian likelihood for calibrated log visibility amplitudes is then expressed as

\begin{eqnarray}
p(\lvarv | \lgarv) & = & \frac{1}{\sqrt{\text{det}\left( 2 \pi \textbf{S}_{\textbf{\textit{a}}} \right)}}
\exp\left[-\frac{1}{2} \left( \lvarv - \lvad \lgarv \right)^{\transpose} \textbf{S}_{\textbf{\textit{a}}}^{-1} \left( \lvav - \lvad \lgarv \right) \right] , \nonumber \\ \label{eqn:LVAGaussianLikelihood}
\end{eqnarray}

\noindent where we note that $\textbf{S}_{\textbf{\textit{a}}}$ contains only baseline thermal noise and is diagonal.

If we further impose independent zero-mean Gaussian priors on each of the model gain correction factors, we can express the joint prior as

\begin{equation}
p(\lgarv) = \frac{1}{\sqrt{\text{det}\left( 2 \pi \lgacov \right)}} \exp\left[-\frac{1}{2} \lgarv^{\transpose} \lgacov^{-1} \lgarv \right] .
\end{equation}

\noindent We can use this prior to marginalize the log visibility amplitudes over the log gain amplitudes, 

\begin{equation}
\mathcal{L}_a = \int p(\lvarv | \lgarv) p(\lgarv) \, \text{d}\lgarv , \label{eqn:MarginalizedLikelihoodIntegral}
\end{equation}

\noindent where $\mathcal{L}_a$ represents the marginalized likelihood.

The integrand in \autoref{eqn:MarginalizedLikelihoodIntegral} is a product of exponentials, which together contain several terms that depend on $\lgarv$.  To evaluate the integral, we would like to consolidate these terms.  By defining

\begin{eqnarray}
\textbf{M} = \textbf{M}^\transpose & \equiv & \lvad^{\transpose} \textbf{S}_{\textbf{\textit{a}}}^{-1} \lvad + \lgacov^{-1} \label{eqn:Sigma} \\
\boldsymbol{\mu} & \equiv & \lvad^{\transpose} \textbf{S}_{\textbf{\textit{a}}}^{-1} \lvarv,
\label{eqn:IntermediateMun}
\end{eqnarray}

\noindent completing the square, and then pulling terms that do not depend on $\lgarv$ out of the integral, we obtain

\begin{eqnarray}
\mathcal{L}_a & = & \frac{1}{\sqrt{\text{det}\left( 2 \pi \textbf{S}_{\textbf{\textit{a}}} \right) \text{det}\left( 2 \pi \lgacov \right)}} \nonumber \\
&& \times \exp\left[ - \frac{1}{2} \left( \lvarv^{\transpose} \textbf{S}_{\textbf{\textit{a}}}^{-1} \lvarv - \boldsymbol{\mu}^{\transpose} \textbf{M}^{-1} \boldsymbol{\mu} \right) \right] \\
&& \times \int \exp\left[ -\frac{1}{2}  \left( \lgarv - \textbf{M}^{-1} \boldsymbol{\mu} \right)^{\transpose} \textbf{M} \left( \lgarv - \textbf{M}^{-1} \boldsymbol{\mu} \right)  \right] \text{d}\lgarv . \nonumber
\end{eqnarray}

\noindent The integrand now contains only a single multivariate Gaussian in $\lgarv$, with mean $\textbf{M}^{-1} \boldsymbol{\mu}$ and covariance $\textbf{M}^{-1}$
Integrating over all $\lgarv$ thus yields the volume $\sqrt{\text{det} \big( 2 \pi \textbf{M}^{-1} \big)}$, so that

\begin{eqnarray}
\mathcal{L}_a & = & \frac{1}{\sqrt{\text{det}\left( 2 \pi \textbf{S}_{\textbf{\textit{a}}} \right) \text{det}\left( 2 \pi \lgacov \right) \text{det}\left( \textbf{M} / 2 \pi \right)}} \nonumber \\
&& \times \exp\left[ - \frac{1}{2} \left( \lvarv^{\transpose} \textbf{S}_{\textbf{\textit{a}}}^{-1} \lvarv - \boldsymbol{\mu}^{\transpose} \textbf{M}^{-1} \boldsymbol{\mu} \right) \right] . \nonumber \\ \label{eqn:IntermediateLa}
\end{eqnarray}

\noindent Upon expanding $\boldsymbol{\mu}^{\transpose} \textbf{M}^{-1} \boldsymbol{\mu}$ and directly applying the Woodbury matrix inverse identity, we obtain

\begin{eqnarray}
\mathcal{L}_a & = & \frac{1}{\sqrt{\text{det}\left( 2 \pi \textbf{S}_{\textbf{\textit{a}}} \right) \text{det}\left( 2 \pi \lgacov \right) \text{det}\left(\textbf{M} / 2 \pi \right)}} \nonumber \\
&& \times \exp\left[ - \frac{1}{2}  \lvarv^{\transpose} \left( \textbf{S}_{\textbf{\textit{a}}} + \lvad \lgacov \lvad^\transpose \right)^{-1} \lvarv \right] \label{eqn:IntermediateLa2} ,
\end{eqnarray}

\noindent where we have obtained the log visibility amplitude covariance $\lvacov = \textbf{S}_{\textbf{\textit{a}}} + \lvad \lgacov \lvad^\transpose$ analogous to \autoref{eqn:VisibilityCovarianceMatrix}.

For the determinants in the normalization constant,
\begin{eqnarray}
&& 
\text{det}\left(\textbf{M} / 2 \pi \right)
 \text{det}\left( 2 \pi \lgacov \right) 
 \text{det}\left( 2 \pi \textbf{S}_{\textbf{\textit{a}}} \right) \\
&=& \text{det}\left(
\lvad^{\transpose} \textbf{S}_{\textbf{\textit{a}}}^{-1} \lvad \lgacov + \textbf{I}\right)
 \text{det}\left( 2 \pi \textbf{S}_{\textbf{\textit{a}}} \right)\\
&=& \text{det}\left(
 \lvad \lgacov \lvad^{\transpose} \textbf{S}_{\textbf{\textit{a}}}^{-1} + \textbf{I}\right)
 \text{det}\left( 2 \pi \textbf{S}_{\textbf{\textit{a}}} \right)\\
&=& \text{det}\left( 2 \pi 
(\lvad \lgacov \lvad^{\transpose} + \textbf{S}_{\textbf{\textit{a}}} )
\right) .
\end{eqnarray}

\noindent Here, we have used
the Weinstein--Aronszajn matrix identity $\det\left( \textbf{I} + \textbf{X} \textbf{Y} \right) = \det\left( \textbf{I} + \textbf{Y} \textbf{X} \right)$. The marginalized likelihood can thus be written as
\begin{equation}
\mathcal{L}_a = \frac{1}{\sqrt{\text{det}\left( 2 \pi \lvacov \right)}} \exp\left[ - \frac{1}{2}\lvarv ^{\transpose} \lvacov^{-1} \lvarv \right] ,
\end{equation}
showing that the marginalization over Gaussian priors in log gain amplitude is fully captured through the use of visibility covariance as in \autoref{eqn:VPhaseLikelihood_app}. The derivation applies to any linear transformation of independent Gaussian observables. In particular, it is the same for partially known visibility phases under the substitution $(\lvav, \lvad, \lgav) \to (\vpv, \vpd, \gpv)$.

\onecolumngrid
\vspace{4em}
\section{Notation}

\autoref{tab:notation} lists the notation used throughout this manuscript. Vector quantities reflect values taken over a common set of recorded signals at $\gcn$ antennas. Throughout, we distinguish the measured values (no accent) with a forward model (breve accent) as well as the model residual (measured minus model values, tilde accent).

\begin{table*}[h]
\centering{}
\begin{tabular}{lccccccccccccc}
\hline \hline
 & & \multicolumn{2}{c}{Measured value} & & \multicolumn{2}{c}{Model parameter} & & \multicolumn{2}{c}{Residual} & & \multicolumn{2}{c}{Residual error} \\ \cline{3-4} \cline{6-7} \cline{9-10} \cline{12-13}
Quantity       & number & single & vector & & single & vector & & single & vector & & variance & covariance & design \\
\hline
complex gain         & \gcn  & \gcs  & \gcv  & & \gcms   & \gcmv  & & \gcrs   & \gcrv  & & \gcvar  & \gccov  & \gcdm \\
complex visibility   & \vcn  & \vcs  & \vcv  & & \vcms   & \vcmv  & & \vcrs   & \vcrv  & & 2\,\vcvar (thermal only) & \vccov & \vcd \\
\hline
gain phase           & \gpn  & \gps  & \gpv  & & \gpms   & \gpmv  & & \gprs   & \gprv  & & \gpvar  & \gpcov  & \gpd  \\
visibility phase     & \vpn  & \vps  & \vpv  & & \vpms   & \vpmv  & & \vprs   & \vprv  & & \vpvar[ij] + \gpvar[i] + \gpvar[j] & \vpcov & \vpd \\
closure phase        & \cpn  & \cps  & \cpv  & & \cpms   & \cpmv  & & \cprs   & \cprv  & & \cpvar  & \cpcov  & \cpd  \\
\hline
gain amplitude       & \gan  & \gas  & \gav  & & \gams   & \gamv  & & \gars   & \garv  & & \gavar  & \gacov  & \gad  \\
visibility amplitude & \van  & \vas  & \vav  & & \vams   & \vamv  & & \vars   & \varv  & & \vavar  & \vacov  & \vad  \\
closure amplitude    & \can  & \cas  & \cav  & & \cams   & \camv  & & \cars   & \carv  & & \cavar  & \cacov  & \cad  \\
log gain amplitude   & \lgan & \lgas & \lgav & & \lgams  & \lgamv & & \lgars  & \lgarv & & \lgavar & \lgacov & \lgad \\
log visibility amplitude & \lvan & \lvas & \lvav & & \lvams  & \lvamv & & \lvars  & \lvarv & & \lvavar[ij] + \lgavar[i] + \lgavar[j] & \lvacov & \lvad \\
log closure amplitude & \lcan & \lcas & \lcav & & \lcams  & \lcamv & & \lcars  & \lcarv & & \lcavar & \lcacov & \lcad \\
\hline
\end{tabular}
\caption{Notation used in this paper. Measured values reflect contributions from thermal noise and systematic errors. The residual error represents the expected covariance across a set of simultaneously observed residual quantities (measured minus model values). For visibilities, the residual covariance includes contributions from both thermal (statistical) and gain (systematic) errors in the general case. At high S/N, the variance for phase and log amplitude are equal, thus we use the same symbol $\vpvar \approx \vcvar / \vas$ for notational simplicity. $\vcvar$ reflects the thermal noise in one component of the complex visibility, and it is known a~priori to high precision.
}
\label{tab:notation}
\end{table*}

\label{app:notation}

\end{document}